\documentclass[aps,prb,twocolumn,superscriptaddress,longbibliography]{revtex4-2}
\usepackage[colorlinks=true,linkcolor=blue,citecolor=blue,urlcolor=blue]{hyperref}
\usepackage{amsmath,amsfonts,amssymb,color}
\usepackage{amsthm}
\usepackage{leftidx}
\usepackage{graphicx}
\usepackage{xcolor}
\usepackage{cleveref}
\usepackage{colortbl}
\usepackage{tikz}
\usepackage{multirow}
\footnotesize
\scriptsize

\usepackage{subcaption} 

\usepackage{bm}
\usepackage{epstopdf}
\usepackage{epsfig}
\usepackage{mathdots} 
\usepackage{mathrsfs} 

\def\RB{\textcolor{black}}
\def\MG{\textcolor{black}}

\usepackage{caption}
\usepackage{graphicx}
\usepackage{subcaption}
\usepackage{braket}
\usepackage{amssymb}
\usepackage{mathrsfs}
\usepackage{cleveref}
\creflabelformat{equation}{(#2#1#3)}
\crefrangeformat{equation}{(#3#1#4--#5#2#6)}
\usepackage{amssymb}

\usepackage{chngcntr}
\usepackage{float}

\usepackage{lipsum} 
\usepackage{subcaption}
\usepackage{placeins} 
\usepackage{tikz}  
\usepackage{array} 

\usepackage{multirow}  
\usepackage{array}     
\usepackage{colortbl}  
\usepackage[utf8]{inputenc}

\usepackage{titlesec}
\usepackage{lipsum} 

\newcommand{\cmark}{\ding{51}} 
\newcommand{\xmark}{\ding{55}} 
\usepackage{colortbl}
\usepackage{pifont} 

\begin{document}
	\title{Zero and Nonzero Energy Majorana Modes in an Extended Kitaev Chain} 
    
        \author{Mohammad Ghuneim}
        \email{mohammad.ghuneim@ufl.edu}
        \affiliation{Department of Physics, University of Florida, Gainesville, Florida 32611, USA
}
        
        \author{Raditya Weda Bomantara}
        \affiliation{%
        Department of Physics, Interdisciplinary Research Center for Intelligent Secure Systems, King Fahd University of Petroleum and Minerals, 31261 Dhahran, Saudi Arabia}

	\date{\today}
	
	
	\vspace{2cm}
	
\begin{abstract}
 This paper studies an extended Kitaev chain with three sublattices per unit cell. This extended version is obtained by hybridizing a modified Su-Schrieffer-Heeger model featuring trimerized unit cells with the standard Kitaev chain, resulting in a hexamer structure on the Majorana basis. Due to the interplay between the sublattice configuration and the $p$-wave superconducting pairing, a rich structure of edge modes beyond the expected Majorana zero modes is obtained. The various Majorana edge modes are further found to demonstrate considerable robustness against some generic perturbations and disorder. \RB{The presence of the non-zero Majorana edge modes in our system has the potential advantage that they could, in principle, be more unambiguously detected as compared to their zero energy counterparts, the detection of which remains an open problem. Therefore, while our system does not directly solve this open problem, it potentially offers a route to an intermediate solution that involves unambiguously detecting non-zero energy Majorana edge modes instead.} 

\end{abstract}

\maketitle
\section{Introduction}
Topological superconductivity is a paradigmatic phase of matter that arises when superconducting pairing induces characteristics of topological phenomena \cite{kitaev2001unpaired, PhysRevLett.86.268, PhysRevB.61.10267}. Characterized by their ability to host robust edge modes that are their own antiparticles, i.e., the so called Majorana zero modes (MZMs), topological superconductors have attracted significant interest due to their potential application in fault-tolerant quantum computation \cite{RevModPhys.80.1083}. Indeed, the MZMs, which are zero energy excitations that live at the boundaries of topological superconductors, may form a nonlocal qubit. Moreover, exchanging MZMs around one another, a process formally referred to as braiding, enables the execution of some quantum gate operations in a topologically protected manner \cite{lahtinen2017short}. Quantum computing with MZMs has since formed an active research area on its own \cite{freedman2003topological, PhysRevB.82.020509, sarma2015majorana, lahtinen2017short, PhysRevB.97.205404, PhysRevB.101.085401}.

Motivated by theoretical proposals \cite{PhysRevLett.105.177002, PhysRevLett.105.077001, PhysRevB.84.201308, PhysRevB.88.020407}, several experimental studies have reported some signatures of MZMs \cite{das2012zero, 10.1126/science.1222360, deng2012anomalous, PhysRevB.87.241401, nadj2014observation, manna2020signature, liu2024signatures}, though such signatures could also be attributed to other mechanisms \cite{PhysRevB.98.245407, PhysRevB.97.165302, PhysRevResearch.2.013377, PhysRevLett.109.267002, PhysRevB.96.075161, 10.21468/SciPostPhys.7.5.061}. At present, unambiguously confirming MZMs in a topological superconductor remains an experimental challenge. Nevertheless, theoretical studies of MZMs and variations of topological superconductors that host them are still actively conducted. For example, despite being Hermitian in nature, MZMs are also found to emerge in a range of non-Hermitian systems \cite{PhysRevA.93.062130, PhysRevB.110.094203, PhysRevB.106.064505,PhysRevB.101.014306}. Moreover, periodically driven systems offer a richer playground that enables a different type of Majorana edge modes to emerge in addition to MZMs \cite{PhysRevLett.106.220402, PhysRevB.87.201109, zhou2022generating, wu2023floquet, bomantara2020combating, PhysRevB.106.L060305}. It is expected that theoretically exploring these variations of topological superconductors may eventually lead to models whose MZMs could be more unequivocally detected experiments, which therefore serves as an important motivation for such studies.

In view of the above, this work presents another variation of topological superconductors that is based on the combination of the Kitaev chain \cite{kitaev2001unpaired} and an extended Su-Schrieffer-Heeger (SSH) model \cite{PhysRevB.22.2099}. The Kitaev chain is a one-dimensional (1D) model of spinless fermions, proposed as a minimal platform to demonstrate the emergence of MZMs. Despite its simplicity, the Kitaev chain often serves as a basis for developing a more sophisticated and/or realistic topological superconducting model. Recently, a diverse range of experimental efforts has been concentrated on various platforms capable of simulating Kitaev chain under appropriate conditions, including, but not limited to, quantum dots \cite{ten2025observation, bordin2025enhanced, ten2024two, dvir2023realization}, electrical circuits \cite{iizuka2023experimental}, mechanical systems \cite{allein2023strain}, and bosonic systems \cite{slim2024optomechanical, busnaina2024quantum}. 

The SSH model is one of the simplest toy models that supports a topological phase \cite{PhysRevB.22.2099}. It describes a 1D chain with nearest-neighbor coupling that alternates in amplitudes between two different values, resulting in the dimerization of the lattice structure (two sites per unit cell). While mathematically very similar to the Kitaev chain, the edge modes in the SSH model are not Majorana fermionic in nature and are thus not as useful for quantum computing purposes. Nevertheless, its simplicity has led to various extensions of the SSH model that involve longer-range coupling \cite{5rtw-ml8b, du2024one, PhysRevB.109.035114} and/or a larger unit cell \cite{lee2022winding, PhysRevB.106.085109,  ghuneim2024topological, PhysRevB.111.195424}, which often result in richer topological features. For example, an extended SSH model with three sites per unit cell supports nonzero energy edge states that are symmetrically placed about zero energy \cite{PhysRevB.106.085109, ghuneim2024topological}.

Drawing inspiration from the above studies of extended SSH models, efforts have been devoted to extending the Kitaev chain as well, leading to topological superconducting models with long-range hopping/pairing \cite{PhysRevLett.113.156402, PhysRevB.95.195160, PhysRevLett.118.267002, mishra2020disordered, PhysRevB.111.155149} and modulated structures such as dimer chains \cite{PhysRevB.90.014505, yang2025multiple}. Our work presents another extension to the Kitaev chain by modulating all system parameters in space with a period of three lattice sites. In the Majorana basis, this results in a lattice structure that exhibits six sites per unit cell. In our proposed model, the system parameters' modulation is mathematically devised as a tensor product between the regular Kitaev chain \cite{kitaev2001unpaired} and the trimerized SSH chain we recently proposed \cite{ghuneim2024topological}. By construction, the model exhibits features of long-range hopping and pairing while maintaining a relatively simple mathematical structure. In addition, it preserves fundamental symmetries, including chiral, particle-hole, and time-reversal symmetries, which are all important for its characteristics. This tapestry of particle-hole symmetry and superconducting pairing enriches the system with Majorana edge modes beyond the zero energy excitation that arises alone in the Kitaev chain and its existing extensions \cite{PhysRevLett.113.156402, PhysRevB.95.195160, PhysRevLett.118.267002, mishra2020disordered, PhysRevB.111.155149, PhysRevB.90.014505, yang2025multiple}. Moreover, the robustness of these Majorana edge modes is highlighted by their persistence under representative generic perturbations we consider below. Remarkably, we find that some of these perturbations actually enrich the system further in terms of its edge states, and therefore its topology. 

The rest of this paper is organized as follows. In Sec.~\ref{2A}, we introduce the model's Hamiltonian and set up some notations. Sec.~\ref{m} presents the momentum space description of the model, which includes its symmetry characterization, the identification of band touching points, and the construction of a suitable topological invariant. The real space description of the model is detailed in Sec.~\ref{2C}, which includes the numerical calculation of the energy spectra at varying system parameters and the verification of the system's various edge modes. In Sec.~\ref{3Exp}, we briefly discuss an experimental proposal for realizing the extended Kitaev chain. In Sec.~\ref{A5}, we consider several representative perturbations and investigate how each perturbation affects the system. Section~\ref{dis} further investigates the impact of disorder as well as its interplay with each of the previously considered perturbations. Finally, Sec.~\ref{Conc} concludes our work and outlines a number of potential research directions for future studies.

\section{Extended Kitaev chain}
\subsection{Hamiltonian}
\label{2A}
We study an extended Kitaev chain with three sites per unit cell whose Hamiltonian reads as follows:
\begin{equation}
    \mathcal{H} = H_{J_1} + H_{J_2} + H_{\delta} + H_{\Delta} + H_{\mu},
    \label{H}
\end{equation}

\begin{equation}
\begin{aligned}
H_{J_1} &= 2J_{1} \sum_{j=1}^{N} (c_{A,j}^\dagger c_{B,j} + c_{B,j}^\dagger c_{C,j} 
 +\text{\textit{h.c.}}),  \\
H_{J_2} &= 2J_{2}\sum_{j=1}^{N-1}  (c_{A,j+1}^\dagger c_{B,j} + c_{B,j+1}^\dagger c_{C,j} 
 +\text{\textit{h.c.}}), \\
 H_{\delta} &= 2i\delta \sum_{j=1}^{N} 
  (c_{B,j}^\dagger c_{A,j}^\dagger + c_{C,j}^\dagger c_{B,j}^\dagger -\text{\textit{h.c.}}),  \\
 H_{\Delta} &= 2i\Delta \sum_{j=1}^{N-1} (c_{A,j+1}^\dagger c_{B,j}^\dagger +  c_{B,j+1}^\dagger c_{C,j}^\dagger
 -\text{\textit{h.c.}}), \\
 H_{\mu} &= 2\mu \sum_{j=1}^{N} (c_{A,j}^\dagger c_{A,j} + c_{B,j}^\dagger c_{B,j} + c_{C,j}^\dagger c_{C,j}),
\end{aligned}
\label{H5}
\end{equation}
where $c_{\xi,j}^\dagger$ ($c_{\xi,j}$) is the fermionic creation (annihilation) operator at sublattice $\xi=A, B, C$ of the $j^\text{th}$ unit cell, $N$ is the number of unit cells, $J_{1}$ is the the intracell hopping parameter, and $J_{2}$ is the intercell hopping parameter. The quantities $\delta$ and $\Delta$ are the intracell and intercell pairing amplitudes, respectively. $\mu$ is the chemical potential. $c_{\xi,j}^\dagger$ and $c_{\xi,j}$ could also be written as
\begin{align*}
c_{\xi,j}^\dagger &= \frac{1}{2} \left( \gamma_{\xi_1,j} - i \gamma_{\xi_2,j} \right),\\
c_{\xi,j} &= \frac{1}{2} \left( \gamma_{\xi_1,j} + i \gamma_{\xi_2,j} \right),
\end{align*}
 where $\gamma_{\xi_1,j}$ and $\gamma_{\xi_2,j}$ are two species of Majorana operators associated with each sublattice site $\xi$. Using these two relations, the Hamiltonian of Eq.~\ref{H} could be written in terms of Majorana operators

\begin{equation}
\mathscr{H} =\mathscr{H}_{J_{1}} + \mathscr{H}_{J_{2}} + \mathscr{H}_{\delta} + \mathscr{H}_{\Delta} + \mathscr{H}_{\mu},
\label{M}
\end{equation}
\begin{equation}
\begin{aligned}
\mathscr{H}_{J_{1}} &= i J_{1} \sum_{j=1}^{N} \left( \gamma_{A_{1},j} \gamma_{B_{2},j} + \gamma_{B_{1},j} \gamma_{A_{2},j} + \gamma_{C_{1},j} \gamma_{B_{2},j} \right. \\
&\quad \left. + \gamma_{B_{1},j} \gamma_{C_{2},j}\right),\\
\mathscr{H}_{J_{2}} &= i J_{2} \sum_{j=1}^{N-1} \left( \gamma_{B_{1},j} \gamma_{A_{2},j+1} + \gamma_{A_{1},j+1} \gamma_{B_{2},j} + \gamma_{C_{1},j} \gamma_{B_{2},j+1} \right. \\
&\quad \left. + \gamma_{B_{1},j+1} \gamma_{C_{2},j} \right),\\
\mathscr{H}_{\delta} &= i \delta \sum_{j=1}^{N} \left( \gamma_{B_{1},j} \gamma_{A_{1},j} - \gamma_{B_{2},j} \gamma_{A_{2},j} + \gamma_{C_{1},j} \gamma_{B_{1},j} \right. \\
&\quad \left. - \gamma_{C_{2},j} \gamma_{B_{2},j}\right),\\
\mathscr{H}_{\Delta} &= i \Delta \sum_{j=1}^{N-1} \left( \gamma_{A_{1},j+1} \gamma_{B_{1},j} - \gamma_{A_{2},j+1} \gamma_{B_{2},j} + \gamma_{B_{1},j+1} \gamma_{C_{1},j} \right. \\
&\quad \left. - \gamma_{B_{2},j+1} \gamma_{C_{2},j} \right),\\
\mathscr{H}_{\mu} &= i \mu \sum_{j=1}^{N} \left( \gamma_{A_{1},j} \gamma_{A_{2},j} + \gamma_{B_{1},j} \gamma_{B_{2},j} + \gamma_{C_{1},j} \gamma_{C_{2},j} \right).
\end{aligned}
\label{HM}
\end{equation}

\begin{center}
\begin{figure}[htpb]
  \includegraphics[width=0.45\textwidth]{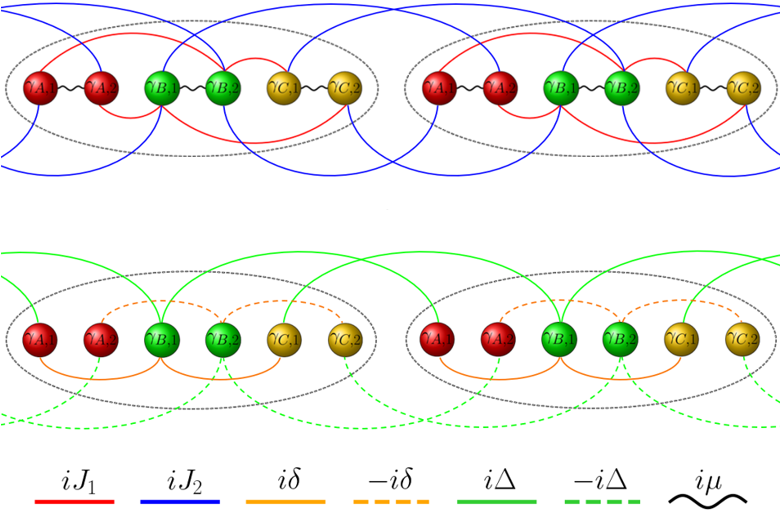}
  \captionof{figure}{Schematic diagram of two adjacent unit cells of the proposed model. Each pair of Majorana sites is represented by a different color for the sake of clarity. (Top) shows the intracell hopping, intercell hopping, and chemical potential. (Bottom) shows the intracell and intercell pairings.}
  \label{fig:figSc}
\end{figure}    
\end{center}

The effect of the various system parameters on the system's Majorana constituents is schematically depicted in Fig.~\ref{fig:figSc}. 

\begin{center}
\begin{figure}[htpb]
\includegraphics[width=0.45\textwidth, height=0.18\textheight]{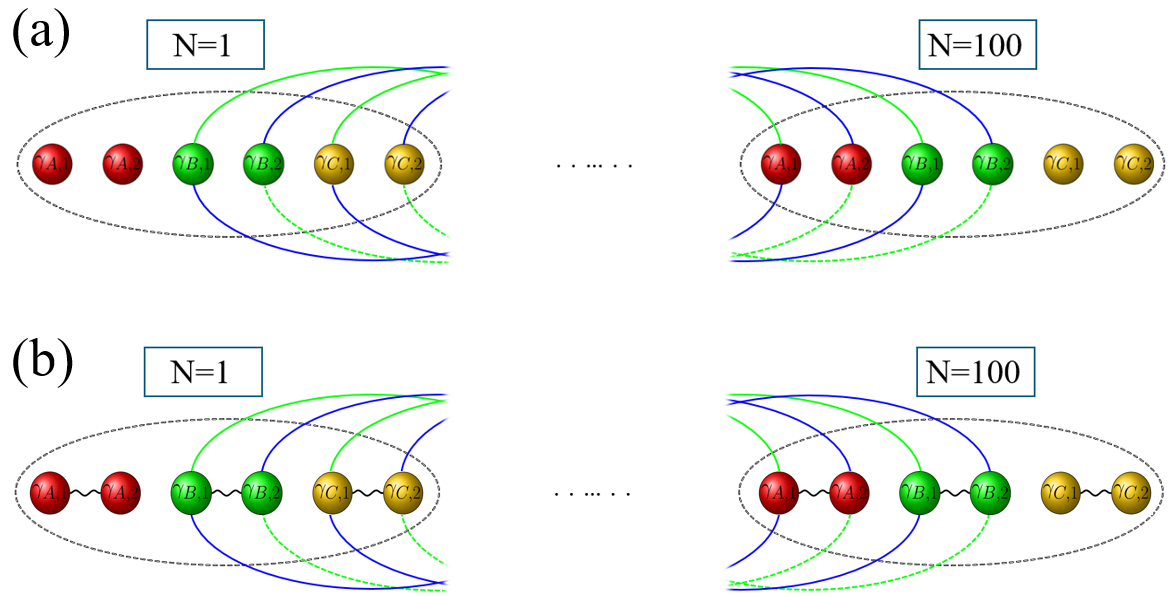}
  \captionof{figure}{\RB{(a) Schematic diagram of the first and last unit cells in a chain of N=100 unit cells at $\mu=J_1=\delta=0$. Note that there are two Majorana modes decoupled at each end of the chain. (b) Schematic diagram of the first and last unit cells in a chain of N=100 unit cells at $J_1=\delta=0$. Note that there are two Majorana modes coupled at each end of the chain, but they are disconnected from the rest of the chain.}}
  \label{fig:fig2MZM}
\end{figure}    
\end{center}

\RB{To intuitively demonstrate the presence of Majorana modes in our system, we first consider the special parameter values $\mu=J_1=\delta=0$. As schematically depicted in Fig.~\ref{fig:fig2MZM}(a), a pair of Majorana fermions at each edge are decoupled from each other and from the rest of the chain, which thus correspond to MZMs. However, by tuning $\mu$ to a finite value while still keeping $J_1=\delta=0$ (see Fig.~\ref{fig:fig2MZM}(b)), the two disconnected MZMs at each edge are now coupled with each other, though they are still decoupled from the rest of the chain. In this case, the linear combination of the two leftmost Majorana operators, i.e., $\gamma_L=a \gamma_{A_1,1}+ b \gamma_{A_2,1}$, is closed under commutator with the system’s Hamiltonian. The equation $[H,\gamma_L]= E \gamma_L$ can then be rewritten as an eigenvalue equation involving a $2\times 2$ matrix, the two solutions of which correspond to non-zero energy edge modes. Therefore, like the regular Kitaev chain, special parameter values exist in which the Majorana edge modes are easily tractable. However, while the regular Kitaev chain only supports MZMs, our model is expected to support non-zero energy edge modes, as we also numerically confirm later on.}


\subsection{Momentum-space analysis and topological invariant}
\label{m}

It is instructive to rewrite the Hamiltonian in momentum space, which enables a better understanding of the band structure and a clearer analysis of the topological properties. The momentum-space representation of Eq.~\eqref{H} takes the form:

\begin{align}
    \mathcal{H} &= \sum_k \psi_k^\dagger H(k) \psi_k , \nonumber \\
    H(k) &= (\mu + (J_1 + J_2 \cos(k)) S_x + J_2 \sin(k) S_y)\tau_{z} \notag \\
            &\quad + ((\delta -\Delta \cos(k)) S_y + \Delta \sin(k) S_x)\tau_x,
    \label{eq:eqH(k)}
\end{align}
where $\psi_k=(c_{A,k},c_{B,k},c_{C,k},c_{A,-k}^\dagger,c_{B,-k}^\dagger,c_{C,-k}^\dagger)^T$. $\tau_{\eta=x,y,z}$ is the well-known Pauli matrices, $S_x$ and $S_y$ are defined as
\[
S_x = \left( \begin{smallmatrix} 
0 & 1 & 0 \\ 
1 & 0 & 1 \\ 
0 & 1 & 0 
\end{smallmatrix} \right), 
\quad
S_y = \left( \begin{smallmatrix} 
0 & -i & 0 \\ 
i & 0 & -i \\ 
0 & i & 0 
\end{smallmatrix} \right).
\]
The fermionic operators in $\psi_k$ are related to their real-space counterparts as follows:
\[
\begin{aligned}
c_{\xi,k} &= \sum_j e^{-ikj} c_{\xi,j}, \quad
c_{\xi,k}^\dagger = \sum_j e^{ikj} c_{\xi,j}^\dagger, \\
c_{\xi,-k} &= \sum_j e^{ikj} c_{\xi,j}, \quad
c_{\xi,-k}^\dagger = \sum_j e^{-ikj} c_{\xi,j}^\dagger.
\end{aligned}
\]

Before proceeding, it is helpful first to identify the symmetries of the system, which are key to understanding its behavior. We find that our model preserves chiral, particle-hole, and time-reversal symmetries, which are expressed as
\begin{equation}
    \begin{aligned}
        \mathcal{C}^{-1} H(k) \mathcal{C} &= -H(k), \\
        \mathcal{P}^{-1} H(k) \mathcal{P} &= -H(-k), \\
        \mathcal{T}^{-1} H(k) \mathcal{T} &= H(-k),
    \end{aligned}
\end{equation}
where the operators $\mathcal{C}$, $\mathcal{P}$, and $\mathcal{T}$ are given as ($\mathcal{K}$ is the complex conjugation operator)
\begin{equation}
  \begin{aligned}
    \mathcal{C} &= \tau_y \otimes \mathbb{I}_{3\times3}, \\
    \mathcal{P} &= (\tau_x \otimes \mathbb{I}_{3\times3}) \mathcal{K}, \\
    \mathcal{T}&=\mathcal{C}\mathcal{P},
  \end{aligned}
\end{equation}
we emphasize that $\mathcal{C}$ is a unitary operator, whereas $\mathcal{P}$ and $\mathcal{T}$ are antiunitary as they involve the complex conjugation operator.

\begin{figure}[htpb]
    \begin{subfigure}[l]{0.14\textwidth}  \includegraphics[width=\linewidth]{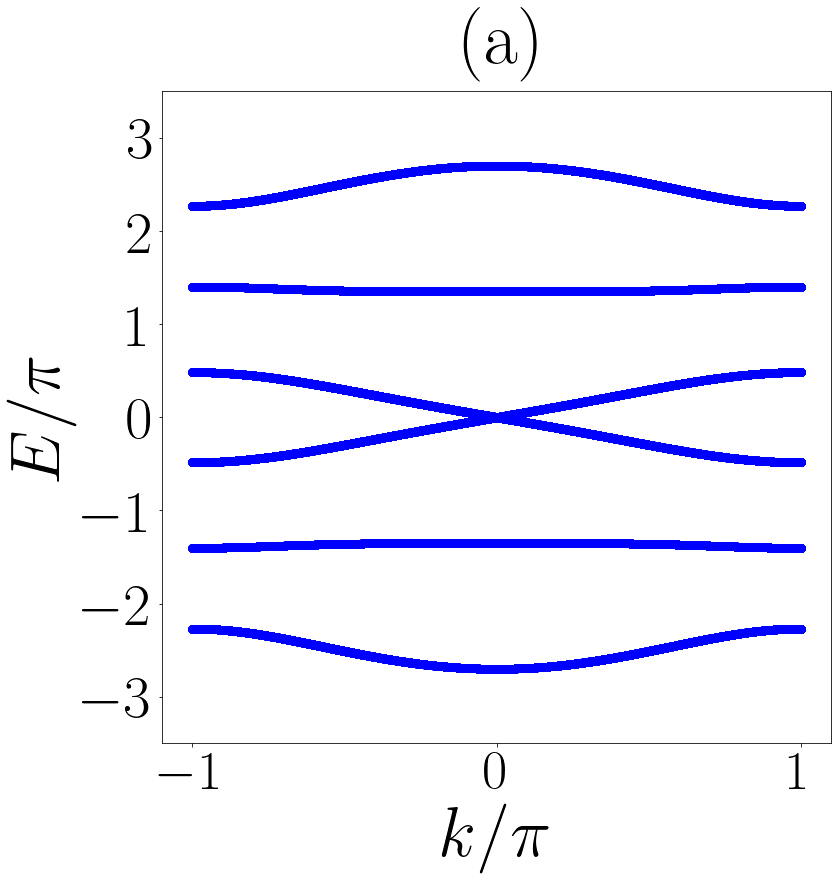}     
    \end{subfigure}%
    \hspace{-0.1cm} 
    \begin{subfigure}[c]{0.125\textwidth}
    \includegraphics[width=\linewidth]{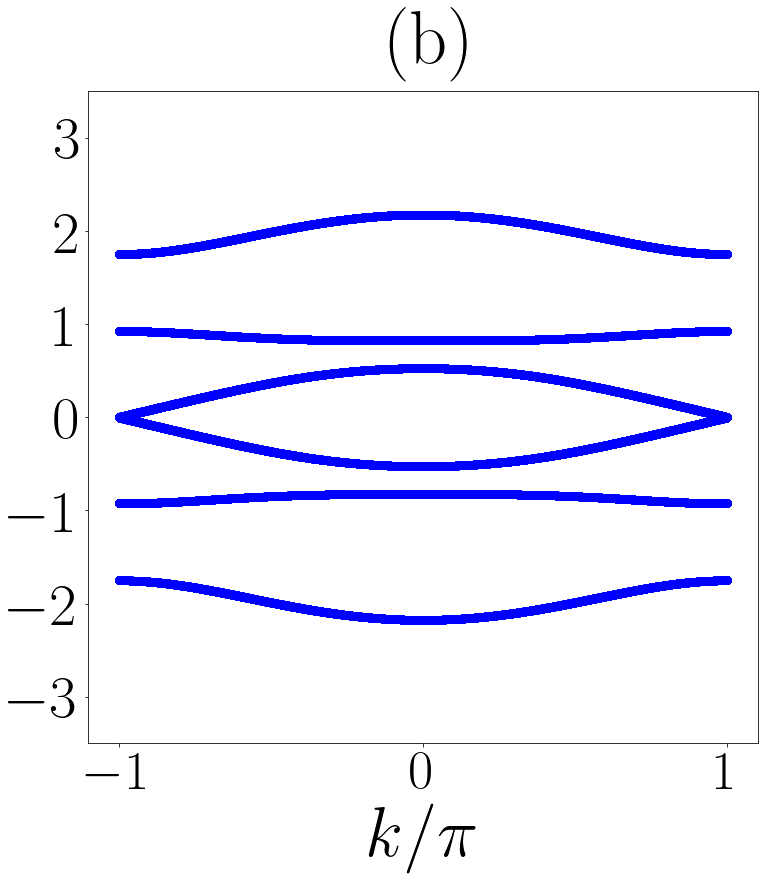}
    \end{subfigure}%
    \hspace{-0.1cm} 
    \begin{subfigure}[r]{0.125\textwidth}
      \includegraphics[width=\linewidth]{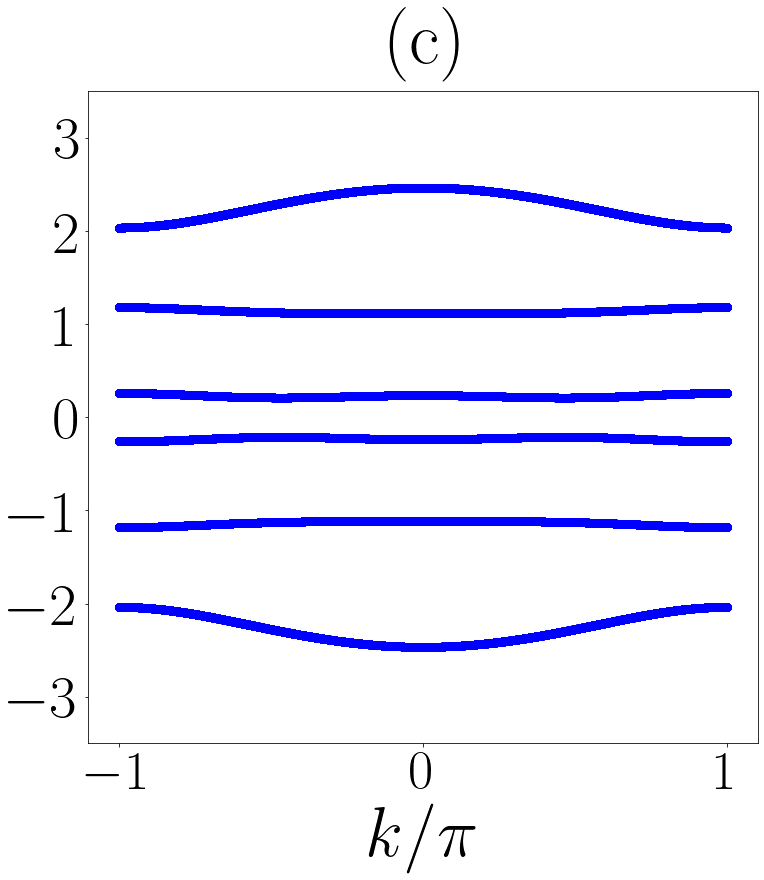}        
    \end{subfigure}
    \caption{Some representative energy spectra corresponding to Eq.~(\ref{eq:eqH(k)}) at $J_1 = 0.5$, $J_2 = 2.5$, and $\delta =\Delta = 0.4$. (a) is at $\mu= 4.24$. (b) is at $\mu=2.59$. (c) at is $\mu=3.5$.}
    \label{fig:fig3}
\end{figure}

Being a combination of the regular Kitaev chain and a trimerized extended SSH model, it is expected that our model supports zero and non-zero energy Majorana edge modes. The emergence and disappearance of these edge modes are followed by an appropriate band gap closing that represents a topological phase transition. In particular, the zero energy Majorana modes (MZMs) are characterized by a band gap closing at zero energy. Such a topological phase transition could be identified by solving
\[
H \ket{\psi} = 0,
\]
this offers a straightforward way for identifying the MZMs. Due to the complexity of dealing with a $6 \times 6$ matrix Hamiltonian (Eq.~\ref{eq:eqH(k)}), we take into account the points of high symmetry, that is, at $k = 0$ and $k = \pi$. This yields the following two conditions for band touching at $E=0$, i.e.,

\begin{eqnarray}
    \mu &=& \sqrt{2 \left( (J_1 + J_2)^2 - (\delta - \Delta)^2 \right)}, \quad \text{at } k = 0 \nonumber , \\ 
    \mu &=& \sqrt{2 \left( (J_1 - J_2)^2 - (\delta + \Delta)^2 \right)}, \quad \text{at } k = \pi
\label{mu}.
\end{eqnarray}
Figure~\ref{fig:fig3} shows some representative energy spectra in momentum space. When the system parameters satisfy Eq.~(\ref{mu}), band touching points at $k=0$ and $k=\pm \pi$, respectively, are observed (see Figs.~\ref{fig:fig3}(a) and~\ref{fig:fig3}(b)). In contrast, at arbitrary parameter values, the system exhibits a fully gapped spectrum (see Fig.~\ref{fig:fig3}(c)).

To identify the topological nature of the MZMs, we now define and compute the associated winding number that exploits the chiral symmetry of the system. To this end, we first apply a unitary transformation that rotates the basis of Pauli matrices such that $\tau_x \rightarrow \tau_y$ and $\tau_z \rightarrow \tau_x$; this transforms the Hamiltonian of Eq.~(\ref{eq:eqH(k)}) into the off-diagonal matrix form 
\[
\tilde{H}(k) =
\begin{bmatrix} 
0 & h \\ 
h^\dagger & 0 
\end{bmatrix},
\]
where
\[
h =
\left(\begin{array}{ccc}
\scriptstyle \mu & \scriptstyle J_1-\delta +(J_2 + \Delta)\,e^{-ik} & \scriptstyle 0\\ 
\scriptstyle J_1+\delta +(J_2-\Delta)\,e^{ik} & \scriptstyle \mu & \scriptstyle J_1-\delta +(J_2 + \Delta)\,e^{-ik} \\
\scriptstyle 0 & \scriptstyle J_1+\delta+(J_2-\Delta)\,e^{ik} & \scriptstyle \mu
\end{array}\right).
\]


In this basis, the winding number $\mathcal{W}$ can be calculated via the following expression:
\begin{align}
\mathcal{W} &= \frac{1}{2 \pi i} \int_{-\pi}^{\pi} 
\text{Tr} \left( h^{-1}(k)\, 
\frac{\partial h(k)}{\partial k} \right) dk \notag \\
&=
\begin{cases}
\pm 1, &   \text{MZMs are present} \\
0    , &   \text{MZMs are absent}
\end{cases},
\label{eq:eqw}
\end{align}
To demonstrate the validity of the above winding number, we shall analytically evaluate $\mathcal{W}$ at special parameter values of $J_1=\delta$ and $J_2=\Delta$. In this case, $\mathcal{W}$ takes the explicit form 
\[
\mathcal{W} = \frac{1}{2\pi i} \oint \frac{1}{  z - \frac{\mu^2}{8 J_1 J_2}}  dz ,
\]
which can be readily evaluated via residue theorem to yield
\begin{align}
    \mathcal{W} = 
\begin{cases}
\pm1 & \text{for } |\mu| < \sqrt{8|J_1 J_2|} \\
0 & \text{for } |\mu| > \sqrt{8|J_1 J_2|} 
\end{cases}.
\end{align}
Note that this is consistent with the band touching conditions of Eq.~(\ref{mu}), which now read $\mu=\sqrt{8|J_1 J_2|}$ at the special parameter values under consideration. In the next section, we further confirm that the numerically calculated winding number above indeed agrees with the existence of MZMs at more general parameter values.

\MG{ As the preceding discussion only presents an invariant that captures the presence/absence of the MZMs, one may wonder if the non-zero energy Majorana edge modes can be characterized by a similar invariant. Unfortunately, developing such an invariant is a highly nontrivial task. In principle, a suitable topological invariant for the non-zero energy Majorana edge modes could be systematically constructed via some transfer matrix method, such as in the spirit of Refs.~\cite{PhysRevLett.110.146404, lima2025interplay}. In these studies, the transfer matrix method was applied to identify MZMs by solving some matrix equation that yields localized, zero energy solutions \cite{PhysRevLett.110.146404, lima2025interplay}. However, adapting the same method to identifying the non-zero energy edge modes in our model is significantly more challenging due to the $6\times 6$ matrix involved here as compared with the $2\times 2$ matrix in previous studies. Moreover, most importantly, the non-zero energy edge modes lack an exact, closed-form expression for their energy within the bulk gap, which is required to extract the conventional transfer matrix. These aspects render any analytical calculation extremely difficult, if not impossible, to carry out. Therefore, the topological nature of the non-zero energy edge modes will be established through another means, i.e., by directly computing their spatial profiles in the following section and comparing them from the expected profiles at special parameter values.}


\subsection{Real-space analysis}
\label{2C}
\begin{center}
\begin{figure}[htpb]
  \includegraphics[width=0.48\textwidth, height=0.5\textheight]{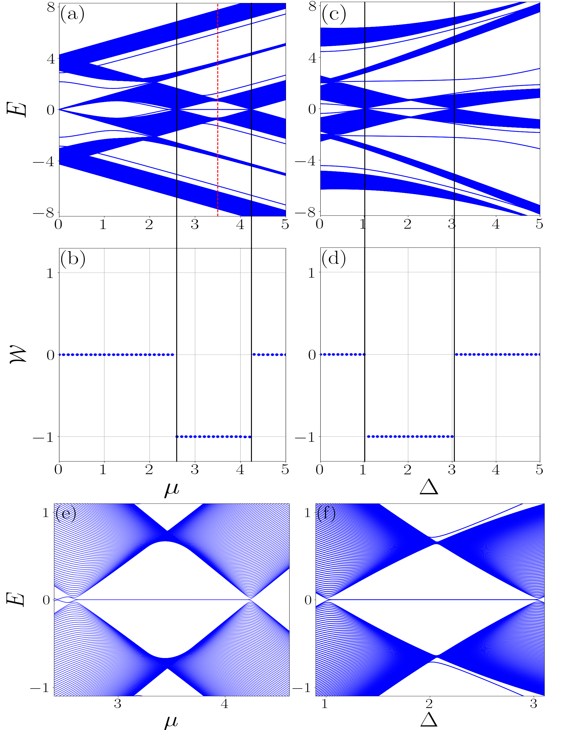}
  \captionof{figure}{(a) and (c) are the energy spectrum as a function of $\mu$ and $\Delta$, respectively, for \MG{$N=100$} unit cells. (b) and (d) show the numerically calculated winding number associated with (a) and (c), respectively. (a) and (b) are evaluated at $J_1 = 0.5$, $J_2 = 2.5$, and $\delta=\Delta=0.4$. (c) and (d) are evaluated at $J_1 = 0.5$, $J_2 = 2.5$, $\delta=0.4$, and $\mu=2$. \MG{(e) and (f) show zoomed-in views of the MZMs from  (a) and (c), respectively.} The black vertical lines mark the locations of the band touching points anticipated by Eq.~(\ref{mu}). The red vertical dashed line indicates the value used in Fig.~\ref{fig:figwf}. }
  \label{fig:figw}
\end{figure}    
\end{center}

We start this section by presenting the energy spectra of the system in the real space. This approach yields direct observation of edge-localized states and their spatial profiles. In Fig.~\ref{fig:figw}(a), we show the real-space energy spectrum, obtained by diagonalizing Eq.~(\ref{H}) under open boundary conditions (OBCs). The figure clearly demonstrates the presence of multiple nonzero energy edge modes at generic parameter values, which come in symmetric pairs at ($E$, $-E$) around zero energy due to the chiral symmetry of the system. Moreover, at some range of $\mu$ values that yield $\mathcal{W}=\pm 1$ (see Fig.~\ref{fig:figw}(b)), MZMs additionally emerge alongside these nonzero edge modes. Such MZMs are the main characteristics of the regular Kitaev chain in the topologically nontrivial regime \cite{kitaev2001unpaired}. On the other hand, nonzero energy edge modes are usually found in extended SSH models with at least three sites per unit cell \cite{ghuneim2024topological}.

That both types of edge modes coexist in our system is a consequence of our construction that mathematically combines the two models. It is also worth noting that, unlike the extended SSH model of Ref.~\cite{ghuneim2024topological}, the nonzero edge modes in our system are also Majorana in nature like their zero energy counterparts. Therefore, such nonzero edge modes may potentially contribute to quantum computing applications or, at the very least, supplement the current experimental efforts in unambiguously detecting Majorana modes.     

\RB{A fundamental characteristic of the Kitaev chain is the emergence of MZMs within the topological phase, which exists for chemical potentials $|\mu|<2|t|$, where $t$ is the hopping parameter. A topological phase transition occurs precisely at $|\mu|=2|t|$, where the pairing parameter does not affect the transition point, as long as it is nonzero. Since there is only a single phase transition point, the MZMs emerge at starting point of $\mu=0$, marking the boundary of this phase, i.e., from $\mu=0$ to $|\mu|=2|t|$. Unlike the regular Kitaev chain, the extended model presented in this paper exhibits a richer phase diagram, characterized by two distinct topological phase transition points, as defined by Eq.~(\ref{mu}). This structure introduces a critical advantage. That is, the region supporting MZMs is not fixed to a specific range of $\mu$. Instead, both of the phase transition points can be shifted to different ranges of $\mu$ and tuned by adjusting the system’s coupling strengths, as the phase transition points depend on all system’s parameters (see Eq.~(\ref{mu})). This feature is absent in the regular Kitaev chain, where the onset of MZMs is fixed at $\mu=0$. 
}

\MG{ The primary experimental signature for MZMs is a zero-bias conductance peak in tunneling spectroscopy \cite{doi:10.1126/science.1222360}. 
At the zero-bias peak, trivial excitations such as the Andreev-bound state \cite{PhysRevLett.109.267002, PhysRevB.96.075161,10.21468/SciPostPhys.7.5.061} and the Kondo effect \cite{PhysRevLett.109.186802, PhysRevB.91.081405} can arise. Such trivial excitations induce zero-bias anomalies that strongly resemble the tunneling signatures often associated with MZMs, posing an experimental challenge to their identification. Such trivial excitations are also expected to take place at finite energy. Likewise, their presence may hamper the detection of the non-zero edge modes. The non-zero edge modes in our system appear as finite-bias peaks in tunneling spectroscopic measurements, and the characteristics of these peaks depend on the system's parameters. Unlike MZMs, which are uniquely identified by a zero-bias peak, the finite-energy Majorana modes provide experimental flexibility by means of their pair of peaks, in principle, that occur at energy $\pm E$ within the superconducting gap. A key strategy is to take advantage of the system's parameter dependency. That is, varying the system parameters will shift or eliminate peaks of trivial origin, while the pair of peaks from finite-energy Majorana modes should evolve coherently and persist across the range of parameters chosen. This enables tracking the non-zero Majorana modes through several experiments across numerous parameter regimes. It is expected that this means of unambiguously detecting non-zero-energy Majorana modes provides an alternative, potentially more practical, pathway to exploring Majorana physics before MZMs unambiguous detection is possible.}


\begin{figure}[htpb]
    \centering
    \begin{subfigure}[b]{0.45\textwidth}
        \includegraphics[width=\textwidth]{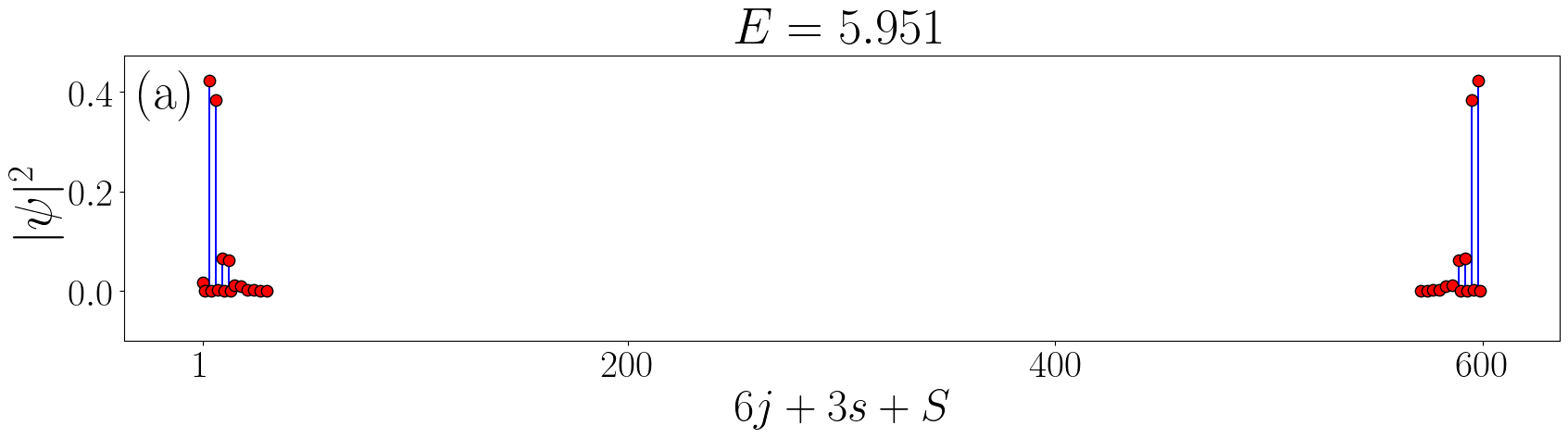} 
        \label{fig:fig1a}
    \end{subfigure}
    \begin{subfigure}[b]{0.45\textwidth}
        \includegraphics[width=\textwidth]{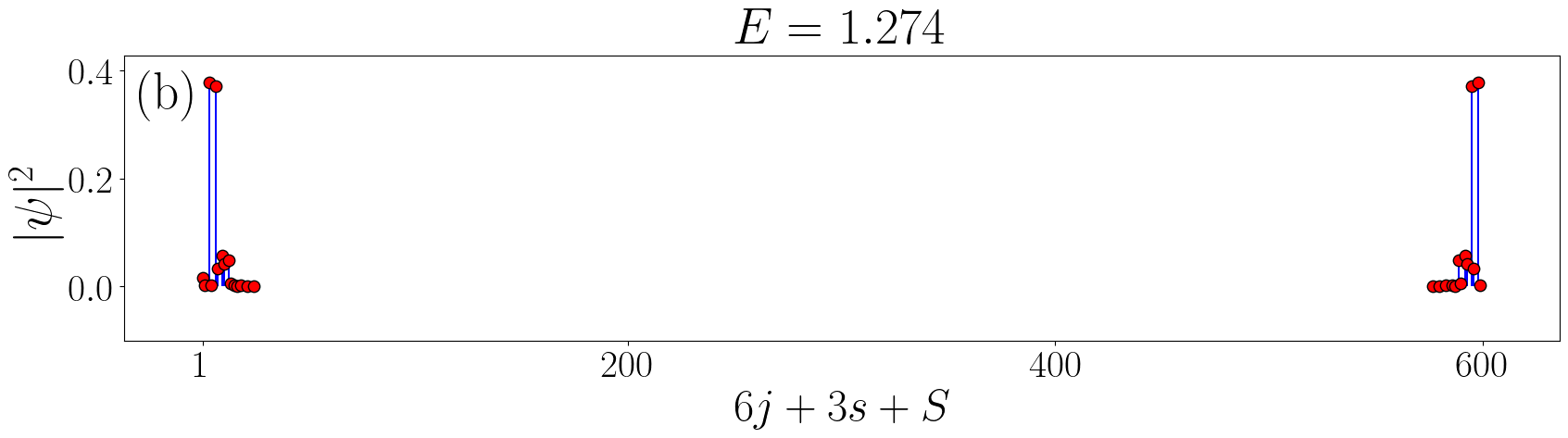}
        \label{fig:fig1b}
    \end{subfigure}
    \begin{subfigure}[b]{0.45\textwidth}
        \includegraphics[width=\textwidth]{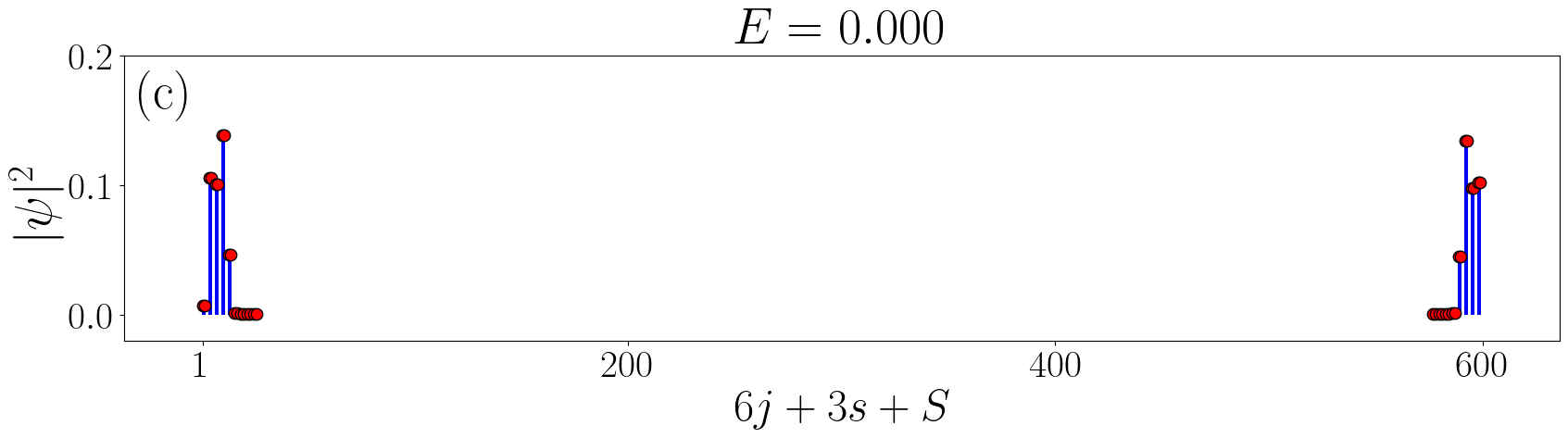} 
        \label{fig:fig1c}
    \end{subfigure}
    \begin{subfigure}[b]{0.45\textwidth}
        \includegraphics[width=\textwidth]{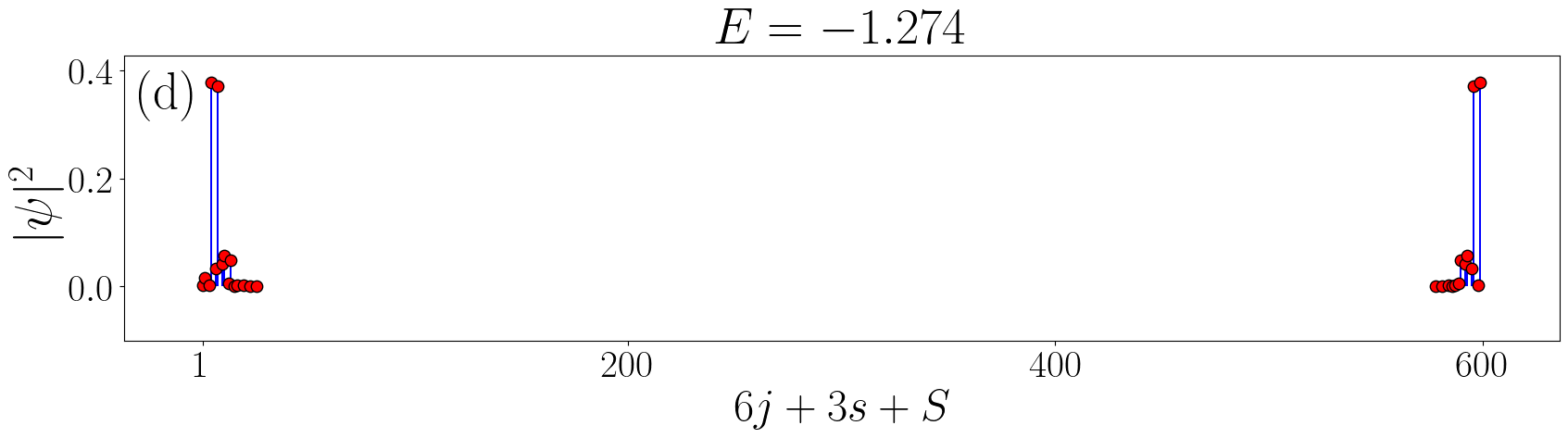}
        \label{fig:fig1d}
    \end{subfigure}
    \begin{subfigure}[b]{0.45\textwidth}
        \includegraphics[width=\textwidth]{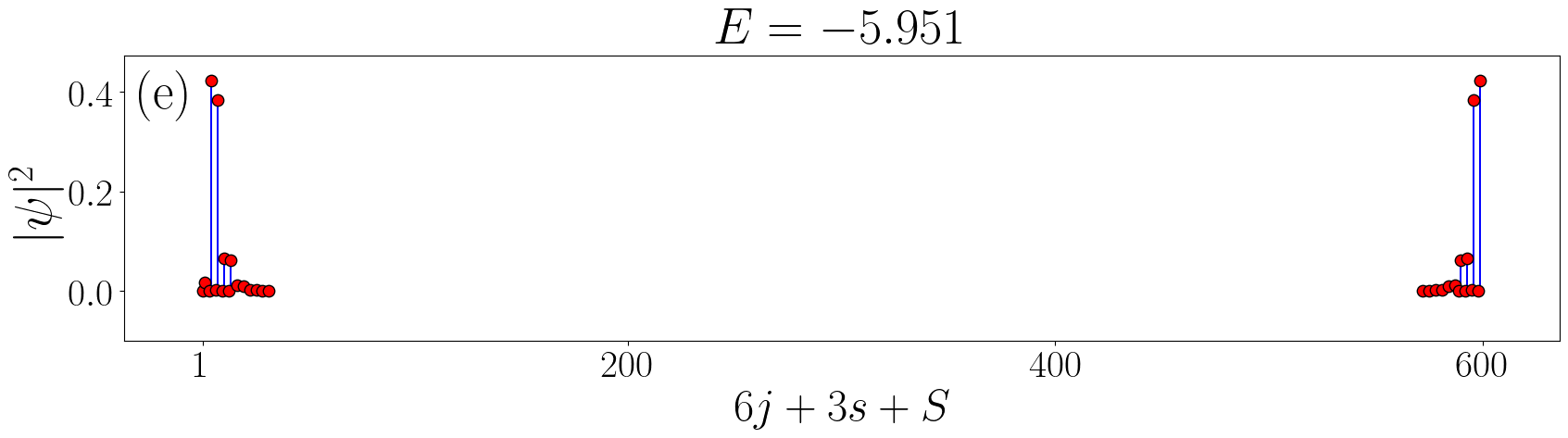}
        \label{fig:fig1e}
    \end{subfigure}
    \caption{Wave function profiles associated with the system's edge states for \MG{$N=100$} unit cells at $J_1 = 0.5$, $J_2 = 2.5$, $\delta=\Delta=0.4$, and $\mu=3.5$. Note that each subplot displays one left-localized edge state and one right-localized edge state.}  
  \label{fig:figwf}
\end{figure}

To gain additional insight into the nature of each Majorana edge mode observed in Fig.~\ref{fig:figw}(a), we show its corresponding wavefunction profile in Fig.~\ref{fig:figwf} at fixed parameter values. There, the wavefunction is defined as
\begin{equation}
    \left| \psi \right|^2 (6j+3s+S) = \sum_{j=0}^{N-1} \sum_{s = 0, 1} \sum_{S = 1,2,3}  |\langle j+1, s, S |\psi \rangle|^2, \label{wfprof}
\end{equation}
where $s \in \{0, 1\}$, with $s = 0$ for $c_{\xi,j}$ and $s = 1$ for $c^\dagger_{\xi,j}$, and $S \in \{1, 2, 3\}$ corresponding to sublattices $A$, $B$, and $C$, respectively. 

Intuitively, Eq.~(\ref{wfprof}) is defined such that each combination of the sublattice, particle-hole, and lattice site degrees of freedom is assigned a unique integer, thereby allowing us to obtain the complete components of each edge mode. 
Figure~\ref{fig:figwf} reveals that while all Majorana edge modes are highly localized, the nonzero energy edge modes only have two dominating peaks near a system's edge (see Fig.~\ref{fig:figwf}(a,b,d,e)). \RB{This feature is consistent with the intuitive picture of the model at special parameter values $J_1=\delta=0$ previously elucidated in Sec.~\ref{2A}, where the first and last two Majorana operators hybridize with each other (while still decoupled from the rest of the chain) and form non-zero energy Majorana edge modes. That the significant two peaks are also observed away from the special parameter values highlights the topological nature of the non-zero energy edge modes.} By contrast, MZMs are found to have significant support on more degrees of freedom near a system's edge. The considerably distinct features of the different types of edge modes in principle enable their separate experimental detection. \RB{To further ensure that these characteristics do not depend on the number of lattice sites in the system, we show the wave function profiles of the edge modes at different system sizes in Appendix~\ref{app:A}.}

\section{Potential experimental realization}
\label{3Exp}
Despite the ongoing challenges in uniquely identifying MZMs, numerous experimental efforts have successfully observed their expected signatures \cite{das2012zero, 10.1126/science.1222360, deng2012anomalous, PhysRevB.87.241401, nadj2014observation, manna2020signature, liu2024signatures}. Recently, quantum dot arrays provide a controllable platform that is expected to host MZMs \cite{ten2025observation, bordin2025enhanced, ten2024two, dvir2023realization}. In this case, a number of quantum dots can be arranged in a one-dimensional chain, with each site in the system formed by a quantum dot with adjustable energy and tunnel coupling. The hopping parameter between the adjacent quantum dots is implemented by a tunneling coupler, which provides a tunable tunnel barrier. Controlling the voltage applied to this barrier allows one to regulate the amplitude of electron tunneling between the quantum dots, thereby deciding the hopping strength \cite{ten2025observation}. Implementing the pairing parameter in quantum dot arrays is also achievable by connecting each dot to a conventional $s$-wave superconductor \cite{deng2016majorana}. This approach allows superconducting correlations to be induced by the proximity effect. Although the parent superconductor provides $s$-wave pairing, the system effectively transforms to exhibit $p$-wave pairing when a combined effect of spin-orbit coupling and an external Zeeman field is applied to the system, which together splits the spin degeneracy \cite{deng2016majorana}. The chemical potential in each quantum dot can be reliably controlled through gate voltages that are capacitively coupled to the dots, which allows local tweaking of the onsite energies \cite{dvir2023realization}. Finally, a robust zero-bias conductance peak in this setup could be taken as a possible signature of the MZMs \cite{das2012zero, 10.1126/science.1222360, deng2012anomalous, PhysRevB.87.241401, nadj2014observation, manna2020signature, liu2024signatures, ten2025observation, bordin2025enhanced, ten2024two, dvir2023realization}, whilst appropriate non-zero-bias peaks could further result from the presence of nonzero energy Majorana edge modes.

The extended Kitaev model presented in this paper is experimentally accessible with minimal adjustments to the current experimental geometry setups of quantum dot systems \cite{ten2025observation, bordin2025enhanced, ten2024two, dvir2023realization} and other platforms \cite{iizuka2023experimental, allein2023strain, slim2024optomechanical, busnaina2024quantum}. In particular, long-range hopping and pairing could be efficiently implemented as nearest-neighbor hopping and pairing by modifying the spatial arrangement of the lattice sites into a ladder-like structure \cite{ghuneim2024topological}.

\section{Discussion}
\subsection{Effects of perturbation}
\label{A5}
The presence and robustness of the edge modes in a noninteracting topological system are usually protected by specific symmetries. To uncover the role of each symmetry in our system, we now consider the effect of some representative perturbations that break and preserve some of these symmetries. To this end, we separately add each of the following terms to the momentum-space Hamiltonian of Eq.~(\ref{eq:eqH(k)}), i.e.,

\begin{eqnarray}
    \alpha(k) \equiv   \tau_z \otimes \alpha_{\text{o}}\scalebox{0.6}{$\begin{pmatrix} 0 & -1 & 0 \\ -1 & 0 & 0 \\ 0 & 0 & 0 \end{pmatrix}$} &,& \beta(k) \equiv \tau_z \otimes\beta_{\text{o}} \scalebox{0.6}{$\begin{pmatrix} 0 & e^{-ik} & 0 \\ e^{ik} & 0 & 0 \\ 0 & 0 & 0 \end{pmatrix}$}  , \nonumber \\
    \zeta(k) \equiv \tau_y \otimes\zeta_{\text{o}} \scalebox{0.6}{$\begin{pmatrix} 0 & -i & 0 \\ i & 0 & 0 \\ 0 & 0 & 0 \end{pmatrix}$}  &,& 
    \vartheta(k) \equiv \tau_y \otimes \vartheta_{\text{o}} \scalebox{0.6}{$\begin{pmatrix} 0 & -i\,e^{-ik} & 0 \\ i\,e^{ik} & 0 & 0 \\ 0 & 0 & 0 \end{pmatrix}$},
    \nonumber \\
    \varrho(k) \equiv \tau_z \otimes \varrho_{\text{o}}\scalebox{0.6}{$\begin{pmatrix} 1 & 0 & 0 \\ 0 & -1 & 0 \\ 0 & 0 & -1 \end{pmatrix}$} ,
\end{eqnarray}
where $\alpha_{\text{o}}$, $\beta_{\text{o}}$, $\zeta_{\text{o}}$, $\vartheta_{\text{o}}$ and $\varrho_{\text{o}}$ represent the corresponding perturbation strengths. The perturbations $\alpha(k)$, $\beta(k)$, and $\varrho(k)$ preserve all chiral, time-reversal, and particle-hole symmetries. Meanwhile, $\zeta(k)$ and $\vartheta(k)$ break chiral and time-reversal symmetries while preserving particle-hole symmetry. The symmetries preserved and broken by these perturbations are also summarized in Table~\ref{tbl:tbl1}.

\begin{table}[ht]
    \centering
    \resizebox{0.35\textwidth}{!}{
    \begin{tabular}{|>{\columncolor{gray!10}}c|c|c|c|c|c|}
        \hline
         \textbf{Symmetry} & $\alpha(k)$ & $\beta(k)$ & $\zeta(k)$ & $\vartheta(k)$ & $\varrho(k)$\\ 
        \hline
        Chiral & \cmark & \cmark & \xmark & \xmark & \cmark \\ \hline
         Particle-hole & \cmark & \cmark & \cmark & \cmark & \cmark  \\ \hline
         Time-reversal & \cmark & \cmark & \xmark & \xmark & \cmark  \\ \hline
    \end{tabular}}
    \caption{A summary of the symmetries preserved and broken by each perturbation operator discussed in Sec.~\ref{A5}}
    \label{tbl:tbl1}
\end{table}

Physically, each of the first four perturbations is designed to introduce imbalance/asymmetry in one of the system parameters over each unit cell throughout the chain. Specifically, the perturbations $\alpha(k)$ and $\beta(k)$ introduce asymmetries in the intracell and intercell hopping parameters, $J_1$ and $J_2$, respectively. Meanwhile, $\zeta(k)$ and $\vartheta(k)$ introduce asymmetries in the intracell and intercell pairing parameters, $\delta$ and $\Delta$, respectively. The term $\varrho(k)$ modifies the chemical potential $\mu$. The corresponding real space expressions of the aforementioned perturbations are obtained as
\begin{eqnarray}
\alpha &=& 2\alpha_{\text{o}} \sum_{j=1}^{N} (c_{A,j} c_{B,j}^\dagger + h.c.),  \nonumber \\
\beta &=& 2\beta_{\text{o}} \sum_{j=1}^{N-1} (c_{A,j+1}^\dagger c_{B,j} + h.c.),  \nonumber \\
\zeta &=& 2\zeta_{\text{o}} \sum_{j=1}^{N} (c_{B,j}^\dagger c_{A,j}^\dagger + h.c.),  \nonumber \\
\vartheta &=& 2\vartheta_{\text{o}} \sum_{j=1}^{N-1} (c_{B,j}^\dagger c_{A,j+1}^\dagger + h.c.),  \nonumber \\
\varrho &=& 2\varrho_{\text{o}} \sum_{j=1}^{N} (c_{A,j}^\dagger c_{A,j} - c_{B,j}^\dagger c_{B,j} - c_{C,j}^\dagger c_{C,j}). \nonumber \\
\label{eq:eq9}
\end{eqnarray}
Finally, it should be noted that while the parameters for intracell and intercell pairing are originally multiplied by the imaginary component $i$ (see Eq.~(\ref{H})), the inclusion of the real functions $\zeta$ and $\vartheta$ has the additional effect of making the effective pairing parameters complex valued.

\begin{center}
\begin{figure*}[htbp]
    \begin{subfigure}{0.1665\textwidth}
        \includegraphics[width=\textwidth]{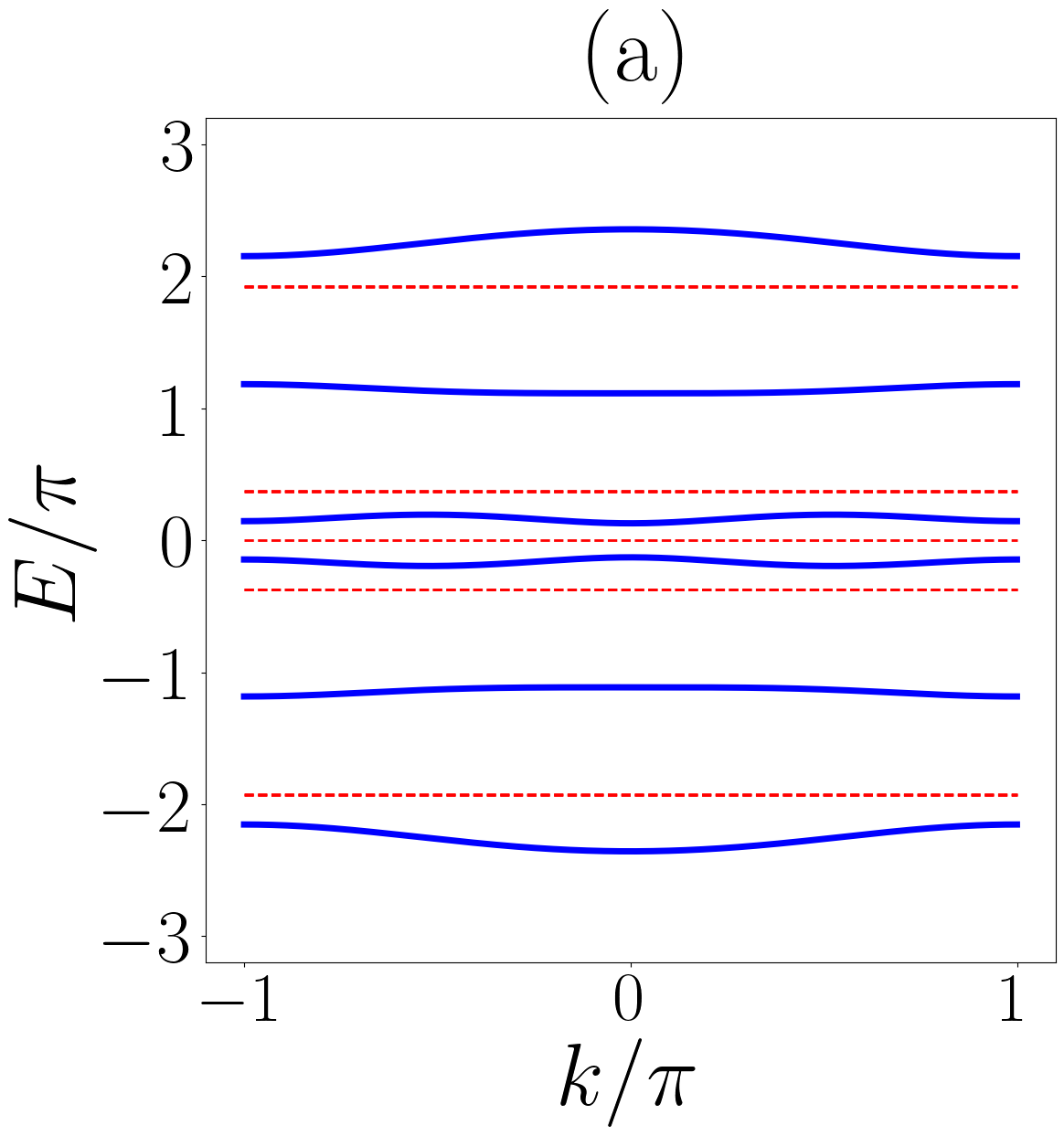}
    \end{subfigure}
    \begin{subfigure}{0.15\textwidth}
        \includegraphics[width=\textwidth]{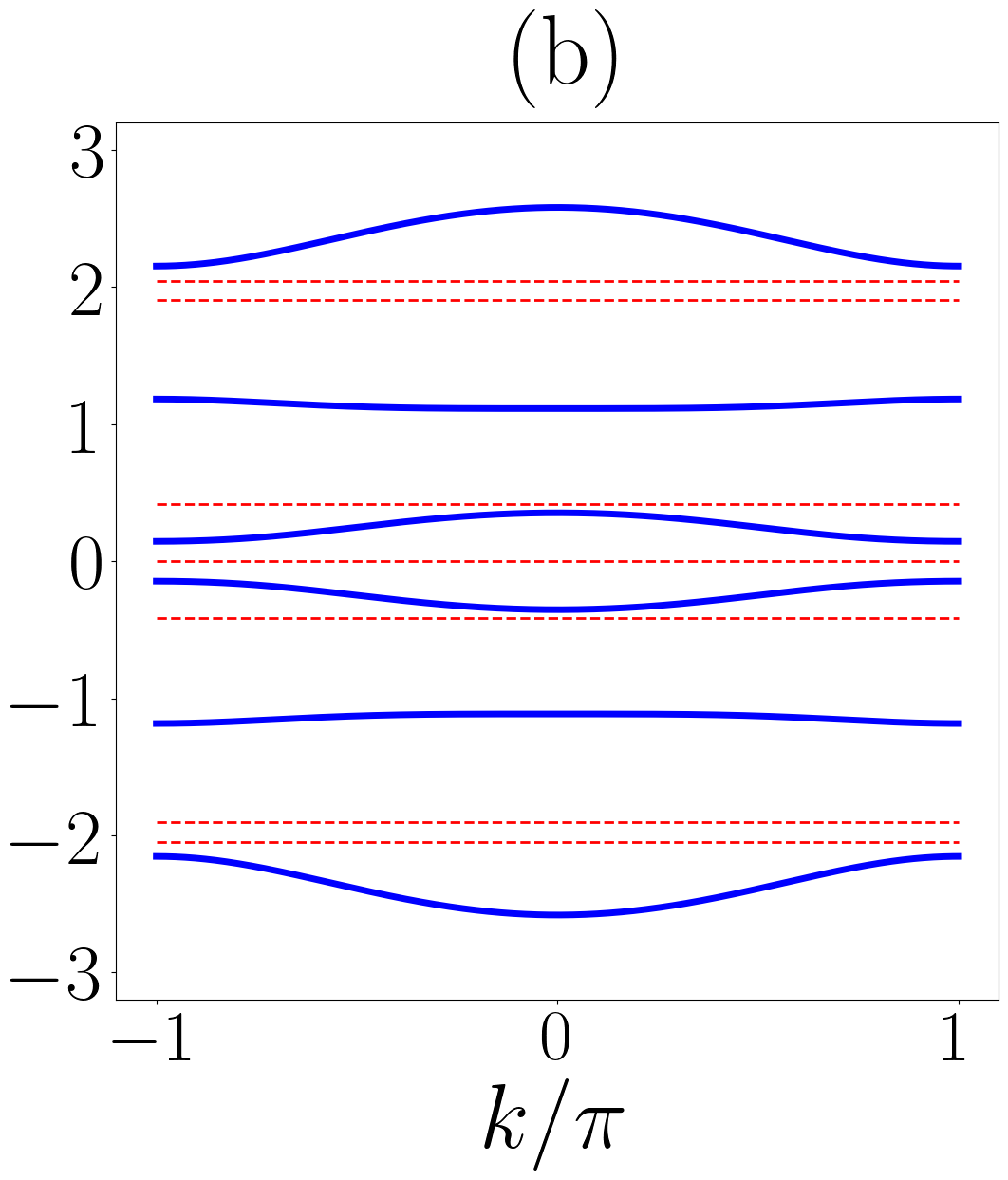}
    \end{subfigure}
    \begin{subfigure}{0.15\textwidth}
        \includegraphics[width=\textwidth]{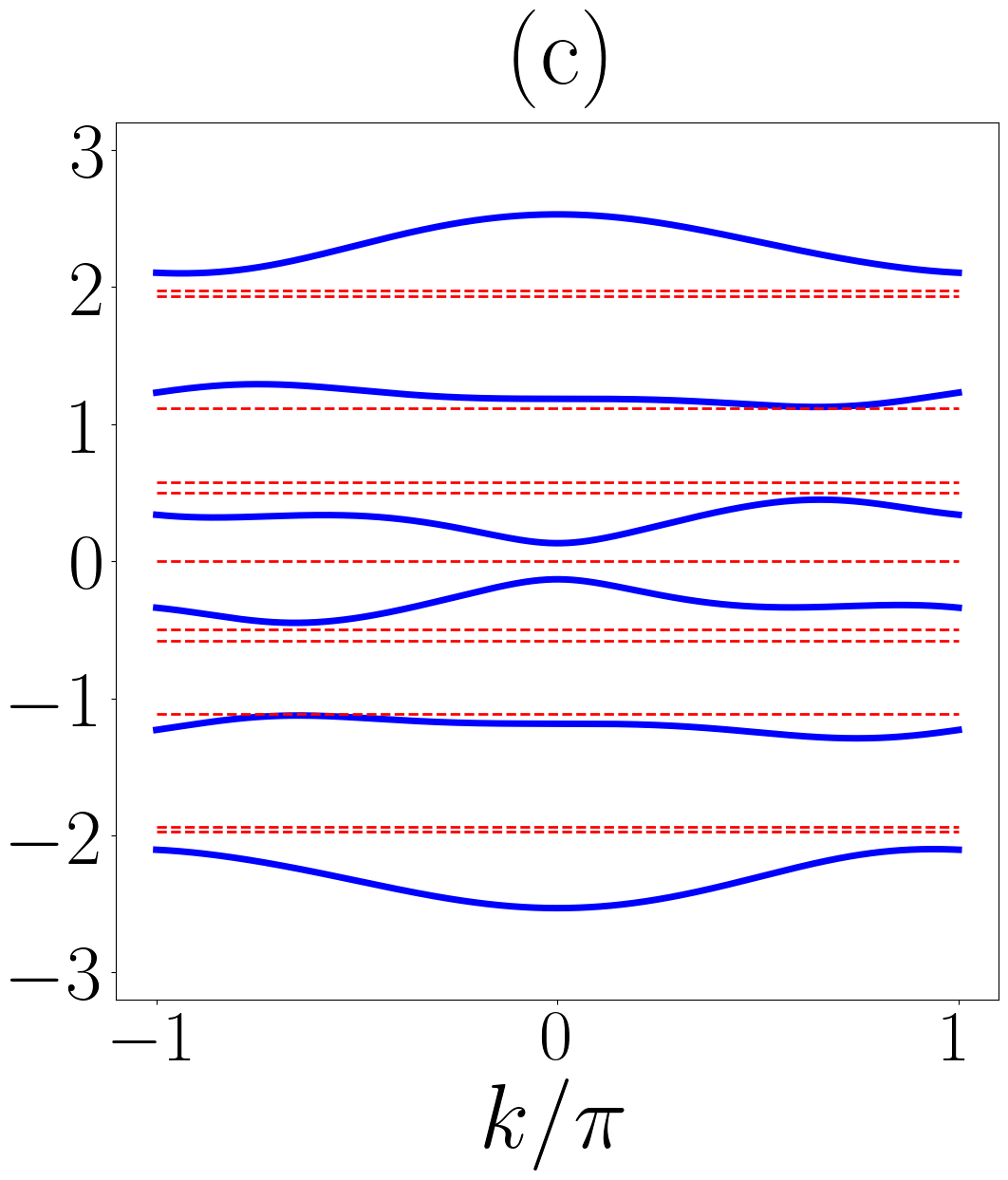}
    \end{subfigure}
    \begin{subfigure}{0.15\textwidth}
        \includegraphics[width=\textwidth]{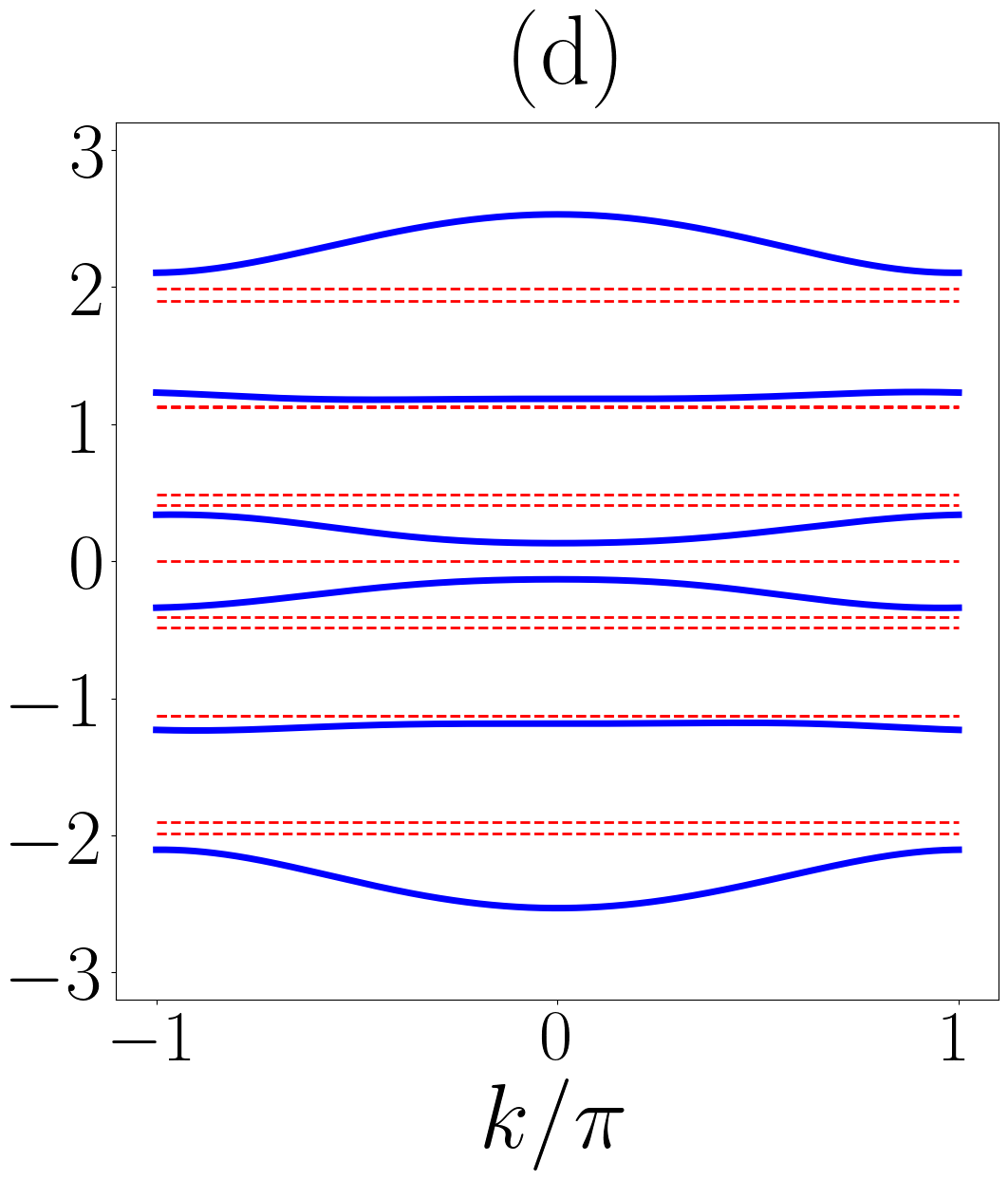}
    \end{subfigure}
    \begin{subfigure}{0.15\textwidth}
        \includegraphics[width=\textwidth]{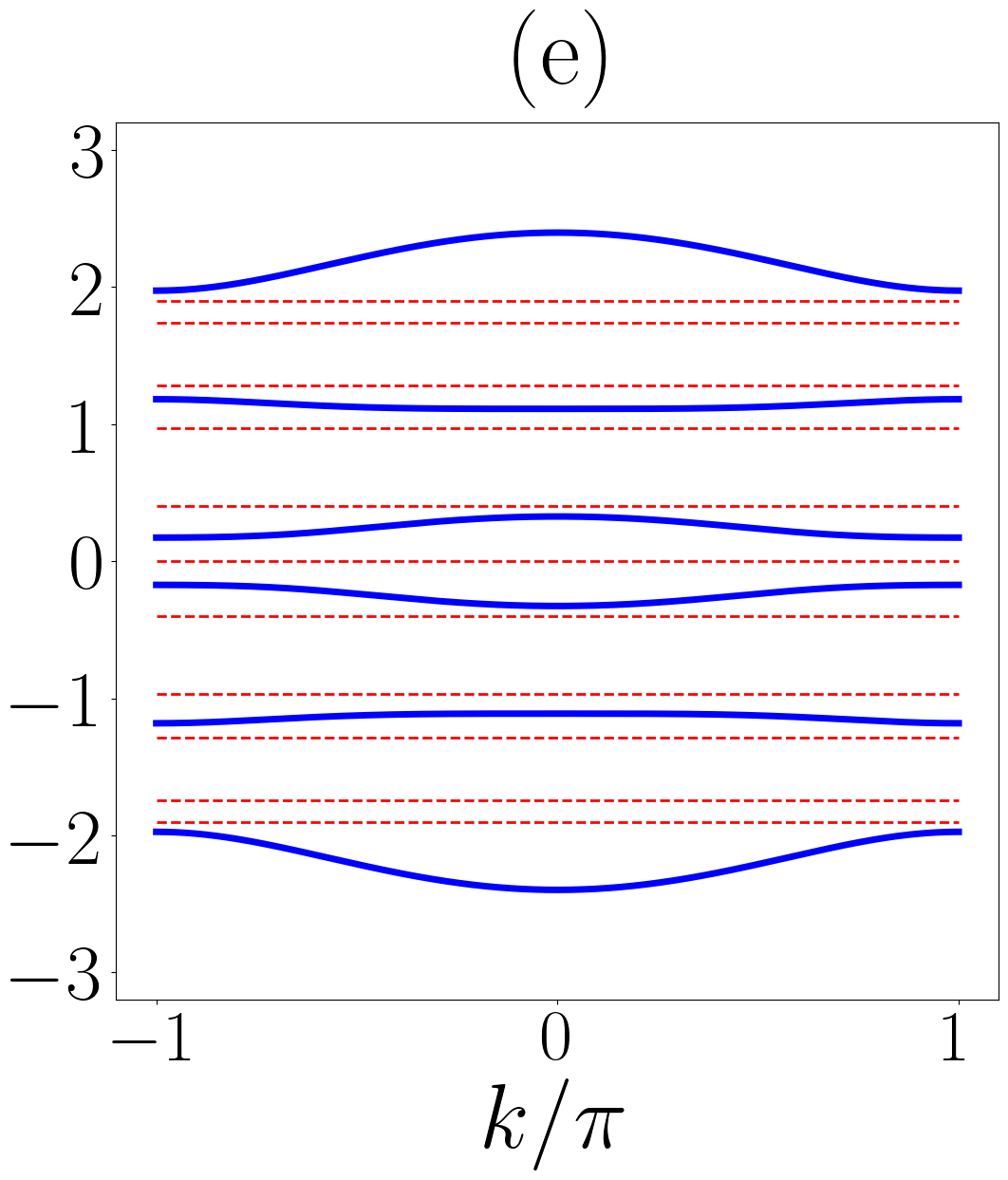}
    \end{subfigure}
    \caption{Energy spectrum in momentum space in the presence of each of the five perturbations, $\alpha(k)$, $\beta(k)$, $\zeta(k)$, $\vartheta(k)$, and $\varrho(k)$. The parameter values for all five cases are $J_1 = 0.5$, $J_2 = 2.5$, $\delta=\Delta=0.4$, and $\mu=3.5$. Subplot (a) is at $\alpha_{o}$ = 0.5, subplot (b) is at $\beta_{o}$ = 0.5, subplot (c) is at $\zeta_{o}$= 1.5, subplot (d) is at $\vartheta_{o}$ = 1.5, and subplot (e) is at $\varrho_{o}= 0.5$. The red dashed horizontal lines mark the energies of the expected edge states when OBC are applied (see Fig.~\ref{pert}).} 
    \label{k5}
\end{figure*}
\end{center}

\begin{center}
\begin{figure*}[htbp]
    \begin{subfigure}{0.1665\textwidth}
        \includegraphics[width=\textwidth]{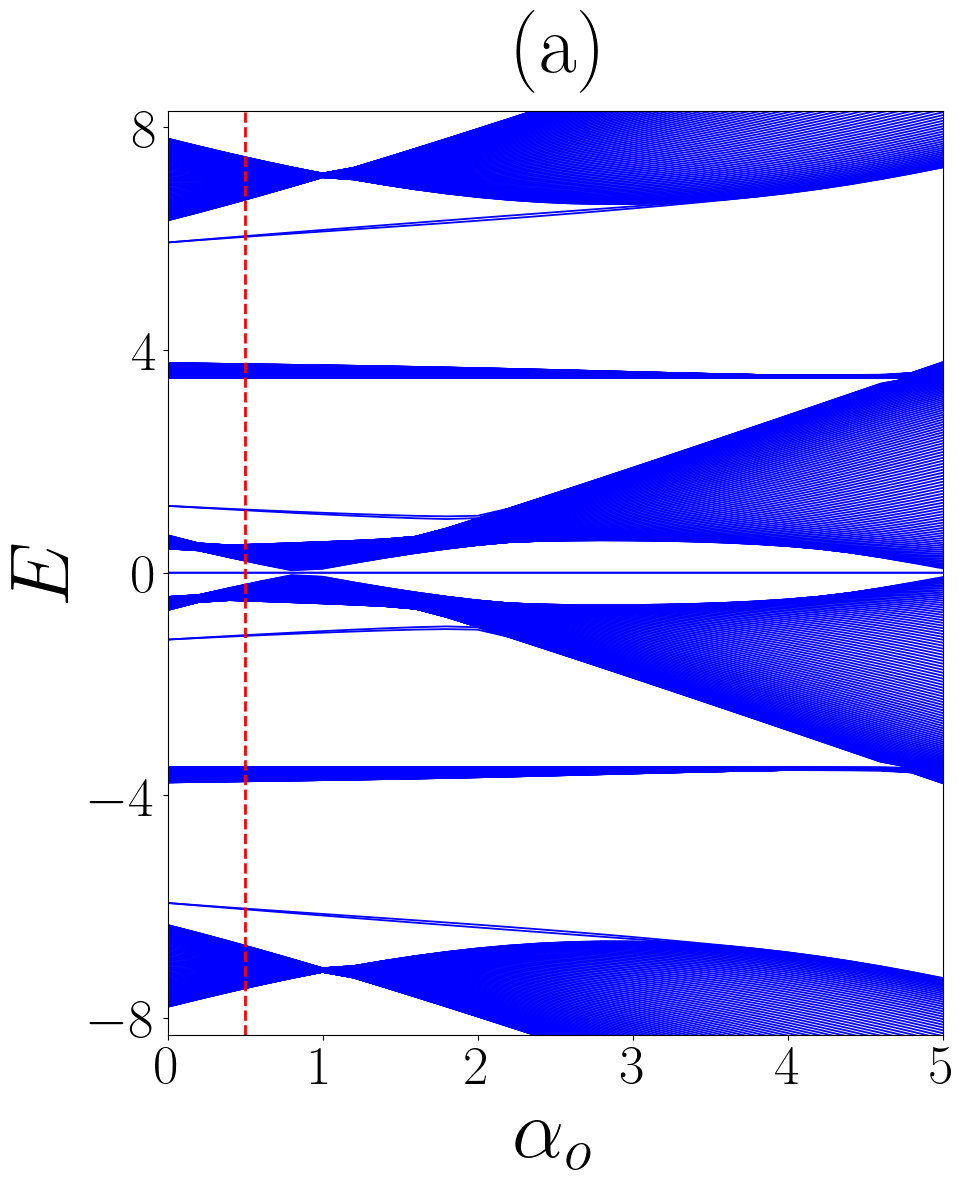}
    \end{subfigure}
    \begin{subfigure}{0.15\textwidth}
        \includegraphics[width=\textwidth]{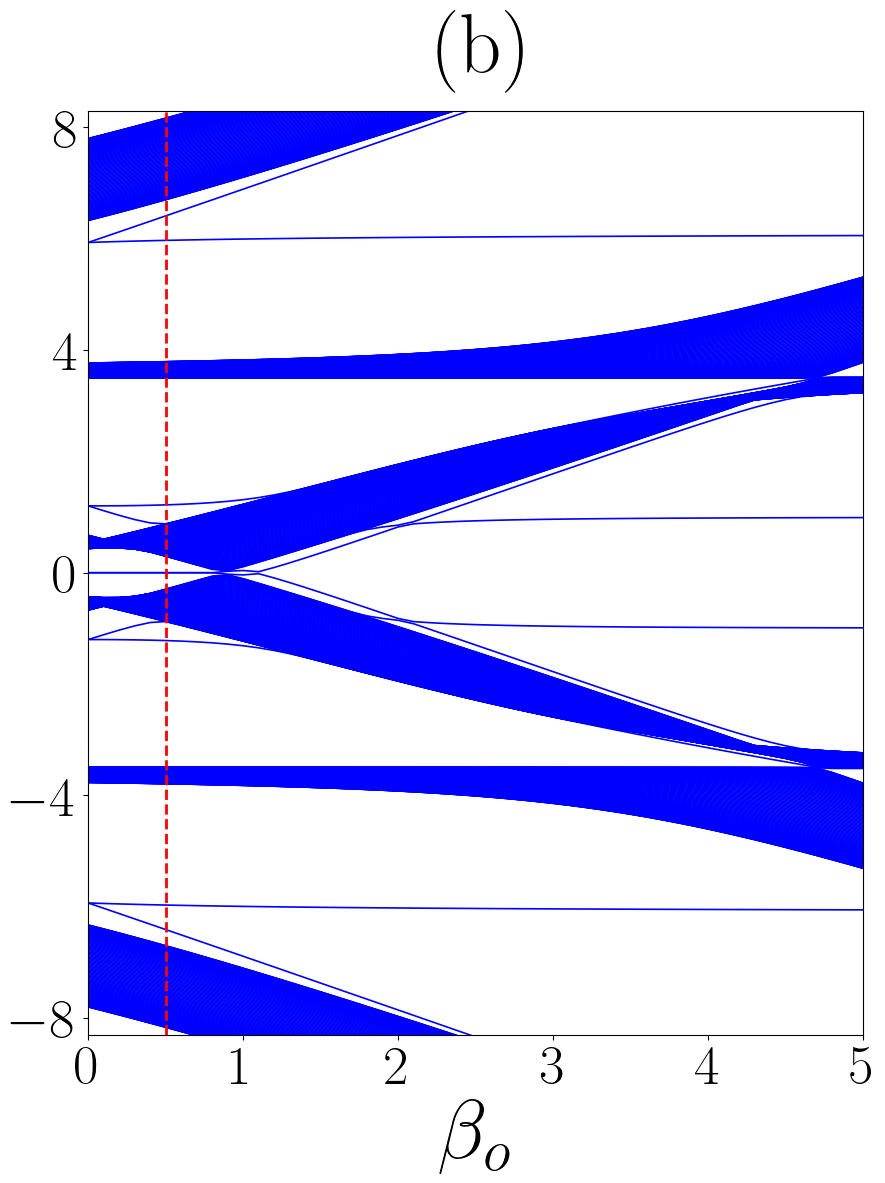}
    \end{subfigure}
    \begin{subfigure}{0.15\textwidth}
        \includegraphics[width=\textwidth]{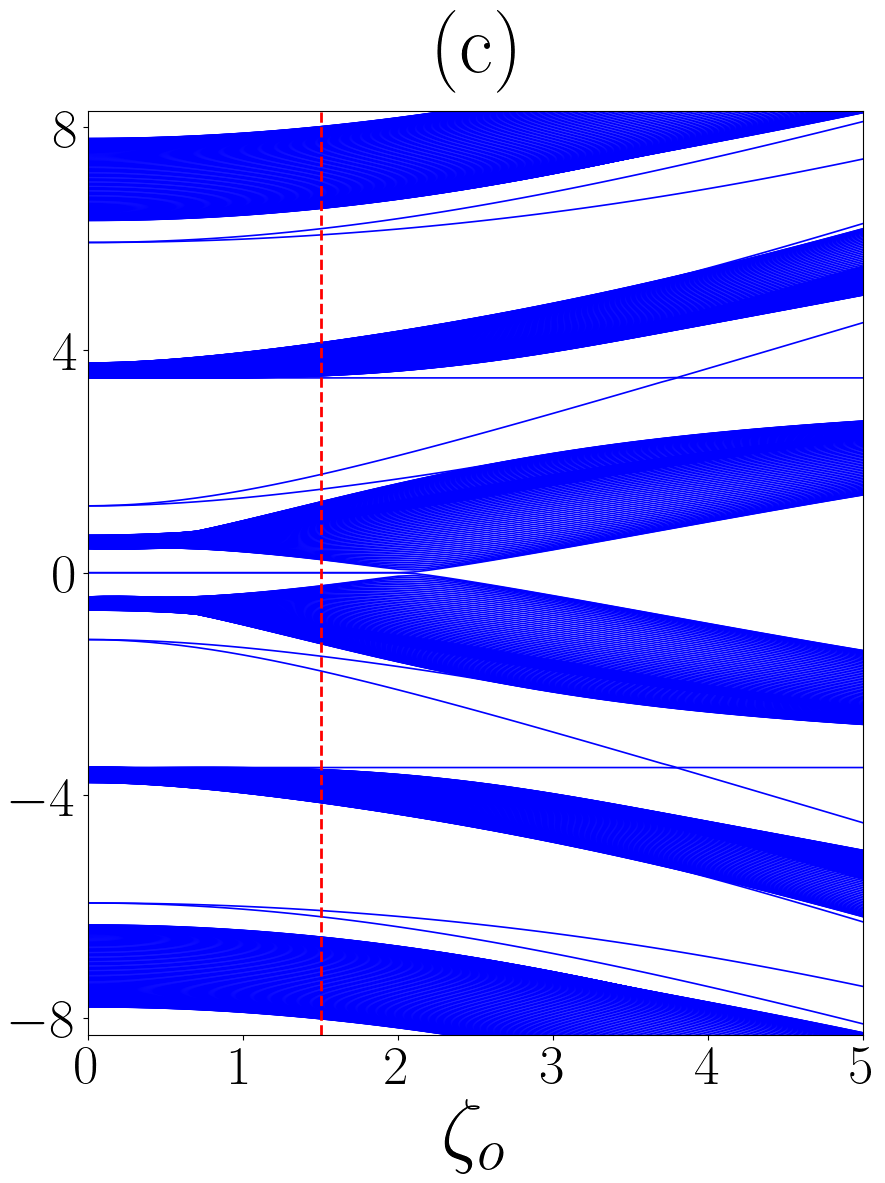}
    \end{subfigure}
    \begin{subfigure}{0.15\textwidth}
        \includegraphics[width=\textwidth]{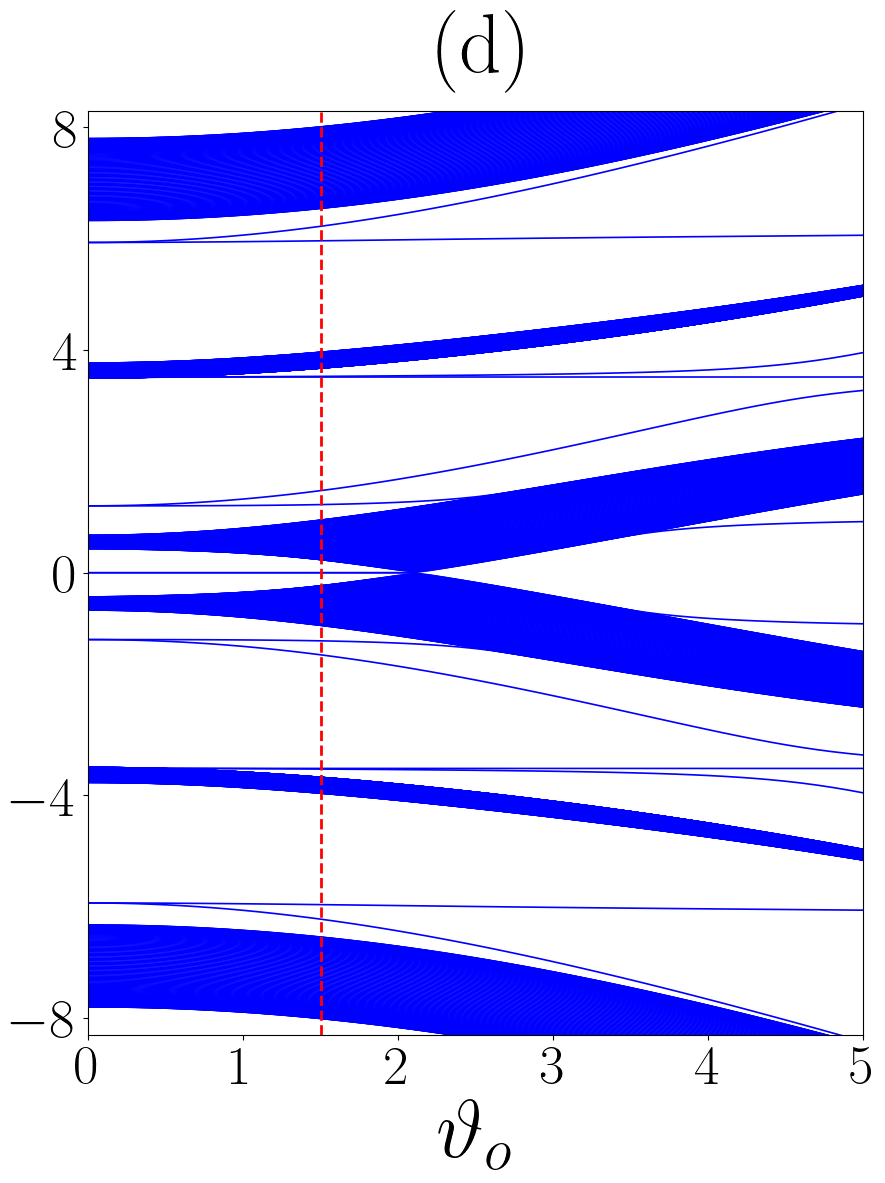}
    \end{subfigure}
    \begin{subfigure}{0.15\textwidth}
        \includegraphics[width=\textwidth]{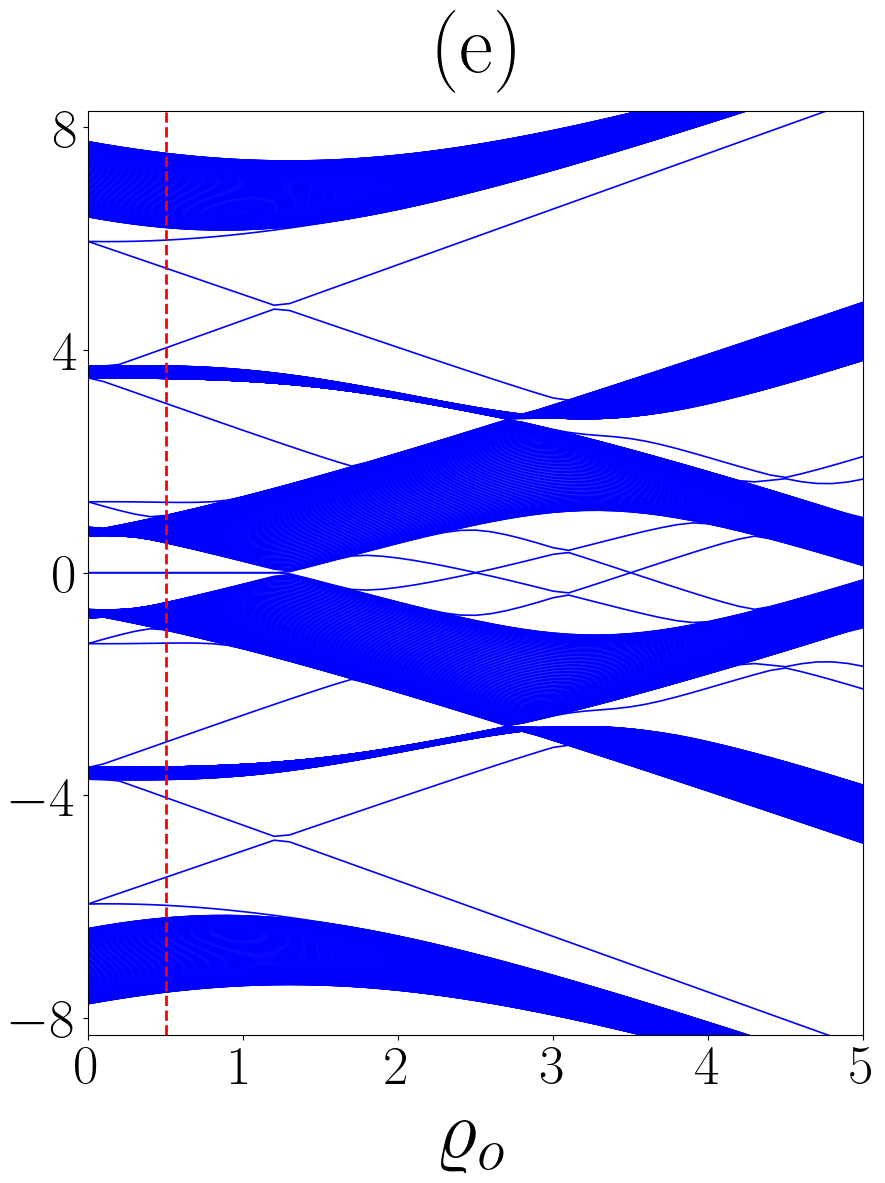}
    \end{subfigure}
    \caption{Energy spectrum with respect to all five perturbation strengths for \MG{$N=100$} unit cells. The parameter values for all five cases are $J_1 = 0.5$, $J_2 = 2.5$, $\delta=\Delta=0.4$, and $\mu=3.5$. The red vertical dashed lines mark the parameter values used in Fig.~\ref{k5}.}
    \label{pert}
\end{figure*}
\end{center}

Figures \ref{k5} and \ref{pert} summarize our findings by respectively plotting the energy spectrum in momentum space and real space in the presence of the aforementioned perturbations. In the momentum space spectrum of Fig.~\ref{k5}, the expected energies of the edge modes when OBC are applied are also plotted to highlight the spectral separation of the edge states from the bulk bands. Meanwhile, the real space spectrum of Fig.~\ref{pert} explicitly highlights the fate of the system's edge states as the perturbation strengths are varied. Together, both figures not only demonstrate the topological robustness of our system, but also highlight specific features that emerge as a result of the applied perturbations. 

In Fig.~\ref{pert}(a), the system is under the effect of perturbation $\alpha$. As the perturbation strength $\alpha_{o}$ increases, the nonzero energy edge modes progressively merge into the bulk bands and eventually disappear. In contrast, despite the presence of zero energy band touching at some value of $\alpha_{o}$, the MZMs remain otherwise present throughout all $\alpha_{o}$ values under consideration. In Fig.~\ref{pert}(b), the system is under the effect of perturbation $\beta$. In particular, such a perturbation is found to yield energy splitting among the nonzero edge modes, such that some remain pinned at a fixed energy throughout the entire range of the applied perturbation strength, whilst some others quickly merge with the bulk bands. On the other hand, the MZMs remain pinned at zero energy until band closing occurs at $\beta_{o} \approx 1$, after which MZMs are absent. 

In Figs.~\ref{pert}(c) and (d), the system is under the perturbations $\zeta$ and $\vartheta$, respectively, which both break the chiral symmetry. Qualitatively similar features are observed in both panels. First, MZMs remain present at small perturbation strengths before band closing occurs at zero energy and induces a topological phase transition. This result demonstrates that MZMs in our system are protected by the particle-hole rather than the chiral symmetry. Apart from the robustness of the MZMs, Figs.~\ref{pert}(c) and (d) also reveal the presence of additional nonzero edge modes that seemingly emerge from two of the bulk bands ($E \sim \pm 4$ in both panels). Such edge modes could be understood as remnants of the zero energy edge modes in the extended SSH component \cite{ghuneim2024topological} of our model that are shifted to nonzero energy by the Kitaev component of the model. In the absence of perturbations, such edge modes overlap with some bulk bands, rendering them invisible and not topologically protected. In this case, the presence of either $\zeta$ or $\vartheta$ induces energy difference between the edge modes and the corresponding bulk bands. 

Finally, Fig.~\ref{pert}(e) shows the energy spectrum of the system under the perturbation $\varrho$. Although the system in this case features a more complex edge mode structure, the origin of these edge modes could be easily tracked from the unperturbed case. In particular, the MZMs remain pinned at zero energy and are present before band closing at zero energy occurs.

\subsection{Effects of disorder}
\label{dis}
The disorder effect is unavoidable in real-world physical systems. From an experimental standpoint, it may originate from minor manufacturing flaws in the materials used or ambient noise that naturally arises due to experimental settings. In topological systems, it is typically expected that edge modes demonstrate some robustness against disorder, as long as the latter does not induce a topological phase transition due to some gap closing. This section aims to explicitly verify this. 

\begin{center}
\begin{figure*}[htbp]
    \begin{subfigure}{0.1665\textwidth}
        \includegraphics[width=\textwidth]{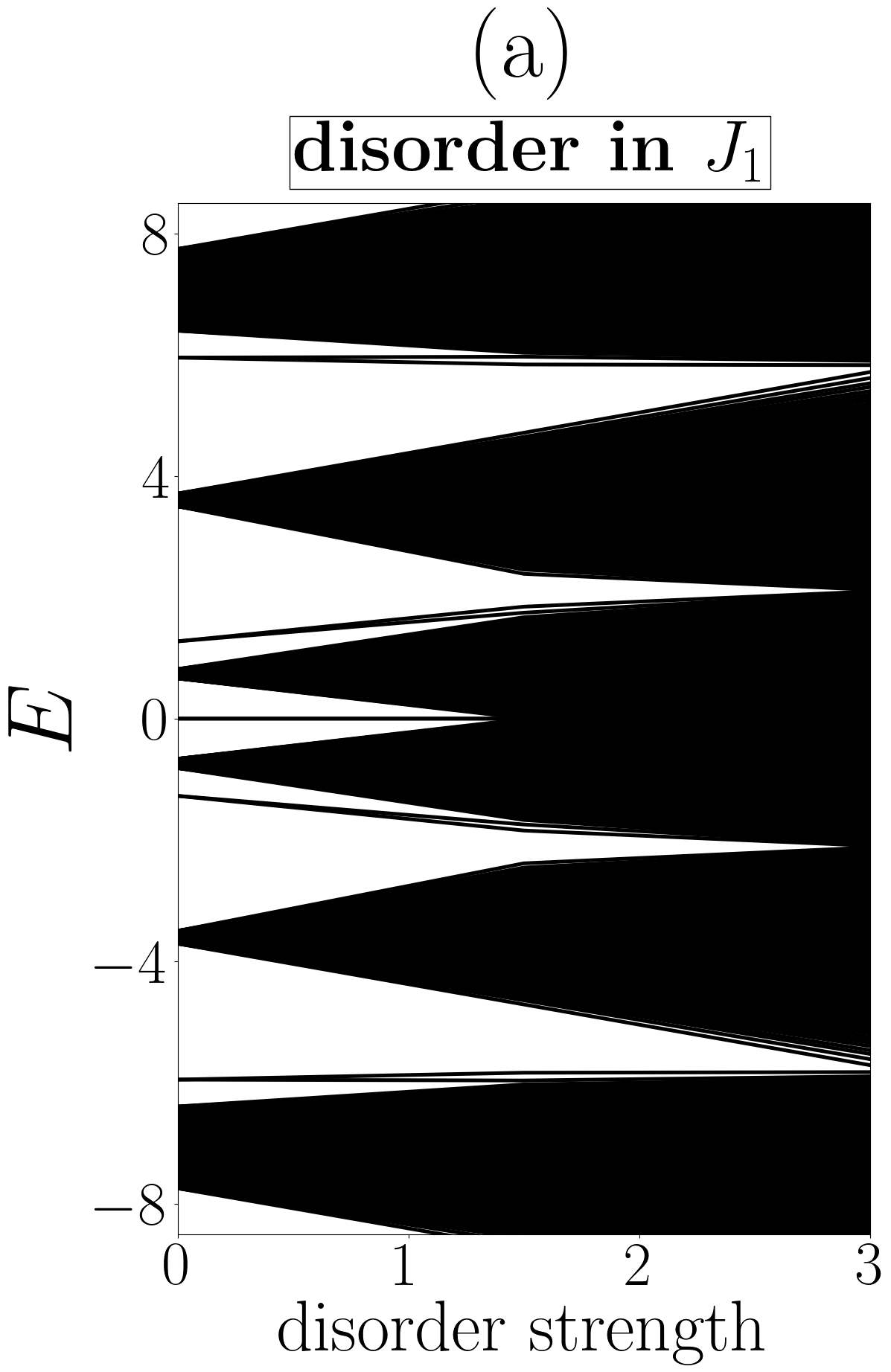}
    \end{subfigure}
    \begin{subfigure}{0.15\textwidth}
        \includegraphics[width=\textwidth]{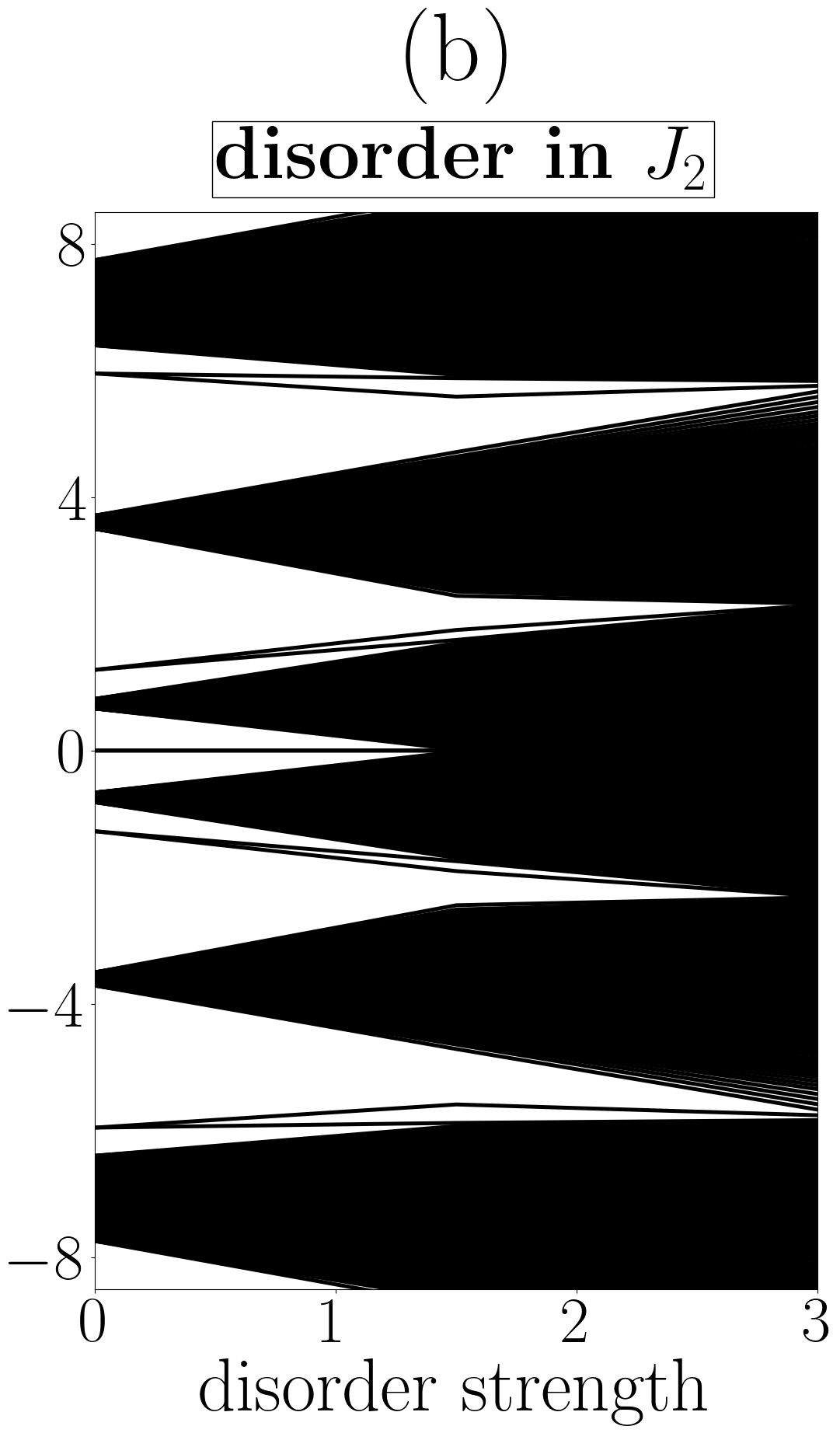}
    \end{subfigure}
    \begin{subfigure}{0.15\textwidth}
        \includegraphics[width=\textwidth]{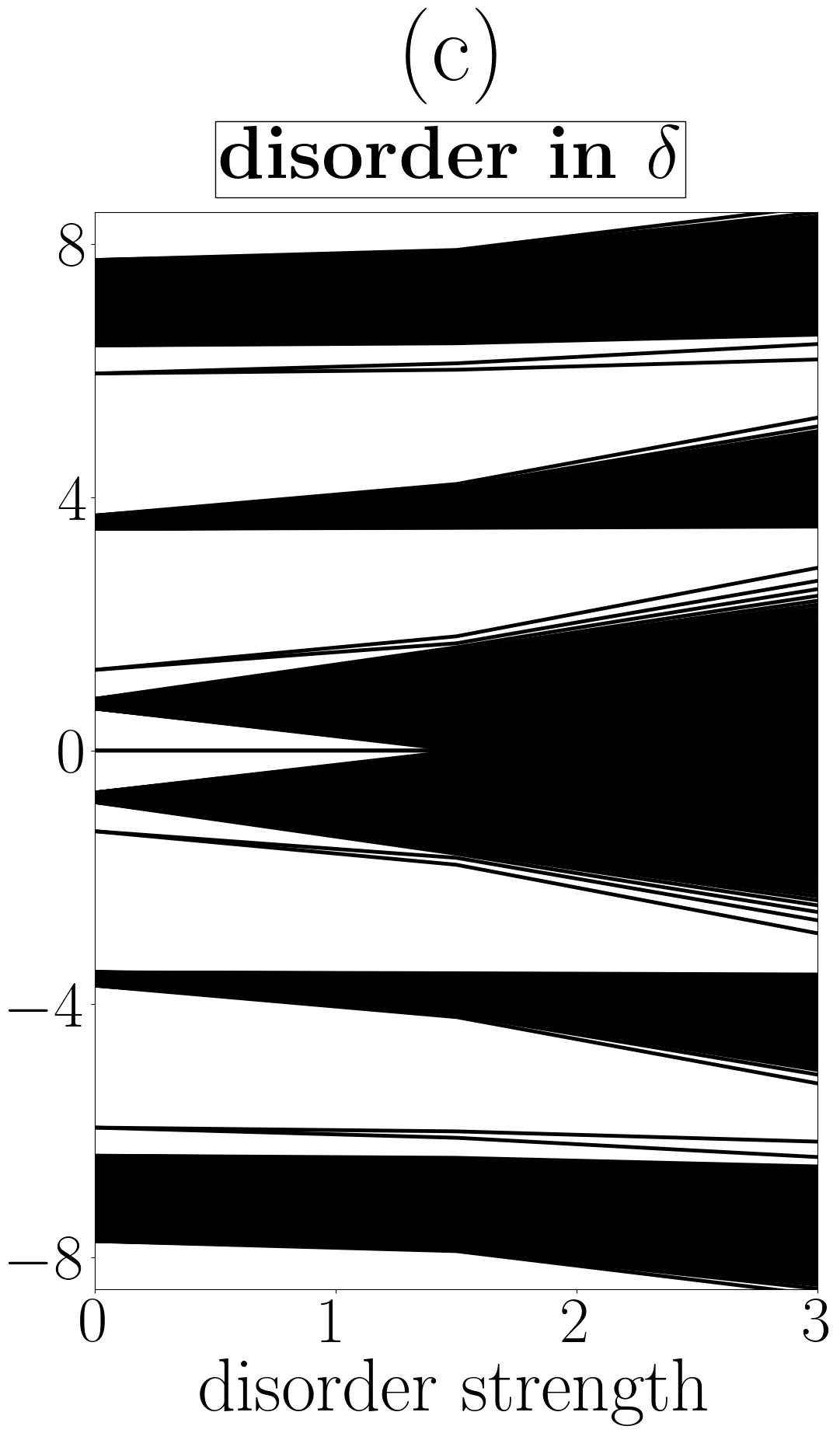}
    \end{subfigure}
    \begin{subfigure}{0.15\textwidth}
        \includegraphics[width=\textwidth]{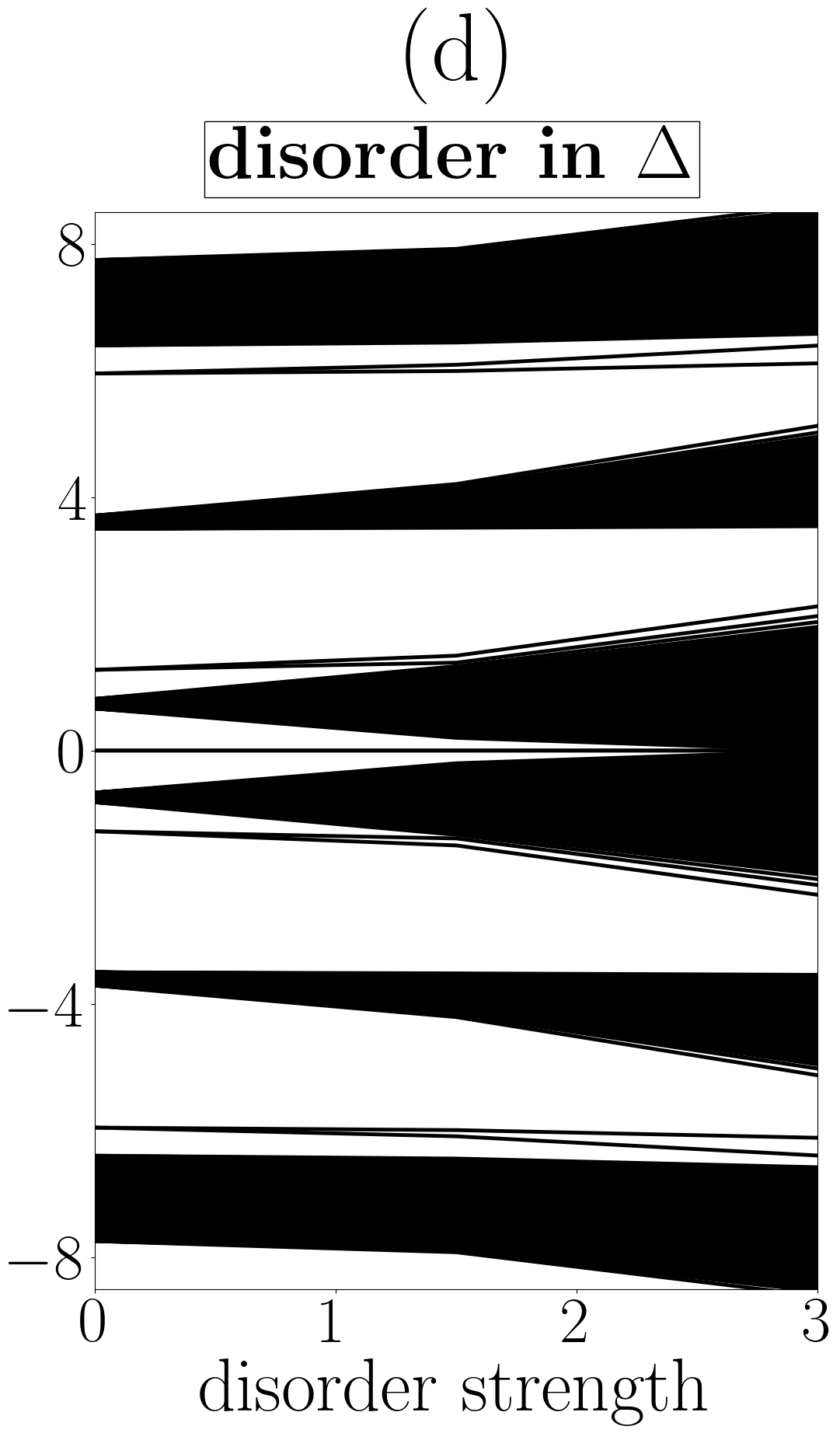}
    \end{subfigure}
    \begin{subfigure}{0.15\textwidth}
        \includegraphics[width=\textwidth]{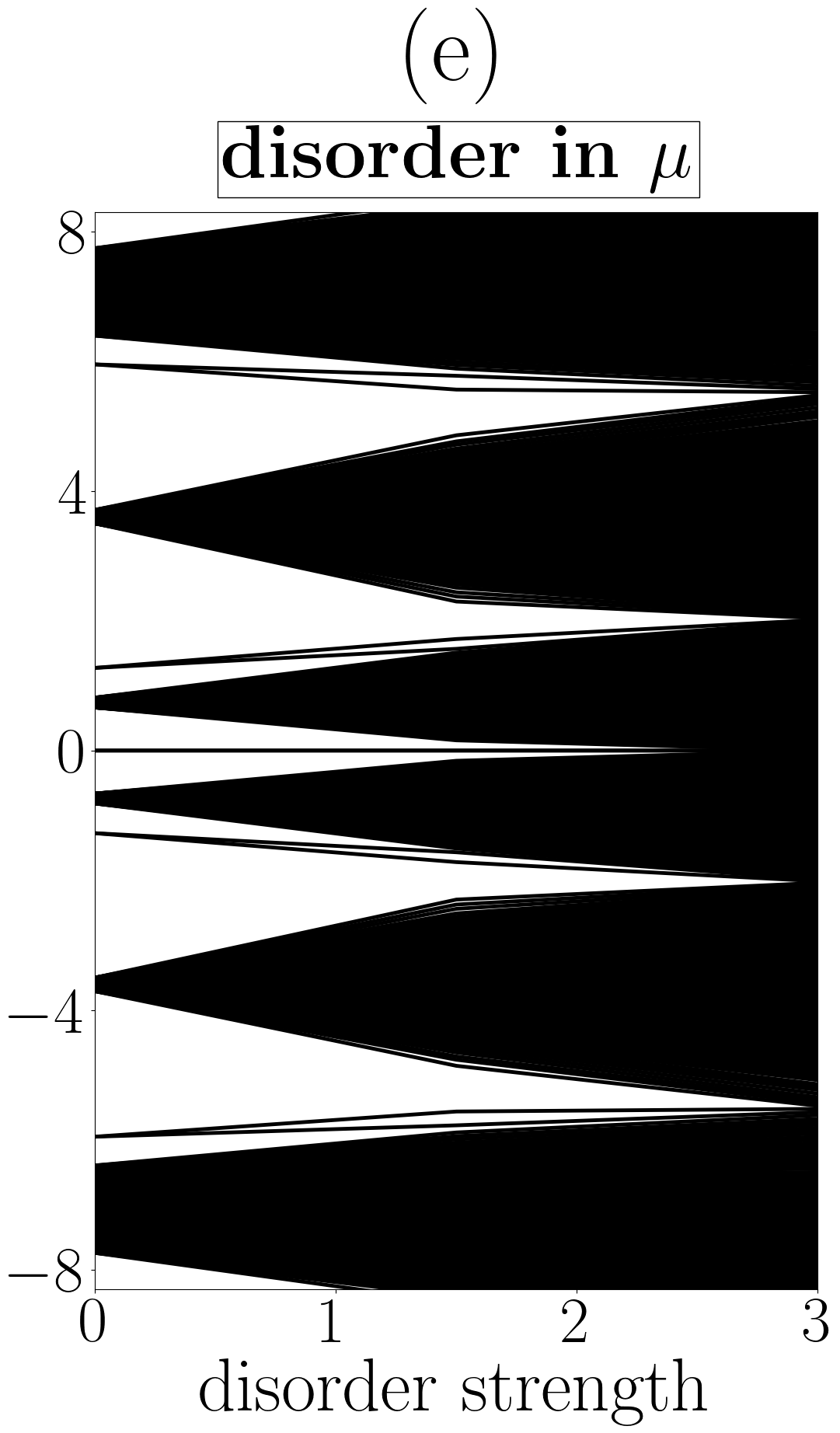}
    \end{subfigure}
    \caption{Energy spectrum with respect to the disorder strength for \MG{$N=100$} unit cells. The starting parameter values for all five cases are $J_1 = 0.5$, $J_2 = 2.5$, $\delta=\Delta= 0.4$ and $\mu= 3.5$. (a) Disorder acts on $J_{1}$, (b) Disorder acts on $J_2$, (c)  Disorder acts on $\delta$, (d) Disorder acts on $\Delta$, and (e) Disorder acts on $\mu$. Each data point is averaged over 100 disorder realizations.}
    \label{fig:figdis1}
\end{figure*}
\end{center}

Here, we theoretically simulate the presence of disorder by making each parameter in the system site-dependent such that $t_{j} \rightarrow t + \sigma_{j}$ (with $t=J_1, J_2, \delta, \Delta, \mu$), $j=1,2,\cdots, N$ and $\sigma_{j}$ are randomly taken from [$-X$, $X$] ($X$ being the strength of the disorder). To this end, we introduce the disorder in each system's parameter separately, which allows us to analyze the effect on each parameter individually. Fig.~\ref{fig:figdis1} summarizes our results, which indeed confirm the robustness of the edge modes in the system against the various disorders. 

\begin{center}
\begin{figure*}[htbp]
    \begin{subfigure}{0.1665\textwidth}
        \includegraphics[width=\textwidth]{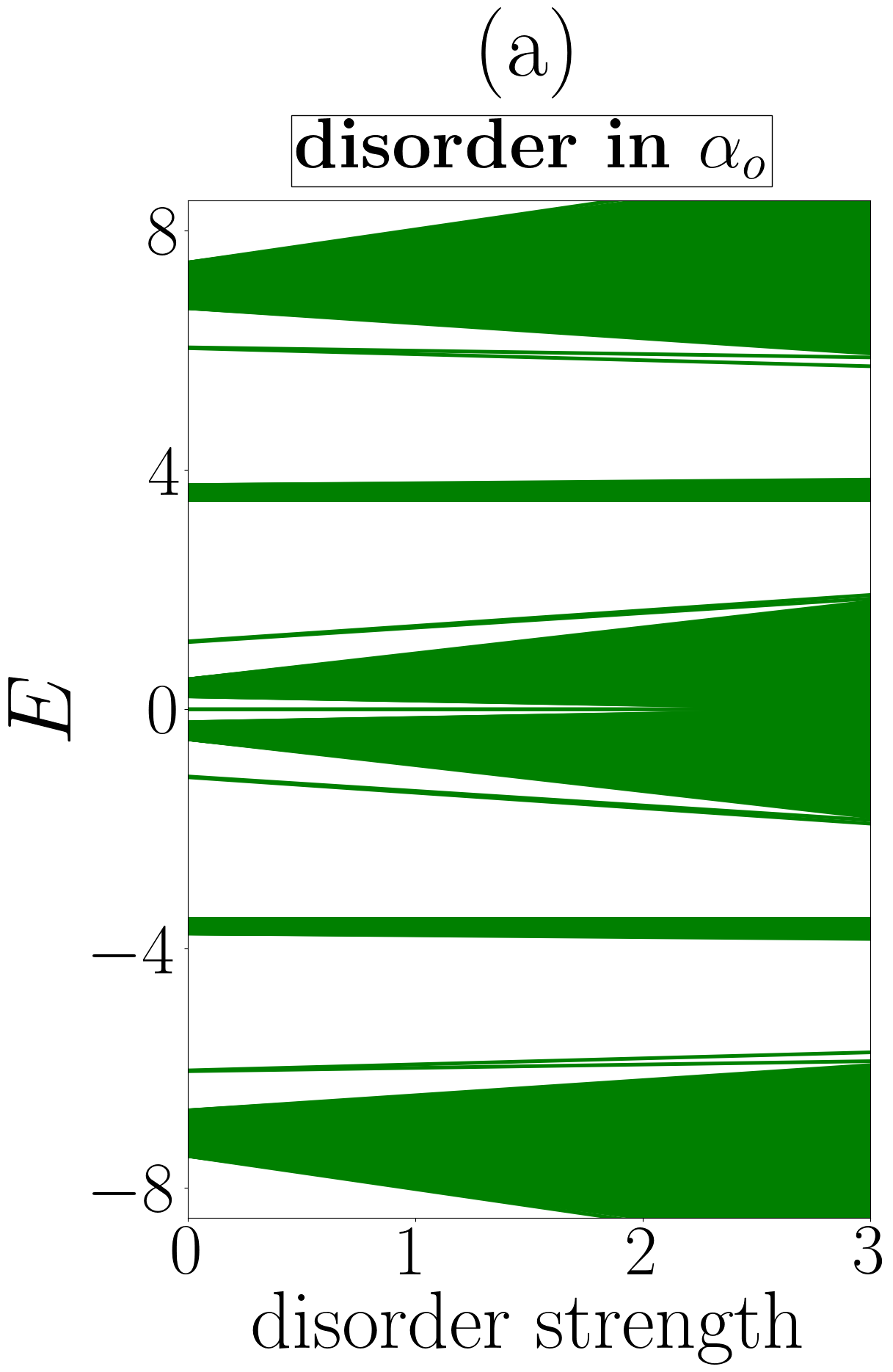}
    \end{subfigure}
    \begin{subfigure}{0.15\textwidth}
        \includegraphics[width=\textwidth]{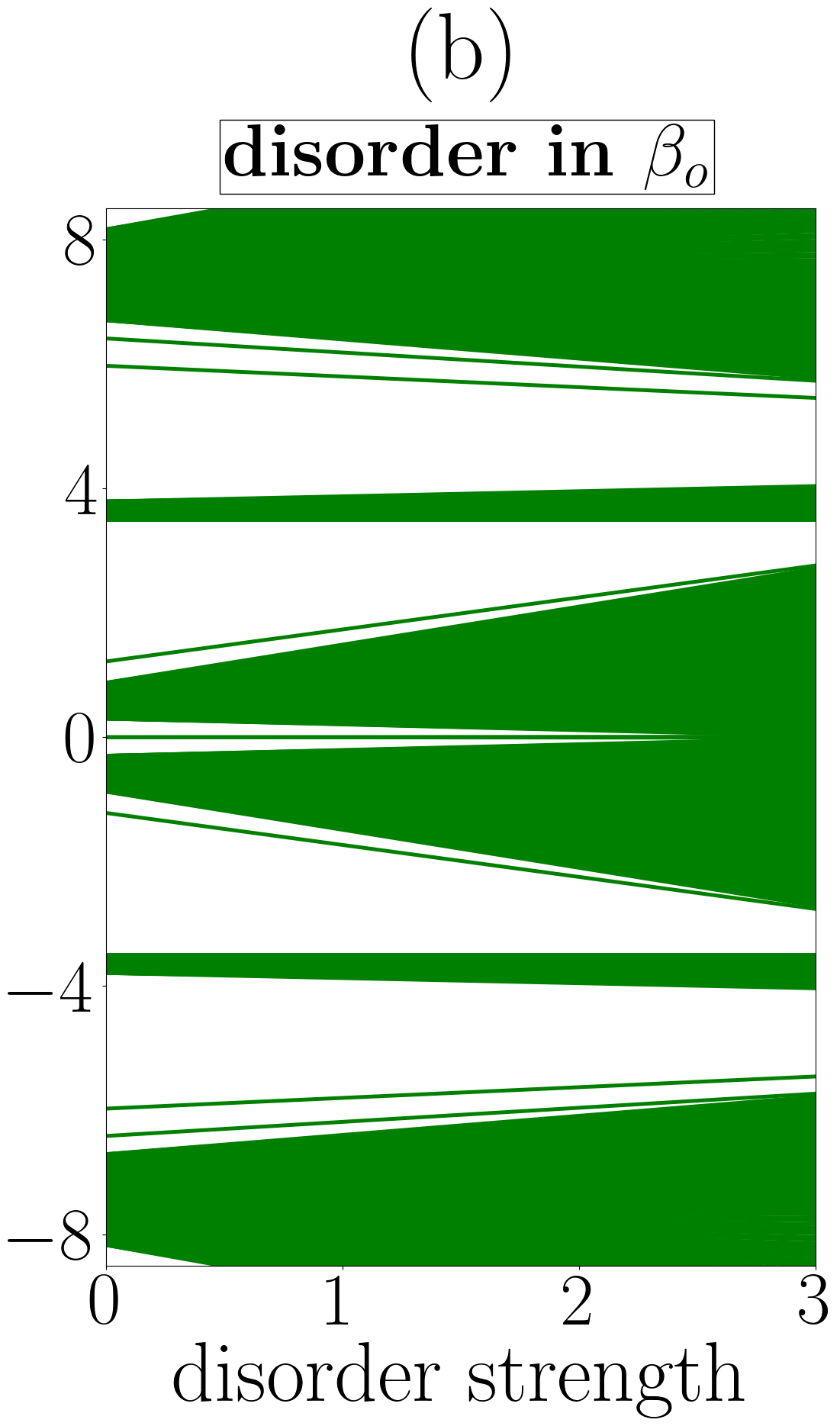}
    \end{subfigure}
    \begin{subfigure}{0.15\textwidth}
        \includegraphics[width=\textwidth]{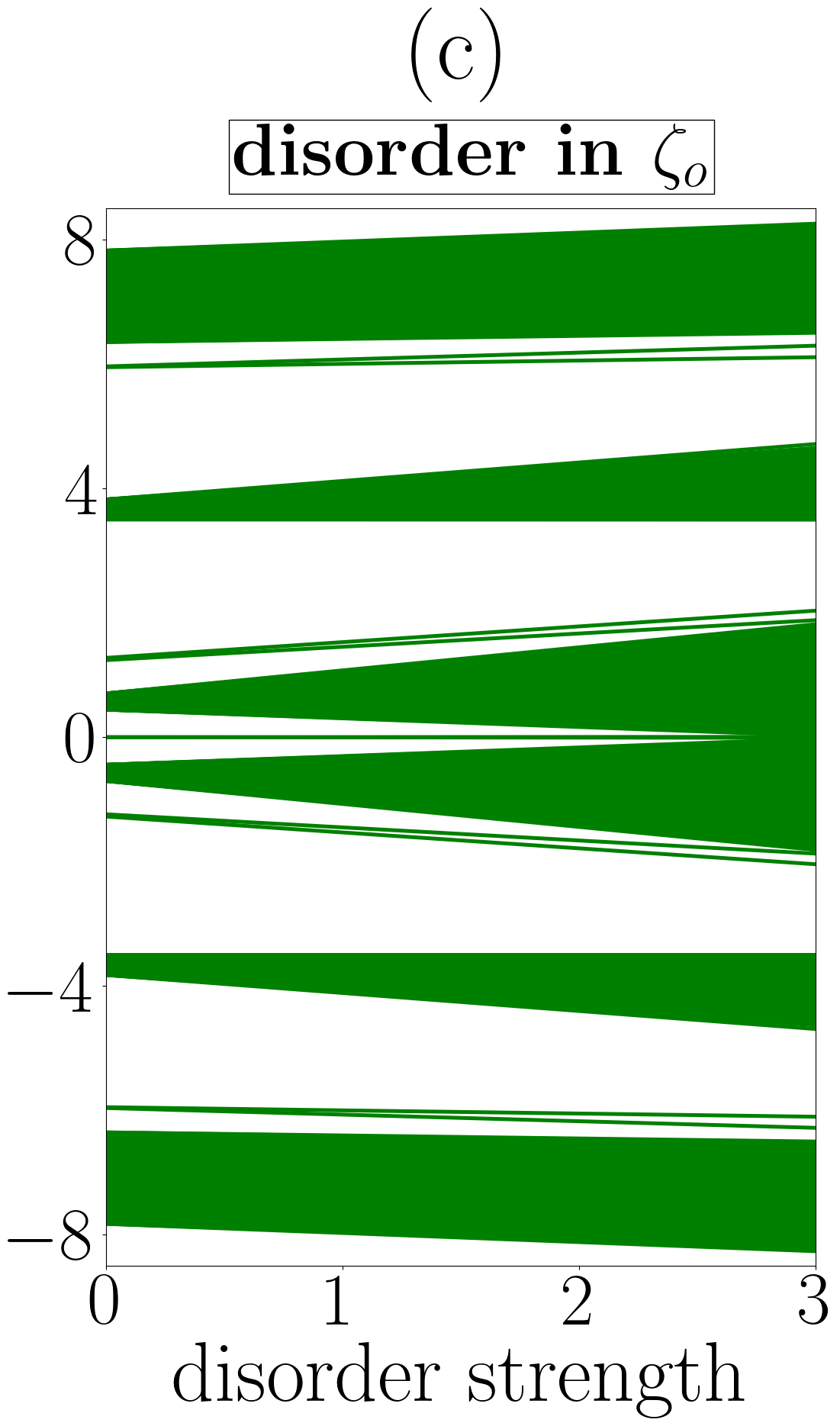}
    \end{subfigure}
    \begin{subfigure}{0.15\textwidth}
        \includegraphics[width=\textwidth]{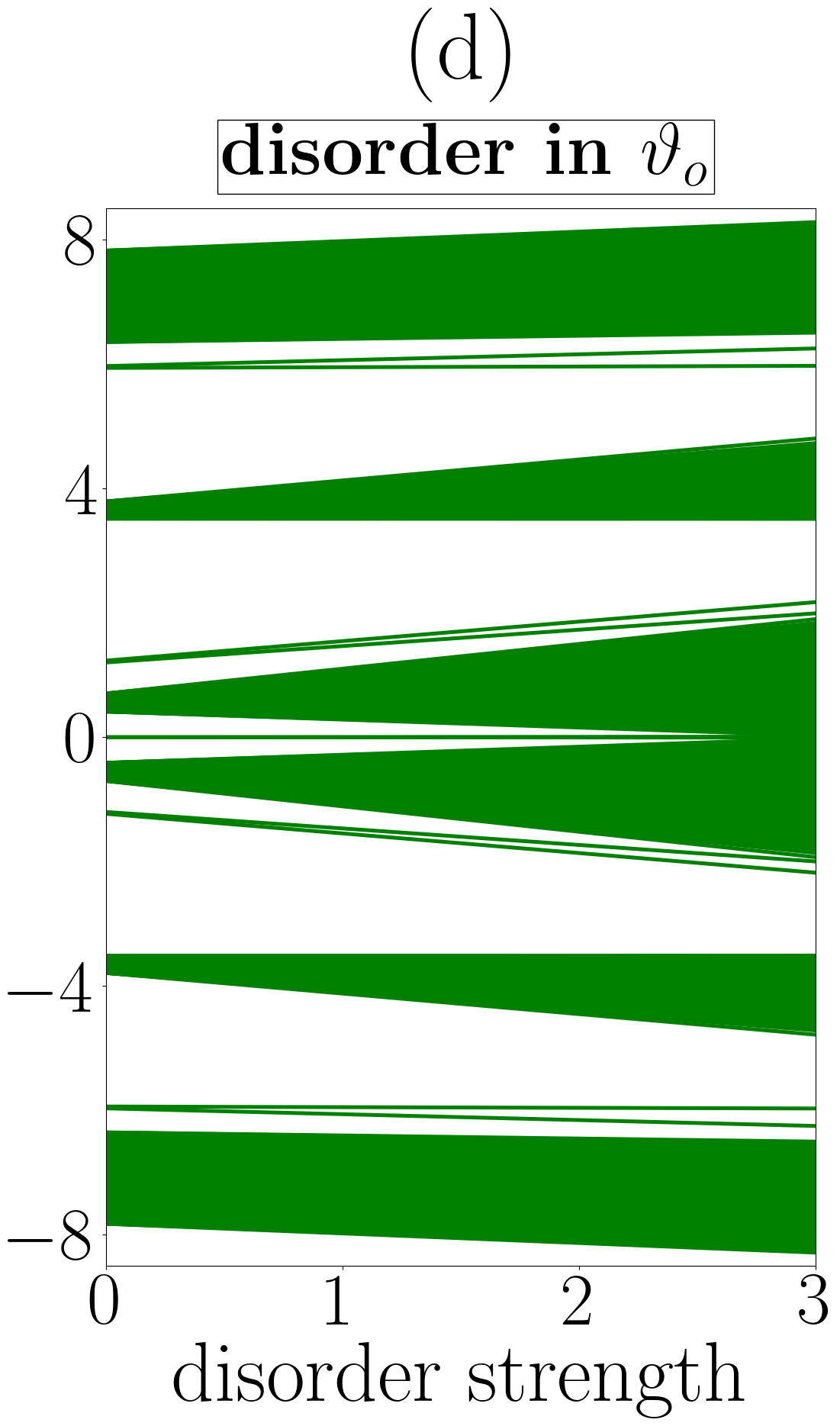}
    \end{subfigure}
    \begin{subfigure}{0.15\textwidth}
        \includegraphics[width=\textwidth]{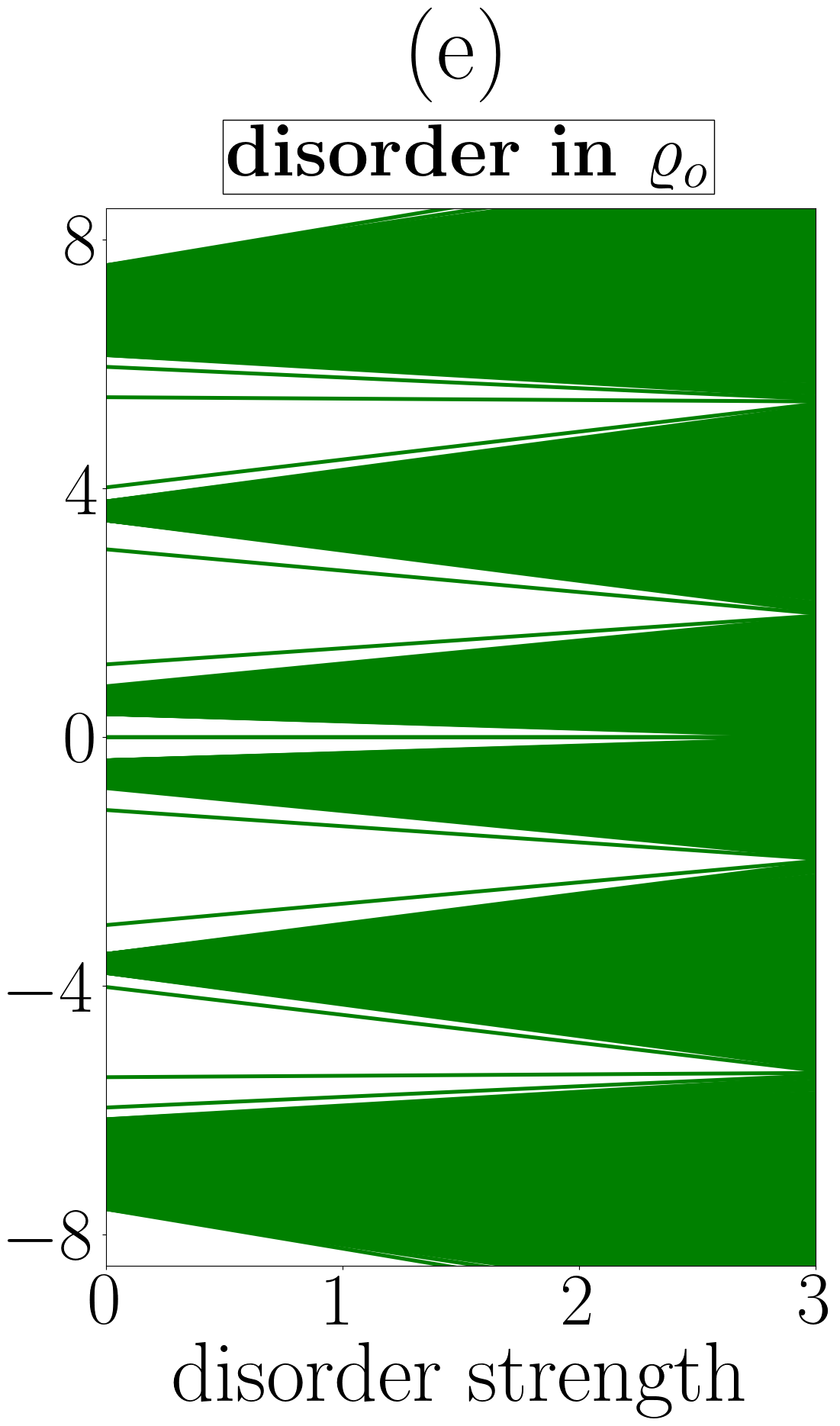}
    \end{subfigure}
    \caption{Energy spectrum with respect to the disorder strength for \MG{$N=100$} unit cells. The starting parameter values for all five cases are $J_1 = 0.5$, $J_2 = 2.5$, $\delta=\Delta=0.4$, $\mu=3.5$, and $\alpha_{o}$ = $\beta_{o}$ = $\zeta_{o}$= $\vartheta_{o}$ = $\varrho_{o}$ = 0.5. (a) Disorder acts on $\alpha_{o}$, (b) Disorder acts on $\beta_{o}$, (c) Disorder acts on $\zeta_{o}$, (d) Disorder acts on $\vartheta_{o}$, and (e) Disorder acts on $\varrho_{o}$. Each data point is averaged over 100 disorder realizations.}
    \label{fig:figdis2}
\end{figure*}
\end{center}

For completeness, in Fig.~\ref{fig:figdis2} we again consider the presence of each perturbation in Eq.~(\ref{eq:eq9}), the strength of which is taken as site-dependent to investigate the simultaneous effect of generic perturbations and disorder. In all cases, the edge modes are found to be robust.

\section{Conclusion}
\label{Conc}
In conclusion, we have investigated an extended Kitaev chain that contains three sites per unit cell, resulting from the hybridization of a trimer SSH model \cite{ghuneim2024topological} with the standard Kitaev chain \cite{kitaev2001unpaired}, which leads to a total of six Majorana sites per unit cell. This hexamer lattice provides a richer edge state structure, which manifests itself compared to existing extended Kitaev models \cite{PhysRevLett.113.156402, PhysRevB.95.195160, PhysRevLett.118.267002, mishra2020disordered, PhysRevB.111.155149, PhysRevB.90.014505, yang2025multiple}. The system is found to preserve chiral, time-reversal, and particle-hole symmetries. To analyse the role of these symmetries in protecting the edge states,  a set of generic perturbations has been applied, some of which preserve all symmetries, whereas the others break some of the symmetries. The edge modes' resilience has been confirmed by this investigation. Furthermore, we considered the effect of spatial disorder on the system. In particular, we apply the disorder individually to each of the system parameters, including the previously introduced perturbations. As a result, we find that the system demonstrates the expected topological protection across all cases, maintaining robust edge modes.

Even though the Kitaev chain has been thoroughly studied and successfully implemented on a variety of physical platforms \cite{ten2025observation, bordin2025enhanced, ten2024two, dvir2023realization, iizuka2023experimental,allein2023strain, slim2024optomechanical, busnaina2024quantum}, it remains a fundamental concept that inspires a considerable range of ongoing research. We expect that this extended version of the Kitaev chain will garner substantial attention in both theoretical and experimental contexts. From the experimental perspective, the inclusion of both robust MZMs and Majorana nonzero energy edge modes in our model provides richer experimental signatures. This coexistence of different edge modes can improve detection techniques by offering more diverse experimental signals, thereby contributing to progress in the experimental identification of Majorana modes. 

On the theoretical side, the model presented in this paper is amenable to further modification for further exploration of exotic topological phases. \RB{For instance, a simplified model that supports more analytically tractable non-zero energy edge modes is worth considering, as it allows the application of the transfer matrix method of Refs.~\cite{PhysRevLett.110.146404, lima2025interplay} to fully characterize their topology.} Moreover, the addition of non-Hermiticity \cite{PhysRevB.104.205131, rahul2022topological, PhysRevB.108.014204, PhysRevB.111.174210, song2024band, LIU2025170073, yb8n-2l9v}, periodic driving \cite{zhou2025topological, PhysRevB.108.245153, PhysRevB.111.195424, PhysRevLett.129.254301, PhysRevB.104.L121410, 10.21468/SciPostPhys.13.2.015, PhysRevB.99.045441, PhysRevE.93.022209, PhysRevE.93.022209, PhysRevResearch.2.033495, PhysRevB.106.195122, jangjan2020floquet}, and/or interaction \cite{PhysRevB.107.L060304, PhysRevB.111.195117}, is expected to further enrich the system. 

Another promising future aspect is to consider replacing the trimerized SSH model \cite{ghuneim2024topological} with one that includes more sites per unit cell and similarly combining it with the Kitaev chain \cite{kitaev2001unpaired}, which we expect to further enrich the edge modes' profiles and potentially lead to a different topological structure. Finally, the technique employed for constructing our model is likely to pique the researchers' curiosity in exploring hybridizations beyond the SSH and Kitaev paradigms, which would give rise to other families of superconductors featuring enlarged unit cells and/or different topological characteristics. 

 \begin{acknowledgements}
 R.~W.~B was supported by the Deanship of Research Oversight and Coordination (DROC) and the Interdisciplinary Research Center (IRC) for Intelligent Secure Systems (ISS) at King Fahd University of Petroleum \& Minerals (KFUPM) through internal research Grant No. INSS2507.
 \end{acknowledgements}

\appendix

\section{Appendix section} 
\label{app:A}
\begin{figure}[htpb]
    \centering
    \begin{subfigure}[b]{0.45\textwidth}
        \includegraphics[width=\textwidth]{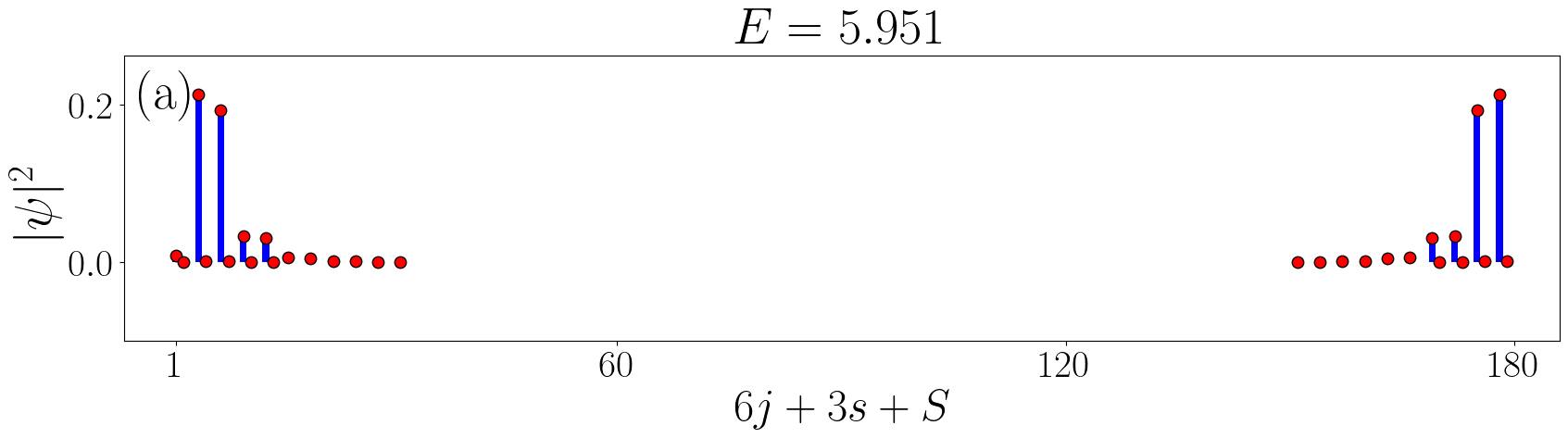} 
        \label{fig:fig1a}
    \end{subfigure}
    \begin{subfigure}[b]{0.45\textwidth}
        \includegraphics[width=\textwidth]{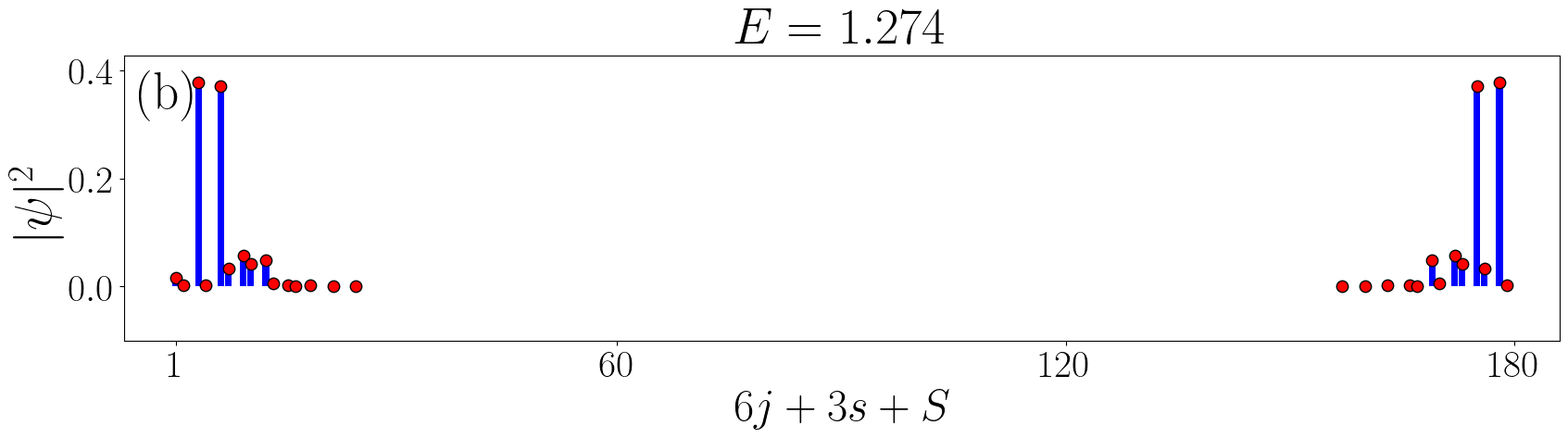}
        \label{fig:fig1b}
    \end{subfigure}
    \begin{subfigure}[b]{0.45\textwidth}
        \includegraphics[width=\textwidth]{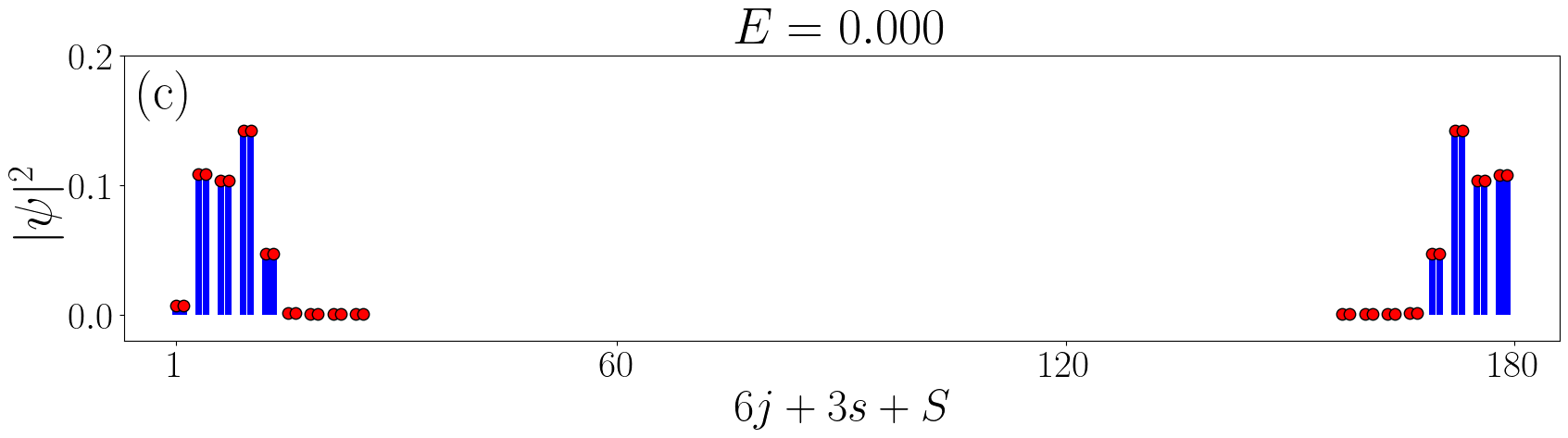} 
        \label{fig:fig1c}
    \end{subfigure}
    \begin{subfigure}[b]{0.45\textwidth}
        \includegraphics[width=\textwidth]{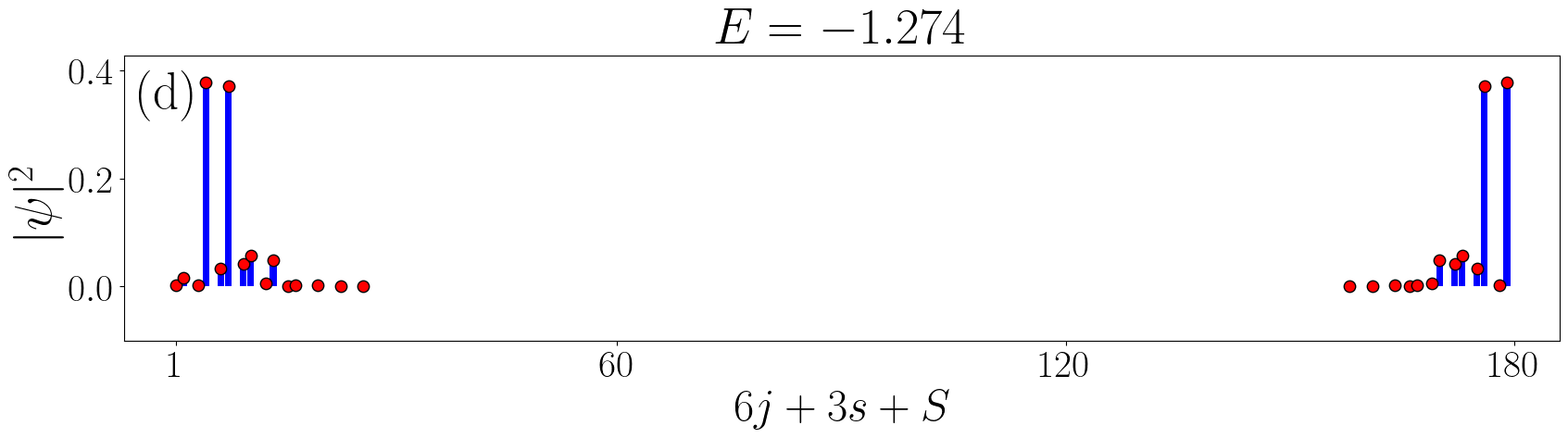}
        \label{fig:fig1d}
    \end{subfigure}
    \begin{subfigure}[b]{0.45\textwidth}
        \includegraphics[width=\textwidth]{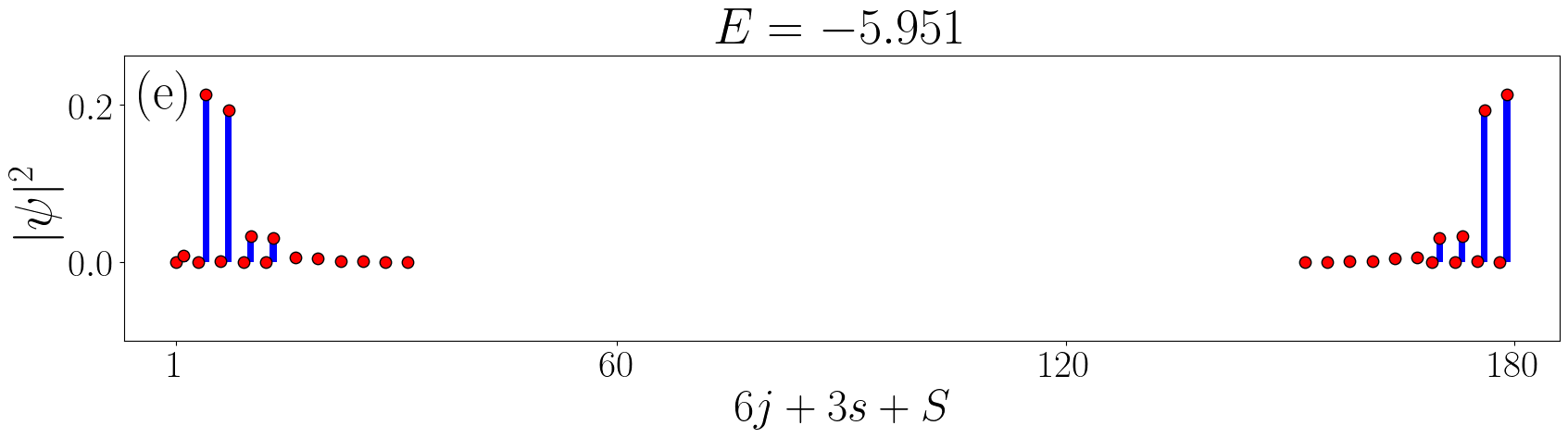}
        \label{fig:fige1}
    \end{subfigure}
    \caption{\RB{Wave function profiles associated with the system's edge states for $N=30$ unit cells at $J_1 = 0.5$, $J_2 = 2.5$, $\delta=\Delta=0.4$, and $\mu=3.5$. Note that each subplot displays one left-localized edge state and one right-localized edge state.} } 
  \label{fig:A1}
\end{figure}

\begin{figure}[htpb]
    \centering
    \begin{subfigure}[b]{0.45\textwidth}
        \includegraphics[width=\textwidth]{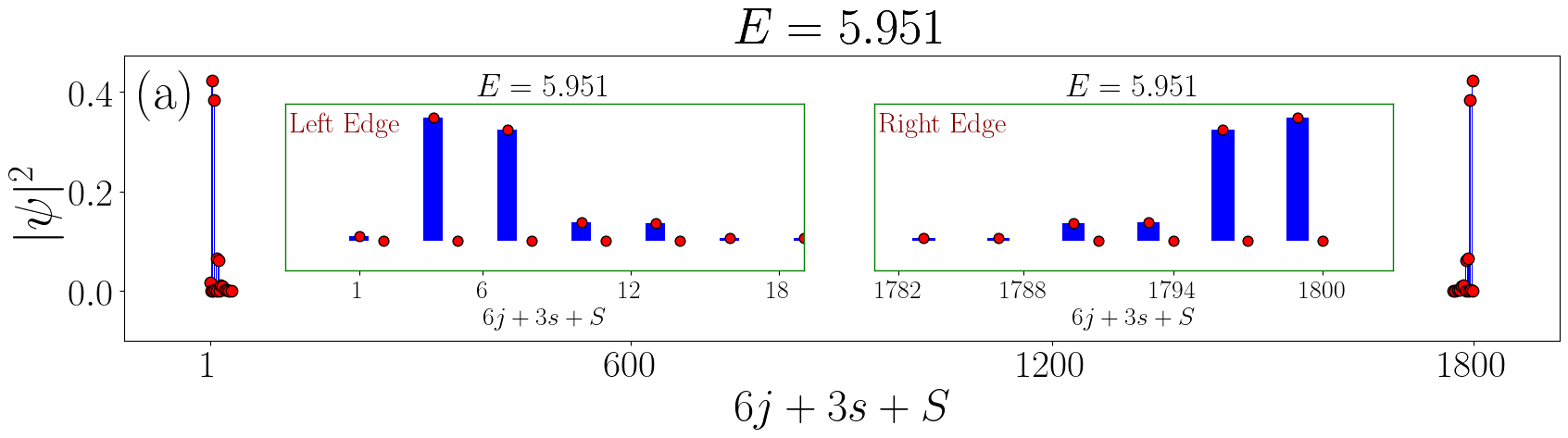} 
        \label{fig:fig1a}
    \end{subfigure}
    \begin{subfigure}[b]{0.45\textwidth}
        \includegraphics[width=\textwidth]{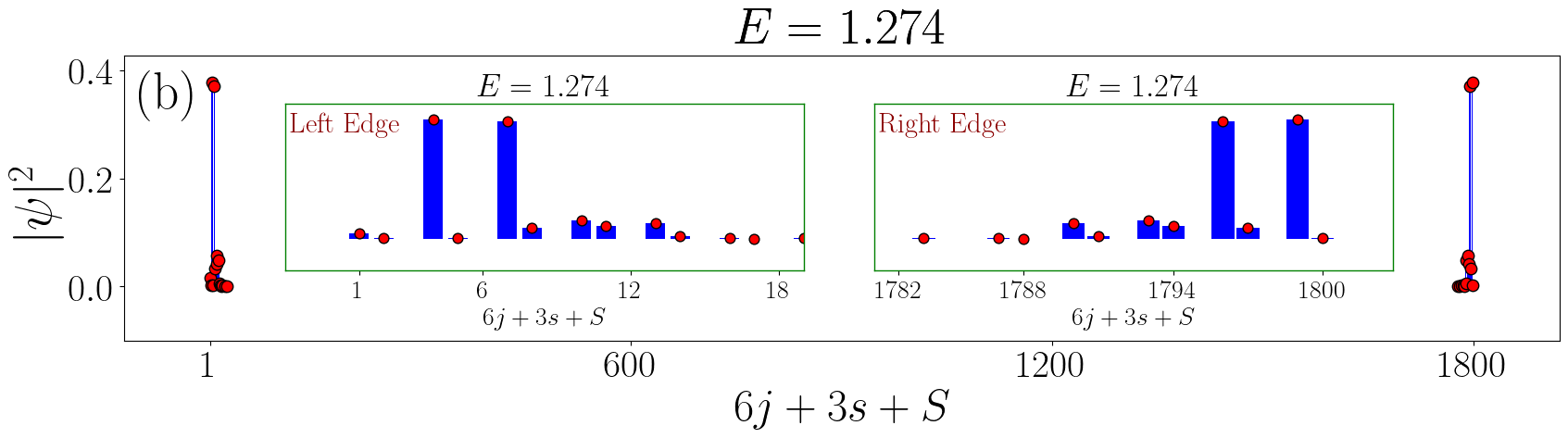}
        \label{fig:fig1b}
    \end{subfigure}
    \begin{subfigure}[b]{0.45\textwidth}
        \includegraphics[width=\textwidth]{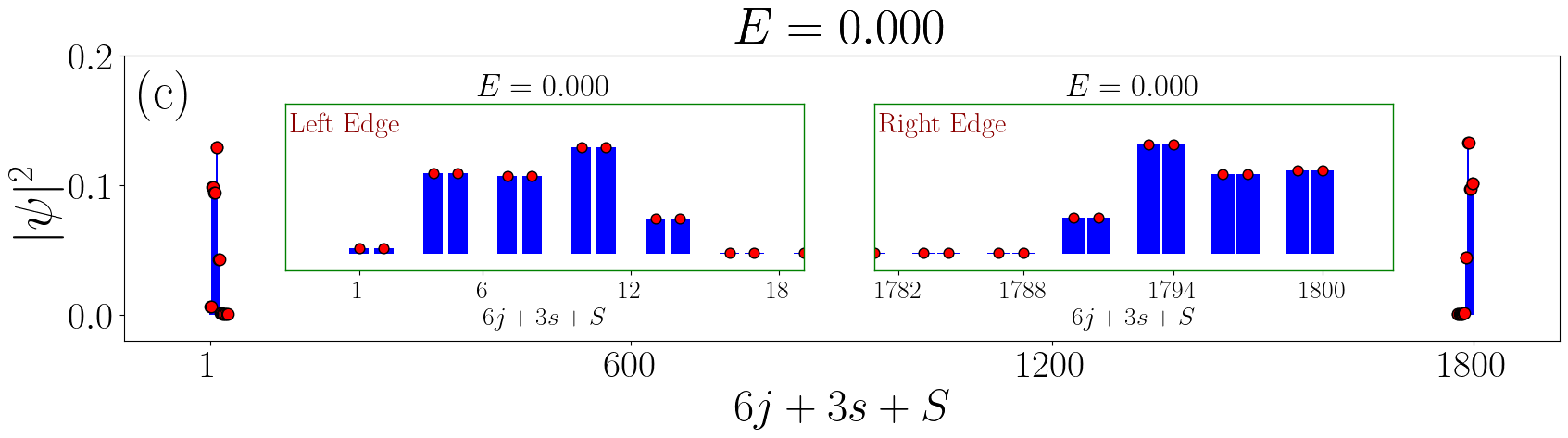} 
        \label{fig:fig1c}
    \end{subfigure}
    \begin{subfigure}[b]{0.45\textwidth}
        \includegraphics[width=\textwidth]{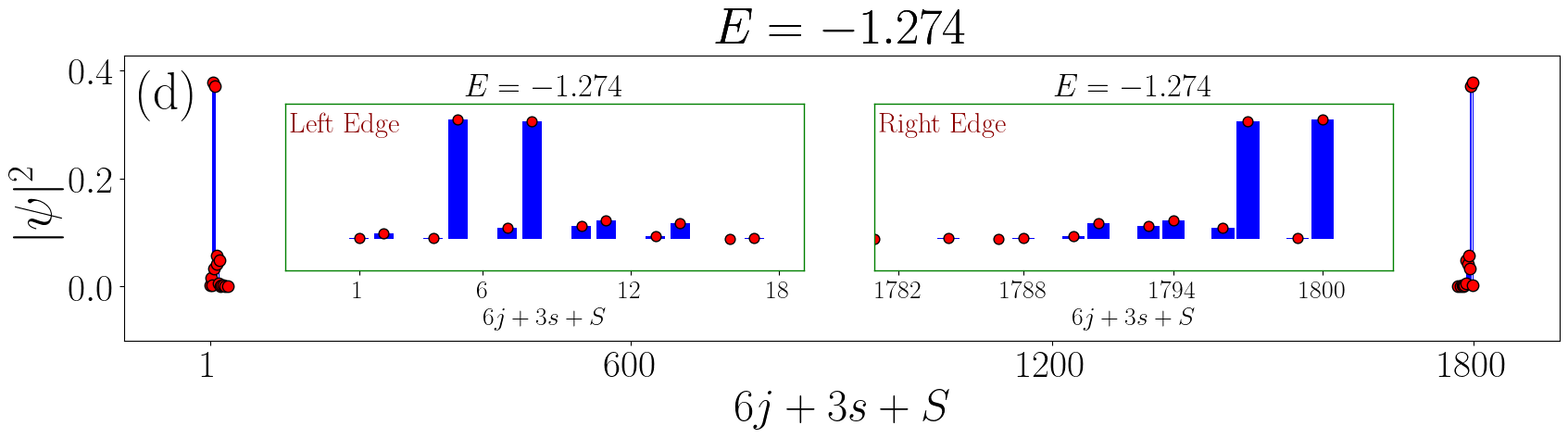}
        \label{fig:fig1d}
    \end{subfigure}
    \begin{subfigure}[b]{0.45\textwidth}
        \includegraphics[width=\textwidth]{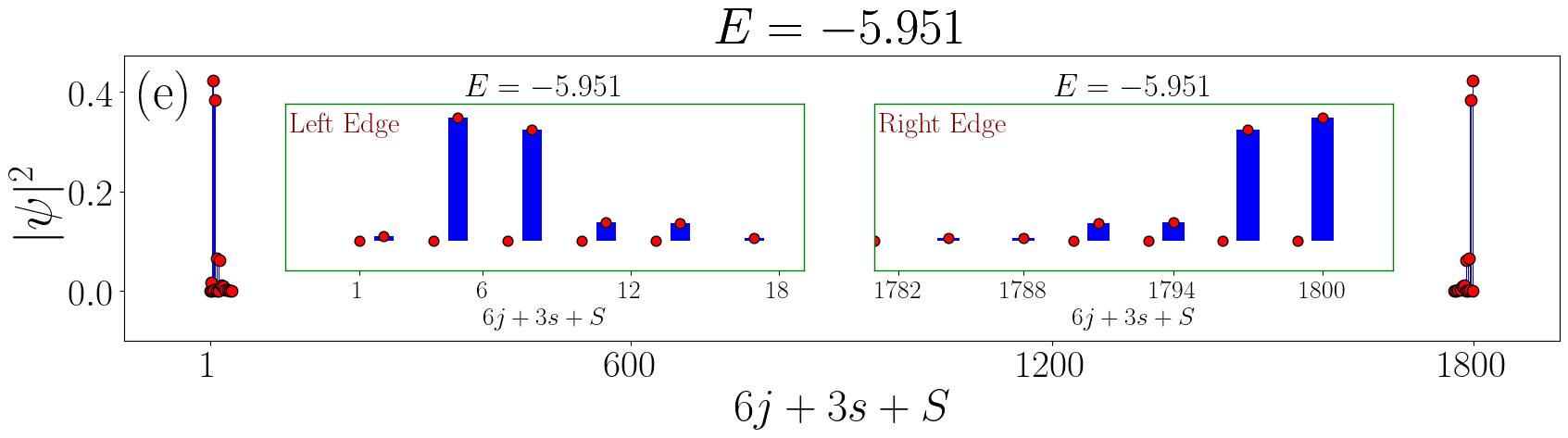}
        \label{fig:fig1e}
    \end{subfigure}
    \caption{\RB{Wave function profiles associated with the system's edge states for $N=300$ unit cells at $J_1 = 0.5$, $J_2 = 2.5$, $\delta=\Delta=0.4$, and $\mu=3.5$. Note that each subplot displays one left-localized edge state and one right-localized edge state. Each subplot contains inset zooms highlighting the peaks at both edges.}}  
  \label{fig:A2}
\end{figure}

\MG{In Figs.~\ref{fig:A1} and~\ref{fig:A2} we show the wavefunction profiles of the edge modes in our system at $N=30$ unit cells and $N=300$ unit cells, respectively. In both figures, the number of peaks for each edge mode remains unchanged despite using a different number of lattice sites. In particular, there are two dominant peaks at each of the finite-energy edge modes, and this number of peaks is invariant and consistent with the profiles shown in the main text for $N=100$ unit cells (Fig.~\ref{fig:figwf}), which confirms their system-size-independent nature.}

\clearpage
\bibliography{references}

@article{kitaev2001unpaired,
  title={Unpaired Majorana fermions in quantum wires},
  author={Kitaev, A Yu},
  journal={Physics-uspekhi},
  volume={44},
  number={10S},
  pages={131},
  year={2001},
  publisher={IOP Publishing},
  doi = {10.1070/1063-7869/44/10S/S29},
  URL ={https://iopscience.iop.org/article/10.1070/1063-7869/44/10S/S29/meta}
}

@article{PhysRevLett.86.268,
  title = {Non-Abelian Statistics of Half-Quantum Vortices in $\mathit{p}$-Wave Superconductors},
  author = {Ivanov, D. A.},
  journal = {Phys. Rev. Lett.},
  volume = {86},
  issue = {2},
  pages = {268--271},
  numpages = {0},
  year = {2001},
  month = {Jan},
  publisher = {American Physical Society},
  doi = {10.1103/PhysRevLett.86.268},
  url = {https://link.aps.org/doi/10.1103/PhysRevLett.86.268}
}

@article{PhysRevB.61.10267,
  title = {Paired states of fermions in two dimensions with breaking of parity and time-reversal symmetries and the fractional quantum Hall effect},
  author = {Read, N. and Green, Dmitry},
  journal = {Phys. Rev. B},
  volume = {61},
  issue = {15},
  pages = {10267--10297},
  numpages = {0},
  year = {2000},
  month = {Apr},
  publisher = {American Physical Society},
  doi = {10.1103/PhysRevB.61.10267},
  url = {https://link.aps.org/doi/10.1103/PhysRevB.61.10267}
}

@article{RevModPhys.80.1083,
  title = {Non-Abelian anyons and topological quantum computation},
  author = {Nayak, Chetan and Simon, Steven H. and Stern, Ady and Freedman, Michael and Das Sarma, Sankar},
  journal = {Rev. Mod. Phys.},
  volume = {80},
  issue = {3},
  pages = {1083--1159},
  numpages = {0},
  year = {2008},
  month = {Sep},
  publisher = {American Physical Society},
  doi = {10.1103/RevModPhys.80.1083},
  url = {https://link.aps.org/doi/10.1103/RevModPhys.80.1083}
}

@article{lahtinen2017short,
  title={A short introduction to topological quantum computation},
  author={Lahtinen, Ville and Pachos, Jiannis},
  journal={SciPost Physics},
  volume={3},
  number={3},
  pages={021},
  year={2017},
 doi={10.21468/SciPostPhys.3.3.021},
 url={https://scipost.org/SciPostPhys.3.3.021}
}

@article{freedman2003topological,
  title={Topological quantum computation},
  author={Freedman, Michael and Kitaev, Alexei and Larsen, Michael and Wang, Zhenghan},
  journal={Bulletin of the American Mathematical Society},
  volume={40},
  number={1},
  pages={31--38},
  year={2003},
  url={https://doi.org/10.1090/S0273-0979-02-00964-3}
}

@article{PhysRevB.82.020509,
  title = {Topological quantum computation away from the ground state using Majorana fermions},
  author = {Akhmerov, A. R.},
  journal = {Phys. Rev. B},
  volume = {82},
  issue = {2},
  pages = {020509},
  numpages = {3},
  year = {2010},
  month = {Jul},
  publisher = {American Physical Society},
  doi = {10.1103/PhysRevB.82.020509},
  url = {https://link.aps.org/doi/10.1103/PhysRevB.82.020509}
}

@article{sarma2015majorana,
  title={Majorana zero modes and topological quantum computation},
  author={Sarma, Sankar Das and Freedman, Michael and Nayak, Chetan},
  journal={npj Quantum Information},
  volume={1},
  number={1},
  pages={1--13},
  year={2015},
  publisher={Nature Publishing Group},
  url={https://www.nature.com/articles/npjqi20151}
}

@article{PhysRevB.97.205404,
  title = {Quantum computing with Majorana fermion codes},
  author = {Litinski, Daniel and von Oppen, Felix},
  journal = {Phys. Rev. B},
  volume = {97},
  issue = {20},
  pages = {205404},
  numpages = {27},
  year = {2018},
  month = {May},
  publisher = {American Physical Society},
  doi = {10.1103/PhysRevB.97.205404},
  url = {https://link.aps.org/doi/10.1103/PhysRevB.97.205404}
}

@article{PhysRevB.101.085401,
  title = {Measurement-only quantum computation with Floquet Majorana corner modes},
  author = {Bomantara, Raditya Weda and Gong, Jiangbin},
  journal = {Phys. Rev. B},
  volume = {101},
  issue = {8},
  pages = {085401},
  numpages = {18},
  year = {2020},
  month = {Feb},
  publisher = {American Physical Society},
  doi = {10.1103/PhysRevB.101.085401},
  url = {https://link.aps.org/doi/10.1103/PhysRevB.101.085401}
}

@article{PhysRevLett.105.177002,
  title = {Helical Liquids and Majorana Bound States in Quantum Wires},
  author = {Oreg, Yuval and Refael, Gil and von Oppen, Felix},
  journal = {Phys. Rev. Lett.},
  volume = {105},
  issue = {17},
  pages = {177002},
  numpages = {4},
  year = {2010},
  month = {Oct},
  publisher = {American Physical Society},
  doi = {10.1103/PhysRevLett.105.177002},
  url = {https://link.aps.org/doi/10.1103/PhysRevLett.105.177002}
}

@article{PhysRevLett.105.077001,
  title = {Majorana Fermions and a Topological Phase Transition in Semiconductor-Superconductor Heterostructures},
  author = {Lutchyn, Roman M. and Sau, Jay D. and Das Sarma, S.},
  journal = {Phys. Rev. Lett.},
  volume = {105},
  issue = {7},
  pages = {077001},
  numpages = {4},
  year = {2010},
  month = {Aug},
  publisher = {American Physical Society},
  doi = {10.1103/PhysRevLett.105.077001},
  url = {https://link.aps.org/doi/10.1103/PhysRevLett.105.077001}
}

@article{PhysRevB.84.201308,
  title = {Detecting a Majorana-fermion zero mode using a quantum dot},
  author = {Liu, Dong E. and Baranger, Harold U.},
  journal = {Phys. Rev. B},
  volume = {84},
  issue = {20},
  pages = {201308},
  numpages = {4},
  year = {2011},
  month = {Nov},
  publisher = {American Physical Society},
  doi = {10.1103/PhysRevB.84.201308},
  url = {https://link.aps.org/doi/10.1103/PhysRevB.84.201308}
}

@article{PhysRevB.88.020407,
  title = {Proposal for realizing Majorana fermions in chains of magnetic atoms on a superconductor},
  author = {Nadj-Perge, S. and Drozdov, I. K. and Bernevig, B. A. and Yazdani, Ali},
  journal = {Phys. Rev. B},
  volume = {88},
  issue = {2},
  pages = {020407},
  numpages = {5},
  year = {2013},
  month = {Jul},
  publisher = {American Physical Society},
  doi = {10.1103/PhysRevB.88.020407},
  url = {https://link.aps.org/doi/10.1103/PhysRevB.88.020407}
}

@article{das2012zero,
  title={Zero-bias peaks and splitting in an Al--InAs nanowire topological superconductor as a signature of Majorana fermions},
  author={Das, Anindya and Ronen, Yuval and Most, Yonatan and Oreg, Yuval and Heiblum, Moty and Shtrikman, Hadas},
  journal={Nature Physics},
  volume={8},
  number={12},
  pages={887--895},
  year={2012},
  publisher={Nature Publishing Group UK London},
  url= {https://doi.org/10.1038/nphys2479}
}

@article{10.1126/science.1222360,
author = {V. Mourik  and K. Zuo  and S. M. Frolov  and S. R. Plissard  and E. P. A. M. Bakkers  and L. P. Kouwenhoven },
title = {Signatures of Majorana Fermions in Hybrid Superconductor-Semiconductor Nanowire Devices},
journal = {Science},
volume = {336},
number = {6084},
pages = {1003-1007},
year = {2012},
doi = {10.1126/science.1222360},
url = {https://www.science.org/doi/abs/10.1126/science.122236}
}

@article{deng2012anomalous,
  title={Anomalous zero-bias conductance peak in a Nb--InSb nanowire--Nb hybrid device},
  author={Deng, MT and Yu, CL and Huang, GY and Larsson, Marcus and Caroff, Philippe and Xu, HQ},
  journal={Nano letters},
  volume={12},
  number={12},
  pages={6414--6419},
  year={2012},
  publisher={ACS Publications},
  url={https://pubs.acs.org/doi/10.1021/nl303758w}
}

@article{PhysRevB.87.241401,
  title = {Superconductor-nanowire devices from tunneling to the multichannel regime: Zero-bias oscillations and magnetoconductance crossover},
  author = {Churchill, H. O. H. and Fatemi, V. and Grove-Rasmussen, K. and Deng, M. T. and Caroff, P. and Xu, H. Q. and Marcus, C. M.},
  journal = {Phys. Rev. B},
  volume = {87},
  issue = {24},
  pages = {241401},
  numpages = {6},
  year = {2013},
  month = {Jun},
  publisher = {American Physical Society},
  doi = {10.1103/PhysRevB.87.241401},
  url = {https://link.aps.org/doi/10.1103/PhysRevB.87.241401}
}

@article{nadj2014observation,
  title={Observation of Majorana fermions in ferromagnetic atomic chains on a superconductor},
  author={Nadj-Perge, Stevan and Drozdov, Ilya K and Li, Jian and Chen, Hua and Jeon, Sangjun and Seo, Jungpil and MacDonald, Allan H and Bernevig, B Andrei and Yazdani, Ali},
  journal={Science},
  volume={346},
  number={6209},
  pages={602--607},
  year={2014},
  publisher={American Association for the Advancement of Science},
  url={https://www.science.org/doi/10.1126/science.1259327}
}

@article{manna2020signature,
  title={Signature of a pair of Majorana zero modes in superconducting gold surface states},
  author={Manna, Sujit and Wei, Peng and Xie, Yingming and Law, Kam Tuen and Lee, Patrick A and Moodera, Jagadeesh S},
  journal={Proceedings of the National Academy of Sciences},
  volume={117},
  number={16},
  pages={8775--8782},
  year={2020},
  publisher={National Academy of Sciences},
  url={https://doi.org/10.1073/pnas.1919753117}
}

@article{liu2024signatures,
  title={Signatures of hybridization of multiple Majorana zero modes in a vortex},
  author={Liu, Tengteng and Wan, Chun Yu and Yang, Hao and Zhao, Yujun and Xie, Bangjin and Zheng, Weiyan and Yi, Zhaoxia and Guan, Dandan and Wang, Shiyong and Zheng, Hao and others},
  journal={Nature},
  volume={633},
  number={8028},
  pages={71--76},
  year={2024},
  publisher={Nature Publishing Group UK London},
  url= {https://doi.org/10.1038/s41586-024-07857-4}
}

@article{PhysRevLett.109.267002,
  title = {Zero-Bias Peaks in the Tunneling Conductance of Spin-Orbit-Coupled Superconducting Wires with and without Majorana End-States},
  author = {Liu, Jie and Potter, Andrew C. and Law, K. T. and Lee, Patrick A.},
  journal = {Phys. Rev. Lett.},
  volume = {109},
  issue = {26},
  pages = {267002},
  numpages = {5},
  year = {2012},
  month = {Dec},
  publisher = {American Physical Society},
  doi = {10.1103/PhysRevLett.109.267002},
  url = {https://link.aps.org/doi/10.1103/PhysRevLett.109.267002}
}

@article{PhysRevB.96.075161,
  title = {Andreev bound states versus Majorana bound states in quantum dot-nanowire-superconductor hybrid structures: Trivial versus topological zero-bias conductance peaks},
  author = {Liu, Chun-Xiao and Sau, Jay D. and Stanescu, Tudor D. and Das Sarma, S.},
  journal = {Phys. Rev. B},
  volume = {96},
  issue = {7},
  pages = {075161},
  numpages = {29},
  year = {2017},
  month = {Aug},
  publisher = {American Physical Society},
  doi = {10.1103/PhysRevB.96.075161},
  url = {https://link.aps.org/doi/10.1103/PhysRevB.96.075161}
}

@article{PhysRevB.98.245407,
  title = {Zero-energy Andreev bound states from quantum dots in proximitized Rashba nanowires},
  author = {Reeg, Christopher and Dmytruk, Olesia and Chevallier, Denis and Loss, Daniel and Klinovaja, Jelena},
  journal = {Phys. Rev. B},
  volume = {98},
  issue = {24},
  pages = {245407},
  numpages = {12},
  year = {2018},
  month = {Dec},
  publisher = {American Physical Society},
  doi = {10.1103/PhysRevB.98.245407},
  url = {https://link.aps.org/doi/10.1103/PhysRevB.98.245407}
}

@article{PhysRevB.97.165302,
  title = {Two-terminal Reproducing topological properties heterostructures},
  author = {Moore, Christopher and Stanescu, Tudor D. and Tewari, Sumanta},
  journal = {Phys. Rev. B},
  volume = {97},
  issue = {16},
  pages = {165302},
  numpages = {14},
  year = {2018},
  month = {Apr},
  publisher = {American Physical Society},
  doi = {10.1103/PhysRevB.97.165302},
  url = {https://link.aps.org/doi/10.1103/PhysRevB.97.165302}
}

@article{PhysRevResearch.2.013377,
  title = {Physical mechanisms for zero-bias conductance peaks in Majorana nanowires},
  author = {Pan, Haining and Das Sarma, S.},
  journal = {Phys. Rev. Res.},
  volume = {2},
  issue = {1},
  pages = {013377},
  numpages = {33},
  year = {2020},
  month = {Mar},
  publisher = {American Physical Society},
  doi = {10.1103/PhysRevResearch.2.013377},
  url = {https://link.aps.org/doi/10.1103/PhysRevResearch.2.013377}
}

@article{PhysRevA.93.062130,
  title = {Majorana edge modes with gain and loss},
  author = {Yuce, C.},
  journal = {Phys. Rev. A},
  volume = {93},
  issue = {6},
  pages = {062130},
  numpages = {4},
  year = {2016},
  month = {Jun},
  publisher = {American Physical Society},
  doi = {10.1103/PhysRevA.93.062130},
  url = {https://link.aps.org/doi/10.1103/PhysRevA.93.062130}
}

@article{PhysRevB.110.094203,
  title = {Non-{H}ermitian {A}ubry-{A}ndr\'e-{H}arper model with short- and long-range $p$-wave pairing},
  author = {Gandhi, Shaina and Bandyopadhyay, Jayendra N.},
  journal = {Phys. Rev. B},
  volume = {110},
  issue = {9},
  pages = {094203},
  numpages = {14},
  year = {2024},
  month = {Sep},
  publisher = {American Physical Society},
  doi = {10.1103/PhysRevB.110.094203},
  url = {https://link.aps.org/doi/10.1103/PhysRevB.110.094203}
}

@article{PhysRevB.106.064505,
  title = {Non-{H}ermiticity-stabilized Majorana zero modes in semiconductor-superconductor nanowires},
  author = {Liu, Hongchao and Lu, Ming and Wu, Yijia and Liu, Jie and Xie, X. C.},
  journal = {Phys. Rev. B},
  volume = {106},
  issue = {6},
  pages = {064505},
  numpages = {6},
  year = {2022},
  month = {Aug},
  publisher = {American Physical Society},
  doi = {10.1103/PhysRevB.106.064505},
  url = {https://link.aps.org/doi/10.1103/PhysRevB.106.064505}
}

@article{PhysRevB.101.014306,
  title = {Non-Hermitian Floquet topological superconductors with multiple Majorana edge modes},
  author = {Zhou, Longwen},
  journal = {Phys. Rev. B},
  volume = {101},
  issue = {1},
  pages = {014306},
  numpages = {15},
  year = {2020},
  month = {Jan},
  publisher = {American Physical Society},
  doi = {10.1103/PhysRevB.101.014306},
  url = {https://link.aps.org/doi/10.1103/PhysRevB.101.014306}
}

@article{PhysRevLett.106.220402,
  title = {Majorana Fermions in Equilibrium and in Driven Cold-Atom Quantum Wires},
  author = {Jiang, Liang and Kitagawa, Takuya and Alicea, Jason and Akhmerov, A. R. and Pekker, David and Refael, Gil and Cirac, J. Ignacio and Demler, Eugene and Lukin, Mikhail D. and Zoller, Peter},
  journal = {Phys. Rev. Lett.},
  volume = {106},
  issue = {22},
  pages = {220402},
  numpages = {4},
  year = {2011},
  month = {Jun},
  publisher = {American Physical Society},
  doi = {10.1103/PhysRevLett.106.220402},
  url = {https://link.aps.org/doi/10.1103/PhysRevLett.106.220402}
}

@article{PhysRevB.87.201109,
  title = {Generating many Majorana modes via periodic driving: A superconductor model},
  author = {Tong, Qing-Jun and An, Jun-Hong and Gong, Jiangbin and Luo, Hong-Gang and Oh, C. H.},
  journal = {Phys. Rev. B},
  volume = {87},
  issue = {20},
  pages = {201109},
  numpages = {5},
  year = {2013},
  month = {May},
  publisher = {American Physical Society},
  doi = {10.1103/PhysRevB.87.201109},
  url = {https://link.aps.org/doi/10.1103/PhysRevB.87.201109}
}

@article{PhysRevB.108.245153,
  title = {Generalized {M}ajorana edge modes in a number-conserving periodically driven $p$-wave superconductor},
  author = {Bomantara, Raditya Weda},
  journal = {Phys. Rev. B},
  volume = {108},
  issue = {24},
  pages = {245153},
  numpages = {14},
  year = {2023},
  month = {Dec},
  publisher = {American Physical Society},
  doi = {10.1103/PhysRevB.108.245153},
  url = {https://link.aps.org/doi/10.1103/PhysRevB.108.245153}
}

@Article{10.21468/SciPostPhys.13.2.015,
	title={{$q$th-root non-Hermitian {F}loquet topological insulators}},
	author={Longwen Zhou and Raditya Weda Bomantara and Shenlin Wu},
	journal={SciPost Phys.},
	volume={13},
	pages={015},
	year={2022},
	publisher={SciPost},
	doi={10.21468/SciPostPhys.13.2.015},
	url={https://scipost.org/10.21468/SciPostPhys.13.2.015},
}

@article{zhou2022generating,
  title={Generating many Majorana corner modes and multiple phase transitions in Floquet second-order topological superconductors},
  author={Zhou, Longwen},
  journal={Symmetry},
  volume={14},
  number={12},
  pages={2546},
  year={2022},
  publisher={MDPI},
  url={https://doi.org/10.3390/sym14122546}
}

@article{wu2023floquet,
  title={Floquet topological superconductors with many Majorana edge modes: Topological invariants, entanglement spectrum and bulk-edge correspondence},
  author={Wu, Hailing and Wu, Shenlin and Zhou, Longwen},
  journal={New Journal of Physics},
  volume={25},
  number={8},
  pages={083042},
  year={2023},
  publisher={IOP Publishing},
  url={https://iopscience.iop.org/article/10.1088/1367-2630/acf0e3}
}

@article{bomantara2020combating,
  title={Combating quasiparticle poisoning with multiple Majorana fermions in a periodically-driven quantum wire},
  author={Bomantara, Raditya Weda and Gong, Jiangbin},
  journal={Journal of Physics: Condensed Matter},
  volume={32},
  number={43},
  pages={435301},
  year={2020},
  publisher={IOP Publishing},
  doi ={10.1088/1361-648X/aba291},
  url={https://iopscience.iop.org/article/10.1088/1361-648X/aba291}
}

@article{PhysRevB.106.L060305,
  title = {Square-root Floquet topological phases and time crystals},
  author = {Bomantara, Raditya Weda},
  journal = {Phys. Rev. B},
  volume = {106},
  issue = {6},
  pages = {L060305},
  numpages = {6},
  year = {2022},
  month = {Aug},
  publisher = {American Physical Society},
  doi = {10.1103/PhysRevB.106.L060305},
  url = {https://link.aps.org/doi/10.1103/PhysRevB.106.L060305}
}

@article{PhysRevB.22.2099,
  title = {Soliton excitations in polyacetylene},
  author = {Su, W. P. and Schrieffer, J. R. and Heeger, A. J.},
  journal = {Phys. Rev. B},
  volume = {22},
  issue = {4},
  pages = {2099--2111},
  numpages = {0},
  year = {1980},
  month = {Aug},
  publisher = {American Physical Society},
  doi = {10.1103/PhysRevB.22.2099},
  url = {https://link.aps.org/doi/10.1103/PhysRevB.22.2099}
}

@article{ten2025observation,
  title={Observation of edge and bulk states in a three-site Kitaev chain},
  author={Ten Haaf, Sebastiaan LD and Zhang, Yining and Wang, Qingzhen and Bordin, Alberto and Liu, Chun-Xiao and Kulesh, Ivan and Sietses, Vincent PM and Prosko, Christian G and Xiao, Di and Thomas, Candice and others},
  journal={Nature},
  pages={1--6},
  year={2025},
  publisher={Nature Publishing Group UK London},
  url={https://doi.org/10.1038/s41586-025-08892-5}
}

@article{bordin2025enhanced,
  title={Enhanced Majorana stability in a three-site Kitaev chain},
  author={Bordin, Alberto and Liu, Chun-Xiao and Dvir, Tom and Zatelli, Francesco and Ten Haaf, Sebastiaan LD and van Driel, David and Wang, Guanzhong and Van Loo, Nick and Zhang, Yining and Wolff, Jan Cornelis and others},
  journal={Nature Nanotechnology},
  pages={1--6},
  year={2025},
  publisher={Nature Publishing Group UK London},
  url={https://doi.org/10.1038/s41565-025-01894-4}
}

@article{ten2024two,
  title={A two-site Kitaev chain in a two-dimensional electron gas},
  author={Ten Haaf, Sebastiaan LD and Wang, Qingzhen and Bozkurt, A Mert and Liu, Chun-Xiao and Kulesh, Ivan and Kim, Philip and Xiao, Di and Thomas, Candice and Manfra, Michael J and Dvir, Tom and others},
  journal={Nature},
  volume={630},
  number={8016},
  pages={329--334},
  year={2024},
  publisher={Nature Publishing Group UK London},
  url= {https://doi.org/10.1038/s41586-024-07434-9}
}

@article{dvir2023realization,
  title={Realization of a minimal Kitaev chain in coupled quantum dots},
  author={Dvir, Tom and Wang, Guanzhong and van Loo, Nick and Liu, Chun-Xiao and Mazur, Grzegorz P and Bordin, Alberto and Ten Haaf, Sebastiaan LD and Wang, Ji-Yin and van Driel, David and Zatelli, Francesco and others},
  journal={Nature},
  volume={614},
  number={7948},
  pages={445--450},
  year={2023},
  publisher={Nature Publishing Group UK London},
  url={https://doi.org/10.1038/s41586-022-05585-1}
}

@article{iizuka2023experimental,
  title={Experimental demonstration of position-controllable topological interface states in high-frequency Kitaev topological integrated circuits},
  author={Iizuka, Tetsuya and Yuan, Haochen and Mita, Yoshio and Higo, Akio and Yasunaga, Shun and Ezawa, Motohiko},
  journal={Communications Physics},
  volume={6},
  number={1},
  pages={279},
  year={2023},
  publisher={Nature Publishing Group UK London},
 doi ={10.1038/s42005-023-01404-9},
 URL = {https://www.nature.com/articles/s42005-023-01404-9}
}

@article{allein2023strain,
  title={Strain topological metamaterials and revealing hidden topology in higher-order coordinates},
  author={Allein, Florian and Anastasiadis, Adamantios and Chaunsali, Rajesh and Frankel, Ian and Boechler, Nicholas and Diakonos, Fotios K and Theocharis, Georgios},
  journal={Nature Communications},
  volume={14},
  number={1},
  pages={6633},
  year={2023},
  publisher={Nature Publishing Group UK London},
  doi={10.1038/s41467-023-42321-3},
  URL={https://doi.org/10.1038/s41467-023-42321-3}
}

@article{slim2024optomechanical,
  title={Optomechanical realization of the bosonic Kitaev chain},
  author={Slim, Jesse J and Wanjura, Clara C and Brunelli, Matteo and Del Pino, Javier and Nunnenkamp, Andreas and Verhagen, Ewold},
  journal={Nature},
  volume={627},
  number={8005},
  pages={767--771},
  year={2024},
  publisher={Nature Publishing Group UK London},
  doi={10.1038/s41586-024-07174-w},
  URL={https://doi.org/10.1038/s41586-024-07174-w}
}

@article{busnaina2024quantum,
  title={Quantum simulation of the bosonic Kitaev chain},
  author={Busnaina, Jamal H and Shi, Zheng and McDonald, Alexander and Dubyna, Dmytro and Nsanzineza, Ibrahim and Hung, Jimmy SC and Chang, CW Sandbo and Clerk, Aashish A and Wilson, Christopher M},
  journal={Nature Communications},
  volume={15},
  number={1},
  pages={3065},
  year={2024},
  publisher={Nature Publishing Group UK London},
  url={https://www.nature.com/articles/s41467-024-47186-8}
}

@article{5rtw-ml8b,
  title = {Edge states and persistent current in a $\mathcal{PT}$-symmetric extended Su-Schrieffer-Heeger model with generic boundary conditions},
  author = {Ghosh, Supriyo and Ghosh, Pijush K. and Sil, Shreekantha},
  journal = {Phys. Rev. B},
  volume = {111},
  issue = {24},
  pages = {245428},
  numpages = {18},
  year = {2025},
  month = {Jun},
  publisher = {American Physical Society},
  doi = {10.1103/5rtw-ml8b},
  url = {https://link.aps.org/doi/10.1103/5rtw-ml8b}
}

@article{du2024one,
  title={One-dimensional extended Su--Schrieffer--Heeger models as descendants of a two-dimensional topological model},
  author={Du, Tao and Li, Yue-Xun and Lu, He-Lin and Zhang, Hui},
  journal={New Journal of Physics},
  volume={26},
  number={2},
  pages={023044},
  year={2024},
  publisher={IOP Publishing},
  doi = {10.1088/1367-2630/ad2896},
  url = {https://iopscience.iop.org/article/10.1088/1367-2630/ad2896/meta}
}

@article{PhysRevB.109.035114,
  title = {Fate of high winding number topological phases in the disordered extended Su-Schrieffer-Heeger model},
  author = {Cinnirella, Emmanuele G. and Nava, Andrea and Campagnano, Gabriele and Giuliano, Domenico},
  journal = {Phys. Rev. B},
  volume = {109},
  issue = {3},
  pages = {035114},
  numpages = {17},
  year = {2024},
  month = {Jan},
  publisher = {American Physical Society},
  doi = {10.1103/PhysRevB.109.035114},
  url = {https://link.aps.org/doi/10.1103/PhysRevB.109.035114}
}

@article{lee2022winding,
  title={Winding number and Zak phase in multi-band SSH models},
  author={Lee, Chen-Shen and Io, Iao-Fai and Kao, Hsien-chung},
  journal={Chinese Journal of Physics},
  volume={78},
  pages={96--110},
  year={2022},
  publisher={Elsevier},
  url={https://doi.org/10.1016/j.cjph.2022.05.007}
}

@article{PhysRevB.106.085109,
  title = {Bulk-edge correspondence in the trimer Su-Schrieffer-Heeger model},
  author = {Anastasiadis, Adamantios and Styliaris, Georgios and Chaunsali, Rajesh and Theocharis, Georgios and Diakonos, Fotios K.},
  journal = {Phys. Rev. B},
  volume = {106},
  issue = {8},
  pages = {085109},
  numpages = {14},
  year = {2022},
  month = {Aug},
  publisher = {American Physical Society},
  doi = {10.1103/PhysRevB.106.085109},
  url = {https://link.aps.org/doi/10.1103/PhysRevB.106.085109}
}

@article{ghuneim2024topological,
  title={Topological phases of tight-binding trimer lattice in the BDI symmetry class},
  author={Ghuneim, Mohammad and Bomantara, Raditya Weda},
  journal={Journal of Physics: Condensed Matter},
  volume={36},
  number={49},
  pages={495402},
  year={2024},
  publisher={IOP Publishing},
  url={https://iopscience.iop.org/article/10.1088/1361-648X/ad744c}
}

@article{PhysRevB.111.195424,
  title = {Anomalous topological edge modes in a periodically driven trimer lattice},
  author = {Ghuneim, Mohammad and Bomantara, Raditya Weda},
  journal = {Phys. Rev. B},
  volume = {111},
  issue = {19},
  pages = {195424},
  numpages = {15},
  year = {2025},
  month = {May},
  publisher = {American Physical Society},
  doi = {10.1103/PhysRevB.111.195424},
  url = {https://link.aps.org/doi/10.1103/PhysRevB.111.195424}
}

@article{PhysRevLett.113.156402,
  title = {Kitaev Chains with Long-Range Pairing},
  author = {Vodola, Davide and Lepori, Luca and Ercolessi, Elisa and Gorshkov, Alexey V. and Pupillo, Guido},
  journal = {Phys. Rev. Lett.},
  volume = {113},
  issue = {15},
  pages = {156402},
  numpages = {5},
  year = {2014},
  month = {Oct},
  publisher = {American Physical Society},
  doi = {10.1103/PhysRevLett.113.156402},
  url = {https://link.aps.org/doi/10.1103/PhysRevLett.113.156402}
}

@article{PhysRevB.95.195160,
  title = {Extended Kitaev chain with longer-range hopping and pairing},
  author = {Alecce, Antonio and Dell'Anna, Luca},
  journal = {Phys. Rev. B},
  volume = {95},
  issue = {19},
  pages = {195160},
  numpages = {14},
  year = {2017},
  month = {May},
  publisher = {American Physical Society},
  doi = {10.1103/PhysRevB.95.195160},
  url = {https://link.aps.org/doi/10.1103/PhysRevB.95.195160}
}

@article{PhysRevLett.118.267002,
  title = {Topological Quantum Liquids with Long-Range Couplings},
  author = {Patrick, Kristian and Neupert, Titus and Pachos, Jiannis K.},
  journal = {Phys. Rev. Lett.},
  volume = {118},
  issue = {26},
  pages = {267002},
  numpages = {5},
  year = {2017},
  month = {Jun},
  publisher = {American Physical Society},
  doi = {10.1103/PhysRevLett.118.267002},
  url = {https://link.aps.org/doi/10.1103/PhysRevLett.118.267002}
}

@article{mishra2020disordered,
  title={Disordered Kitaev chain with long-range pairing: Loschmidt echo revivals and dynamical phase transitions},
  author={Mishra, Utkarsh and Jafari, R and Akbari, Alireza},
  journal={Journal of Physics A: Mathematical and Theoretical},
  volume={53},
  number={37},
  pages={375301},
  year={2020},
  publisher={IOP Publishing},
  doi={10.1088/1751-8121/ab97de},
  url={https://iopscience.iop.org/article/10.1088/1751-8121/ab97de}
}

@article{PhysRevB.111.155149,
  title = {Phase diagram of the disordered Kitaev chain with long-range pairing connected to external baths},
  author = {Cinnirella, Emmanuele G. and Nava, Andrea and Campagnano, Gabriele and Giuliano, Domenico},
  journal = {Phys. Rev. B},
  volume = {111},
  issue = {15},
  pages = {155149},
  numpages = {14},
  year = {2025},
  month = {Apr},
  publisher = {American Physical Society},
  doi = {10.1103/PhysRevB.111.155149},
  url = {https://link.aps.org/doi/10.1103/PhysRevB.111.155149}
}

@article{PhysRevB.90.014505,
  title = {Fermion fractionalization to {M}ajorana fermions in a dimerized {K}itaev superconductor},
  author = {Wakatsuki, Ryohei and Ezawa, Motohiko and Tanaka, Yukio and Nagaosa, Naoto},
  journal = {Phys. Rev. B},
  volume = {90},
  issue = {1},
  pages = {014505},
  numpages = {11},
  year = {2014},
  month = {Jul},
  publisher = {American Physical Society},
  doi = {10.1103/PhysRevB.90.014505},
  url = {https://link.aps.org/doi/10.1103/PhysRevB.90.014505}
}

@article{yang2025multiple,
  title={Multiple localization transitions in a dimerized Kitaev superconductor model with random-dimer disorder},
  author={Yang, Yuan and Ding, Zhengxuan and Li, Xiaobing},
  journal={Annals of Physics},
  pages={170070},
  year={2025},
  publisher={Elsevier},
  doi={10.1016/j.aop.2025.170070},
  url={https://doi.org/10.1016/j.aop.2025.170070}
}

@article{PhysRevLett.110.146404,
  title = {Majorana Fermions in Superconducting 1D Systems Having Periodic, Quasiperiodic, and Disordered Potentials},
  author = {DeGottardi, Wade and Sen, Diptiman and Vishveshwara, Smitha},
  journal = {Phys. Rev. Lett.},
  volume = {110},
  issue = {14},
  pages = {146404},
  numpages = {5},
  year = {2013},
  month = {Apr},
  publisher = {American Physical Society},
  doi = {10.1103/PhysRevLett.110.146404},
  url = {https://link.aps.org/doi/10.1103/PhysRevLett.110.146404}
}

@article{lima2025interplay,
  title={Interplay between topology and interactions in superconducting chains},
  author={Lima, ACP and Figueira, MS and Continentino, Mucio A},
  journal={Journal of Physics: Condensed Matter},
  volume={37},
  number={35},
  pages={355601},
  year={2025},
  publisher={IOP Publishing},
  doi={10.1088/1361-648X/adf680},
  url={https://iopscience.iop.org/article/10.1088/1361-648X/adf680/meta}
}

@article{doi:10.1126/science.1222360,
author = {V. Mourik  and K. Zuo  and S. M. Frolov  and S. R. Plissard  and E. P. A. M. Bakkers  and L. P. Kouwenhoven },
title = {Signatures of Majorana Fermions in Hybrid Superconductor-Semiconductor Nanowire Devices},
journal = {Science},
volume = {336},
number = {6084},
pages = {1003-1007},
year = {2012},
doi = {10.1126/science.1222360},
URL = {https://www.science.org/doi/abs/10.1126/science.1222360}
}

@Article{10.21468/SciPostPhys.7.5.061,
	title={{Reproducing topological properties with quasi-Majorana states}},
	author={Adriaan Vuik and Bas Nijholt and Anton R. Akhmerov and Michael Wimmer},
	journal={SciPost Phys.},
	volume={7},
	pages={061},
	year={2019},
	publisher={SciPost},
	doi={10.21468/SciPostPhys.7.5.061},
	url={https://scipost.org/10.21468/SciPostPhys.7.5.061},
}

@article{PhysRevLett.109.186802,
  title = {Zero-Bias Anomaly in a Nanowire Quantum Dot Coupled to Superconductors},
  author = {Lee, Eduardo J. H. and Jiang, Xiaocheng and Aguado, Ram\'on and Katsaros, Georgios and Lieber, Charles M. and De Franceschi, Silvano},
  journal = {Phys. Rev. Lett.},
  volume = {109},
  issue = {18},
  pages = {186802},
  numpages = {5},
  year = {2012},
  month = {Oct},
  publisher = {American Physical Society},
  doi = {10.1103/PhysRevLett.109.186802},
  url = {https://link.aps.org/doi/10.1103/PhysRevLett.109.186802}
}

@article{PhysRevB.91.081405,
  title = {Probing Majorana physics in quantum-dot shot-noise experiments},
  author = {Liu, Dong E. and Cheng, Meng and Lutchyn, Roman M.},
  journal = {Phys. Rev. B},
  volume = {91},
  issue = {8},
  pages = {081405},
  numpages = {5},
  year = {2015},
  month = {Feb},
  publisher = {American Physical Society},
  doi = {10.1103/PhysRevB.91.081405},
  url = {https://link.aps.org/doi/10.1103/PhysRevB.91.081405}
}

@article{deng2016majorana,
  title={Majorana bound state in a coupled quantum-dot hybrid-nanowire system},
  author={Deng, MT and Vaitiek{\.e}nas, S and Hansen, Esben Bork and Danon, Jeroen and Leijnse, M and Flensberg, Karsten and Nyg{\aa}rd, Jesper and Krogstrup, P and Marcus, Charles M},
  journal={Science},
  volume={354},
  number={6319},
  pages={1557--1562},
  year={2016},
  publisher={American Association for the Advancement of Science},
  url={https://www.science.org/doi/10.1126/science.aaf3961}
}

@article{PhysRevB.104.205131,
  title = {Defective {M}ajorana zero modes in a non-{H}ermitian {K}itaev chain},
  author = {Zhao, Xiao-Ming and Guo, Cui-Xian and Kou, Su-Peng and Zhuang, Lin and Liu, Wu-Ming},
  journal = {Phys. Rev. B},
  volume = {104},
  issue = {20},
  pages = {205131},
  numpages = {9},
  year = {2021},
  month = {Nov},
  publisher = {American Physical Society},
  doi = {10.1103/PhysRevB.104.205131},
  url = {https://link.aps.org/doi/10.1103/PhysRevB.104.205131}
}

@article{rahul2022topological,
  title={Topological quantum criticality in non-Hermitian extended Kitaev chain},
  author={Rahul, S and Sarkar, Sujit},
  journal={Scientific Reports},
  volume={12},
  number={1},
  pages={6993},
  year={2022},
  publisher={Nature Publishing Group UK London},
  doi={10.1038/s41598-022-11126-7},
  url={https://doi.org/10.1038/s41598-022-11126-7}
}

@article{PhysRevB.108.014204,
  title = {Topological triple phase transition in non-{H}ermitian quasicrystals with complex asymmetric hopping},
  author = {Gandhi, Shaina and Bandyopadhyay, Jayendra N.},
  journal = {Phys. Rev. B},
  volume = {108},
  issue = {1},
  pages = {014204},
  numpages = {7},
  year = {2023},
  month = {Jul},
  publisher = {American Physical Society},
  doi = {10.1103/PhysRevB.108.014204},
  url = {https://link.aps.org/doi/10.1103/PhysRevB.108.014204}
}

@article{PhysRevB.111.174210,
  title = {Superconducting $p$-wave pairing effects on one-dimensional non-Hermitian quasicrystals with power law hopping},
  author = {Gandhi, Shaina and Bandyopadhyay, Jayendra N.},
  journal = {Phys. Rev. B},
  volume = {111},
  issue = {17},
  pages = {174210},
  numpages = {12},
  year = {2025},
  month = {May},
  publisher = {American Physical Society},
  doi = {10.1103/PhysRevB.111.174210},
  url = {https://link.aps.org/doi/10.1103/PhysRevB.111.174210}
}

@article{song2024band,
  title={Band structure of the one-dimensional tetrameric Kitaev chain induced by imaginary potentials},
  author={Song, Wen-Dan and Li, Jia-Rui and Song, Yan-Song and Chen, Jun-Jie and Gong, Wei-Jiang},
  journal={Annals of Physics},
  volume={467},
  pages={169693},
  year={2024},
  publisher={Elsevier},
  url={https://doi.org/10.1016/j.aop.2024.169693},
}

@article{LIU2025170073,
  title = {Non-Hermiticity enhanced topological immunity of one-dimensional p-wave superconducting chain},
  author = {Min Liu and Yue Zhang and Rui Tian and Xiayao He and Tianhao Wu and Maksims Arzamasovs and Shuai Li and Bo Liu},
  journal = {Annals of Physics},
  volume = {479},
  pages = {170073},
  year = {2025},
  issn = {0003-4916},
  doi = {https://doi.org/10.1016/j.aop.2025.170073},
  url = {https://www.sciencedirect.com/science/article/pii/S000349162500154X}
}

@article{yb8n-2l9v,
  title = {Bulk-boundary correspondence and infernal points in the non-Hermitian trimer Su-Schrieffer-Heeger model},
  author = {Tang, K. J. and Cui, H. T. and Mu, H. F.},
  journal = {Phys. Rev. A},
  volume = {112},
  issue = {6},
  pages = {062204},
  numpages = {8},
  year = {2025},
  month = {Dec},
  publisher = {American Physical Society},
  doi = {10.1103/yb8n-2l9v},
  url = {https://link.aps.org/doi/10.1103/yb8n-2l9v}
}

@article{zhou2025topological,
  title={Topological edge states at Floquet quantum criticality},
  author={Zhou, Longwen and Gong, Jiangbin and Yu, Xue-Jia},
  journal={Communications Physics},
  volume={8},
  number={1},
  pages={1--8},
  year={2025},
  publisher={Nature Publishing Group},
  url={https://www.nature.com/articles/s42005-025-02137-7}
}

@article{PhysRevLett.129.254301,
  title = {Observation of $\ensuremath{\pi}/2$ Modes in an Acoustic Floquet System},
  author = {Cheng, Zheyu and Bomantara, Raditya Weda and Xue, Haoran and Zhu, Weiwei and Gong, Jiangbin and Zhang, Baile},
  journal = {Phys. Rev. Lett.},
  volume = {129},
  issue = {25},
  pages = {254301},
  numpages = {7},
  year = {2022},
  month = {Dec},
  publisher = {American Physical Society},
  doi = {10.1103/PhysRevLett.129.254301},
  url = {https://link.aps.org/doi/10.1103/PhysRevLett.129.254301}
}

@article{PhysRevB.104.L121410,
  title = {${Z}_{4}$ parafermion $\ifmmode\pm\else\textpm\fi{}\ensuremath{\pi}/2$ modes in an interacting periodically driven superconducting chain},
  author = {Bomantara, Raditya Weda},
  journal = {Phys. Rev. B},
  volume = {104},
  issue = {12},
  pages = {L121410},
  numpages = {6},
  year = {2021},
  month = {Sep},
  publisher = {American Physical Society},
  doi = {10.1103/PhysRevB.104.L121410},
  url = {https://link.aps.org/doi/10.1103/PhysRevB.104.L121410}
}

@article{PhysRevB.99.045441,
  title = {Coupled-wire construction of static and Floquet second-order topological insulators},
  author = {Bomantara, Raditya Weda and Zhou, Longwen and Pan, Jiaxin and Gong, Jiangbin},
  journal = {Phys. Rev. B},
  volume = {99},
  issue = {4},
  pages = {045441},
  numpages = {13},
  year = {2019},
  month = {Jan},
  publisher = {American Physical Society},
  doi = {10.1103/PhysRevB.99.045441},
  url = {https://link.aps.org/doi/10.1103/PhysRevB.99.045441}
}

@article{PhysRevE.93.022209,
  title = {Floquet topological semimetal phases of an extended kicked Harper model},
  author = {Bomantara, Raditya Weda and Raghava, Gudapati Naresh and Zhou, Longwen and Gong, Jiangbin},
  journal = {Phys. Rev. E},
  volume = {93},
  issue = {2},
  pages = {022209},
  numpages = {12},
  year = {2016},
  month = {Feb},
  publisher = {American Physical Society},
  doi = {10.1103/PhysRevE.93.022209},
  url = {https://link.aps.org/doi/10.1103/PhysRevE.93.022209}
}

@article{PhysRevResearch.2.033495,
  title = {Time-induced second-order topological superconductors},
  author = {Bomantara, Raditya Weda},
  journal = {Phys. Rev. Res.},
  volume = {2},
  issue = {3},
  pages = {033495},
  numpages = {16},
  year = {2020},
  month = {Sep},
  publisher = {American Physical Society},
  doi = {10.1103/PhysRevResearch.2.033495},
  url = {https://link.aps.org/doi/10.1103/PhysRevResearch.2.033495}
}

@article{PhysRevB.106.195122,
  title = {Symmetry-protected topological corner modes in a periodically driven interacting spin lattice},
  author = {Koor, Kelvin and Bomantara, Raditya Weda and Kwek, Leong Chuan},
  journal = {Phys. Rev. B},
  volume = {106},
  issue = {19},
  pages = {195122},
  numpages = {15},
  year = {2022},
  month = {Nov},
  publisher = {American Physical Society},
  doi = {10.1103/PhysRevB.106.195122},
  url = {https://link.aps.org/doi/10.1103/PhysRevB.106.195122}
}

@article{jangjan2020floquet,
  title={Floquet engineering of topological metal states and hybridization of edge states with bulk states in dimerized two-leg ladders},
  author={Jangjan, Milad and Hosseini, Mir Vahid},
  journal={Scientific Reports},
  volume={10},
  number={1},
  pages={14256},
  year={2020},
  publisher={Nature Publishing Group UK London},
  doi = {10.1038/s41598-020-71196-3},
  url = {https://www.nature.com/articles/s41598-020-71196-3}
}

@article{PhysRevB.111.195117,
  title = {Interaction-induced phase transitions at topological quantum criticality of an extended Su-Schrieffer-Heeger model},
  author = {Zhou, Xiaofan and Jia, Suotang and Pan, Jian-Song},
  journal = {Phys. Rev. B},
  volume = {111},
  issue = {19},
  pages = {195117},
  numpages = {8},
  year = {2025},
  month = {May},
  publisher = {American Physical Society},
  doi = {10.1103/PhysRevB.111.195117},
  url = {https://link.aps.org/doi/10.1103/PhysRevB.111.195117}
}

@article{PhysRevB.107.L060304,
  title = {Out-of-equilibrium Majorana zero modes in interacting Kitaev chains},
  author = {Pandey, Bradraj and Mohanta, Narayan and Dagotto, Elbio},
  journal = {Phys. Rev. B},
  volume = {107},
  issue = {6},
  pages = {L060304},
  numpages = {5},
  year = {2023},
  month = {Feb},
  publisher = {American Physical Society},
  doi = {10.1103/PhysRevB.107.L060304},
  url = {https://link.aps.org/doi/10.1103/PhysRevB.107.L060304}
}

\end{document}